\documentclass[reprint,amsmath,amssymb,aps,prd,nofootinbib,floatfix]{revtex4-1}

\usepackage{hyperref}
\usepackage{amssymb,amsmath,amsfonts}
\usepackage{mathrsfs}
\usepackage{dsfont} 
\usepackage{graphicx,bm}
\usepackage{multirow}
\usepackage{color}
\usepackage{nicematrix}
\usepackage{enumerate}
\usepackage{placeins}
\usepackage{capt-of}
\graphicspath{{./figures/}}

\usepackage{dcolumn}
\usepackage{array}

\newcommand{\overbar}[1]{\mkern 1.5mu\overline{\mkern-1.5mu#1\mkern-1.5mu}\mkern 1.5mu}

\allowdisplaybreaks

\begin{document}

\title{What neutron stars tell about the hadron--quark phase transition: a Bayesian study}

\author{J{\'a}nos Tak{\'a}tsy}
\email{takatsy.janos@wigner.hu}
\affiliation{
Institute for Particle and Nuclear Physics, Wigner Research Centre for Physics, 1121 Budapest, Hungary
}
\affiliation{
Institute of Physics, E\"otv\"os University, 1117 Budapest, Hungary
}
\author{P{\'e}ter Kov{\'a}cs}
\affiliation{
Institute for Particle and Nuclear Physics, Wigner Research Centre for Physics, 1121 Budapest, Hungary
}
\author{J{\"u}rgen Schaffner-Bielich}
\affiliation{
Institut f{\"u}r Theoretische Physik, Goethe Universit{\"a}t Frankfurt, D-60438 Frankfurt, Germany
}
\author{Gy{\"o}rgy Wolf}
\affiliation{
Institute for Particle and Nuclear Physics, Wigner Research Centre for Physics, 1121 Budapest, Hungary
}

\begin{abstract}
The existence of quark matter inside the heaviest neutron stars has been the topic of numerous recent studies, many of them suggesting that a phase transition to strongly interacting conformal matter inside neutron stars is feasible. Here we examine this hybrid star scenario using a soft and a stiff hadronic model, a constituent quark model with three quark flavours, and applying a smooth crossover transition between the two. Within a Bayesian framework, we study the effect of up-to-date constraints from neutron star observations on the equation-of-state parameters and various neutron star observables. Our results show that a pure quark core is only possible if the maximum mass of neutron stars is below $\sim2.35~M_\odot$. However, we also find, consistently with other studies, that a peak in the speed of sound, exceeding $1/3$, is highly favoured by astrophysical measurements, which might indicate the percolation of hadrons at $\sim3-4n_0$. Even though our prediction for the phase transition parameters varies depending on the specific astrophysical constraints utilized, the position of the speed of sound peak only changes slightly, while the existence of pure quark matter below $\sim4 n_0$, using our parameterization, is disfavoured. On the other hand, the preferred range for the EoS shows signs of conformality above $\sim4n_0$. Additionally, we present the difference in the upper bounds of radius estimates using the full probability density data and sharp cut-offs, and stress the necessity of using the former.
\end{abstract}

\maketitle

\section{Introduction}
\label{sec:intro}

Neutron stars (NSs) are one of the endpoints of stellar evolution formed in core-collapse supernovae with a progenitor mass of $8M_\odot$ or more. NSs are so compact that the central energy density can reach several times the one of
nuclear matter at saturation. At these high densities new particles could emerge and/or matter is transformed to a new phase characterised by approximate chiral symmetry restoration, which is dubbed the hadron-quark phase transition (see e.g. Ref.~\cite{Schaffner-Bielich:2020psc} for an introduction).

In the last few years, observations of NSs revealed several breakthrough measurements of their global properties. So it is now well established that pulsars, rotation-powered NSs, can have masses of around two solar masses \cite{Demorest:2010bx,Antoniadis:2013pzd,NANOGrav:2017wvv,Cromartie:2019kug,Fonseca:2021wxt}, as determined from the timing of the pulses of NSs in binary systems with corrections from general relativity, such as the pulsar PSR J0740+6620 with a mass of  $M=2.08\pm 0.07M_\odot$ \cite{Fonseca:2021wxt}. NSs with a low-mass stellar companion, so called black-widow or redback pulsars, can have even higher masses. These masses are extracted from the observation of the stellar companion and amount to $M =2.11\pm 0.04 M_\odot$ for PSR J1810+1744,  $M=2.22\pm 0.10M_\odot$ for PSR 1311-3430 and even $M= 2.35 \pm 0.17 M_\odot$ for PSR J0952-0607 \cite{Romani:2021xmb,Romani:2022jhd,Kandel:2022qor}.

Furthermore, the mass and radius of NSs could be constrained directly with the phase-resolved observations of the hot spots on the surface of the NS with the NICER mission for PSR J0030+0451 \cite{Miller:2019cac,Riley:2019yda,Raaijmakers:2019qny} 
and PSR J0740+6620 \cite{Miller:2021qha,Riley:2021pdl,Raaijmakers:2021uju}.
Analysis of the gravitational wave (GW) event GW170817 of a binary NS merger reveals that NSs must have a rather small radius. The limit on the tidal deformability inferred from the GW has been extracted for low and high spins of the merging NSs and under different assumptions of the property of NS matter, i.e.\ the equation of state (EoS) by the LIGO/Virgo scientific collaboration \cite{LIGOScientific2017,LIGOScientific:2018cki,LIGOScientific:2018hze}. The limit on the radius has been inferred by various groups to be less than $R\leq 13.2$ to 13.7~km for a $1.4M_\odot$ NS (see Refs.~\cite{Annala2017,Most:2018hfd,De:2018uhw,Tews:2018iwm}).

Due to the high densities present in the cores of the most massive NSs it is possible that hybrid stars exist with cores containing deconfined quark matter. The possibility of such a hadron to quark phase transition in NSs has been the topic of numerous recent studies. Many of them have investigated the impact of a strong first-order phase transition on astrophysical observables \cite{Alvarez-Castillo:2017qki,Blaschke:2019tbh}, as such a phase transition is proposed by effective quark--meson models. However, recent astrophysical measurements seem to rule out strong first-order phase transitions at low densities while making their existence unlikely at higher densities as well \cite{Christian:2019qer,Christian:2020xwz,Christian:2021uhd,Gorda:2022lsk}. Another way to look for an indication of deconfinement inside NSs is to investigate if the conformal limit is approached. Multiple studies suggest that the existence of conformal matter inside NSs is feasible, which might indicate the existence of hybrid stars \cite{Annala:2019puf,Marczenko:2022jhl}. Contrary to a first-order phase transition many recent studies propose an alternative scenario with a peak appearing in the speed of sound and reaching above the conformal limit \cite{Ecker:2022xxj,Jiang:2022tps,Marczenko:2022jhl}. This possibility is naturally achieved in models of the so-called quarkyonic matter (e.g. \cite{McLerran:2018hbz,Jeong:2019lhv}). Recent investigations of color superconductivity using functional methods also independently found such a peak in their models \cite{Leonhardt:2019fua,Braun:2021uua,Braun:2022jme}.

Despite recent developments in the fields of the complex Langevin method or alternative expansion schemes \cite{Guenther2020,Attanasio2020,Borsanyi2021}, the sign problem still poses a huge challenge for first-principle calculations of quantum chromodynamics (QCD). Therefore, when trying to describe strongly interacting matter at finite densities and low temperature, an effective treatment of the strong degrees of freedom is reasonable.

The nuclear EoS below saturation density is well established. Here, two- and three-body interactions are determined by experimental data mostly based on nucleon-nucleon scattering and properties of light nuclei (e.g. \cite{Akmal1998,Wiringa1988}). For these microscopic methods the source of uncertainties usually stem from the interactions applied, as well as the calculation methods themselves \cite{Hebeler2010,Gandolfi2013}. In addition to higher-body interaction becoming more important at higher densities, one might also need to account for new degrees of freedom, such as hyperons or quarks. Chiral Effective Field Theory (EFT) provides a robust way to estimate these uncertainties \cite{Lynn2015,Tews2018}. According to state-of-the-art calculations, the uncertainties of the nuclear EoS above $\sim1.1\,n_0$ become increasingly significant, with $n_0$ being the baryon density at nuclear saturation \cite{Tews2012,Tews2019}. Hadron resonance gas models provide a different way to calculate the low density EoS, while accommodating lattice data \cite{Huovinen2009,Bazavov2012}. Another common way to account for the low density behaviour of hadronic matter is to apply a relativistic mean field model with parameters set by nuclear properties at saturation (e.g. \cite{Steiner2013,Hempel2009,Typel2010}).

On the opposite side of the phase diagram, at very high densities, due to the asymptotic freedom, one can resort to perturbative QCD methods \cite{Kurkela2009,Mogliacci2013}. This area, however, despite what its name might suggest, is remarkably challenging due to an infinite number of diagrams that need to be accounted for at a given order. In fact in the past several decades, only few advancements have been reported in this field, with recent studies calculating the leading contributions to N$^3$LO \cite{Gorda2018}. Considering the zero temperature EoS, this method gives reliable results at $\mu_B\gtrsim2.5$~GeV, or equivalently $n_B\gtrsim40\,n_0$.

Therefore, there is more than an order of magnitude in density, where the EoS is largely uncertain. Several approaches exist in this region as well, apart from the ones that extrapolate from saturation properties of nuclear matter \cite{Oertel:2016bki}, one might also use Nambu--Jona-Lasinio (NJL) or linear sigma type models, which are based on the global symmetries of QCD, especially on chiral symmetry and the premise of chiral symmetry restoration at high densities (e.g. \cite{Nambu:1961tp,Nambu:1961fr,Hatsuda:1994pi,Gasiorowicz1969,Ko1994}). In this region the phase boundaries are also ambiguous. Quark deconfinement can occur at virtually any densities in this uncertain domain, while there are also strong indications for the existence of a color superconducting phase at densities reachable inside the cores of massive NSs \cite{Ivanytskyi:2022bjc,Ivanytskyi:2022oxv,Blaschke:2022egm}. Extensive studies of the NJL type models exist in the literature, with non-local interactions also being considered in recent studies \cite{Hell2009,Kondo2010,Radzhabov2010}.

Many recent studies investigate this topic in a model-agnostic way, using general, parameterized EoSs \cite{Annala:2019puf,Jiang:2022tps,Han:2022rug,Annala:2023cwx}, or using a constant speed-of-sound construction, with no underlying microphysical input for the quark part of the EoS \cite{Miao:2020yjk,Xie:2020rwg}. In this paper we instead utilize EoSs derived from a $\mathrm{U_L}(3)\times \mathrm{U_R}(3)$ chirally symmetric constituent quark--meson model developed by our group \cite{Parganlija:2012fy,Kovacs:2016juc,Takatsy:2019vjs,Kovacs:2021kas}. At zero density and finite temperature this model manages to comply with lattice results remarkably well \cite{Kovacs:2016juc}, while the parameters of this model are also carefully determined using meson vacuum phenomenology, which is not typical of similar studies. This way, our approach incorporates physical constraints from vacuum phenomenology and QCD thermodynamics at finite temperature. On the other hand, NS observations provide additional constraints for our model parameters undetermined by the parameterization. We use a hybrid approach in describing NSs with quark cores, utilizing relativistic mean field models at low densities (the SFHo and the DD2 models, \cite{Steiner2013,Hempel2009,Typel2010}), EoSs from our constituent quark model at high densities, while a smooth interpolation is applied between the two parts. Similar studies, using NJL-type models are also available in the literature (see e.g. Refs.~\cite{Ayriyan:2018blj,Ayriyan:2021prr,Pfaff:2021kse,Contrera:2022tqh}). We are going beyond previous works by utilizing a general concatenation scheme and presenting an in-depth analysis of the allowed parameter regions of our constituent quark model, including the most recent astrophysical constraints. We also investigate how our results compare to studies using a model-agnostic approach.

This paper is organized in the following way. In Sec.~\ref{sec:methods} we review the method we used to construct hybrid EoSs for cold NS matter, then after summarizing recent results from NS observations we introduce our Bayesian framework. In Sec.~\ref{sec:results} we show the results from our Bayesian analysis and demonstrate how different astrophysical measurements influence the outcome of these.

\section{Methods}
\label{sec:methods}

In this section we first review how our hybrid EoS is constructed, then after an overview of the calculation of NS observables and recent observations we proceed to describe the details of our Bayesian analysis, devoting special attention to how our posterior probabilities are calculated.

\subsection{Equation of state}
\label{ssec:EoS}

To be able to investigate the effect of variations in the properties of our quark model on stable sequences of NSs, we need to construct a reliable EoS covering a large range in density from below saturation density up to $n_B\approx5-6 \, n_0$ where $n_0$ is the baryon density at nuclear saturation.

As already mentioned in Sec.~\ref{sec:intro} we use a hybrid approach combining EoSs from hadronic and quark matter. For the hadronic part we use two relativistic mean field models, the Steiner-Fischer-Hempel (SFHo) model \cite{Steiner2013} and the density-dependent model of Typel et al. (DD2) \cite{Typel2010,Hempel2009}. We obtained both EoSs from the CompOSE database\footnote{\href{https://compose.obspm.fr}{https://compose.obspm.fr}}. Both models are consistent with chiral EFT calculations at low densities \footnote{see e.g. Fig.~9 of Ref.~\cite{Kruger:2013kua} at the relevant density above $\sim0.5n_0$, below which the EoS is strongly influenced by nuclear clustering, and thus the proper crust EoS has to be used \cite{Baym:1971pw,Negele:1971vb,Okamoto:2011tc,Oertel:2016bki}, which we extracted from the unified EoS by Douchin \& Haensel \cite{Douchin:2001sv}}, but differ in the stiffness at higher densities; while the SFHo EoS is relatively soft, the DD2 EoS is quite stiff, with maximally stable NS masses of $\sim2.05~M_\odot$ and $\sim2.42~M_\odot$, respectively.

For the quark part we utilize the (axial)vector meson extended linear sigma model (eLSM), developed and thoroughly investigated in several previous papers \cite{Parganlija:2012fy,Kovacs:2016juc,Takatsy:2019vjs,Kovacs:2021kas} with investigations about the large-$N_c$ behaviour as well \cite{Kovacs:2022zcl}. This is a three-flavour quark--meson model containing constituent quarks and the complete nonets of (pseudo)scalar and (axial)vector mesons. The advantage of this model -- altogether with the parameterization procedure and the approximations that were used -- is that it reproduces the meson spectrum (and also various decay widths) quite well at $T=\mu_B=0$, and moreover, its finite temperature version also agrees well with various lattice results \cite{Kovacs:2016juc}. The detailed description of the approximation we use in this paper can be found in Ref.~\cite{Kovacs2021}, where we have also already provided a comparison between the sequences of static (non-rotating) NSs predicted by the model and astrophysical observations.

It is worth to note that due to various particle mixings in the scalar sector, there is more than one possibility to assign scalar mesons to experimental resonances \cite{Parganlija:2012fy}. Possibly stemming from this mixing is that the mass of the sigma meson needs to be very low in our model in order to achieve a correct finite temperature behaviour and to reproduce other meson masses correctly. This problem might be resolved by considering additional bound quark states, such as tetraquarks. However, as we showed in Ref.~\cite{Kovacs2021}, this low sigma meson mass, within the framework of this model, is also consistent with astrophysical observations.

Nevertheless, similarly to the analysis in Ref.~\cite{Kovacs2021} we leave $m_\sigma$ as a free parameter and let it vary between $290-700$~MeV. Another important parameter, which is not fixed by experimental data in the approximation we use, is the coupling between vector mesons and constituent quarks. We vary this parameter in the range $g_V\in[0,10]$.

Since the low- and high density models operate with different degrees of freedom, we need to utilize some effective method to arrive from one phase to another. The simplest way to do this, which we will also follow in our paper, is to simply interpolate between the two zero temperature EoSs in some intermediate-density region (see e.g. Refs.~\cite{Abgaryan2018,Baym2019,Masuda2012,Blaschke:2021poc}). Also similar to other studies, we use a polynomial interpolation, however, in contrast to using the pressure, $p(\mu_B)$, as a thermodynamic potential for the interpolation, we use the energy density, $\varepsilon(n_B)$. We do this, since we find that this way the sound speed in the intermediate region shows a less sharp peak, and therefore a larger ensemble of EoSs will turn out to be causal.

In case the hadronic EoS, $\varepsilon_\mathrm{H}(n_B)$, is valid up to $n_{BL}$, and the quark EoS, $\varepsilon_\mathrm{Q}(n_B)$, can be utilized above $n_{BU}$, the interpolating polynomial looks like:
\begin{equation}
    \varepsilon(n_B) = \sum\limits_{k=0}^{N} C_k n_B^k \: , \quad n_{BL}<n_B<n_{BU},
\end{equation}
where the coefficients $C_k$ are determined so that the energy density and several of its derivatives remain continuous in the whole region. In our case, we use a fifth-order polynomial, so we need the energy density and its first and second derivatives to be continuous at the boundaries. The first derivative of the energy density with respect to the baryon number density is the baryon chemical potential, so this condition will ensure that the pressure is also continuous at the boundaries. The condition for the second derivative, in addition, ensures a continuous speed of sound.

From $n_{BL}$ and $n_{BU}$ we define the central density and width of the phase transition as $\bar{n} = (n_{BU}+n_{BL})/2$ and $\Gamma = (n_{BU}-n_{BL})/2$, respectively.

As a further remark, we add that a Maxwell construction is also a common way to get from one phase to the other, however, this method limits the possible range of concatenations by requiring the $p(\mu_B)$ curves of the two phases to cross each other, and also removing the freedom of choosing the density at which the phase transition occurs. From a philosophical point of view one might also argue that since both models have a limited region of validity, at intermediate densities one can only assume some interpolation between the two models.

\subsection{Neutron stars and observations}
\label{ssec:NSobs}

Here we briefly summarize how one can calculate different equilibrium properties of NSs given a specific EoS. For details of these calculations we refer the reader to Ref.~\cite{Kovacs2021}, as well as references therein.

Once we have an EoS, $p(\varepsilon)$, we can obtain mass--radius relations of NSs by solving the Tolmann--Oppenheimer--Volkoff (TOV) equations \cite{Tolman1939,Oppenheimer1939}:
\begin{align}
\frac{\mathrm{d}m(r)}{\mathrm{d}r} &= 4\pi r^2 \varepsilon(r) , \label{eq:tov_m} \\ 
\frac{\mathrm{d}p(r)}{\mathrm{d}r} &= - [\varepsilon(r)+p(r)]\dfrac{m(r)+4\pi r^3 p(r)}{r^2 - 2 m(r) r} ,\label{eq:tov_p}
\end{align}
where $m(r)$ is the gravitational mass enclosed within a sphere with radius $r$, and $p(r)$ is the pressure related to the energy density $\varepsilon(r)$ by the EoS. These equations can usually only be integrated numerically. The total mass ($M$) and radius ($R$) of the NS for a certain central energy density $\varepsilon_c$ is obtained through the boundary conditions, $\varepsilon(r=0)=\varepsilon_c$, $p(R)=0$ and $m(R)=M$.

Another property of NSs that is becoming more and more important due to the recent and future observations of inspirals of NSs with GW detectors, is the $\lambda$ tidal deformability parameter (e.g. \cite{Hinderer2007,Yagi2013,LIGOScientific2017}). This parameter is related to the $k_2$ quadrupole tidal Love number through
\begin{equation}
    k_2 = \frac{3}{2} \lambda R^{-5} ,
\end{equation}
where $k_2$ can usually be obtained by numerical integration (see e.g. Refs.~\cite{Hinderer2007,Damour2009,Postnikov2010,Takatsy2020}). The dimensionless parameter $\Tilde{\Lambda}$ measurable through GW observations of binary NSs can then be calculated by
\begin{equation}
    \Tilde{\Lambda} = \frac{16}{13}\Lambda_1 \frac{M_1^4}{M_\mathrm{tot}^4} \left( 12-11 \frac{M_1}{M_\mathrm{tot}} \right) + 1 \longleftrightarrow 2 ,
\end{equation}
where $M_\mathrm{tot}=M_1+M_2$, and where $\Tilde{\Lambda}$ is directly determined by the phase shift in the GW signal of circular NS binaries due to tidal effects \cite{Flanagan2007}. Here $\Lambda_i$ is the dimensionless tidal parameter of component $i$, which can be obtained through
\begin{equation}
    \Lambda_i = \frac{\lambda_i}{M_i^5} .
\end{equation}

There are already several stringent observational constraints on how the EoS should look like, with more expected to come in the near future. These constraints stem from various sources, such as electromagnetic, GW, or combined, multi-messenger observations.

Masses of NSs, in case they form a binary with an other object, might be measured with remarkable precision, using, for example, the Shapiro time delay effect. In the past decade, multiple highly massive NSs have been observed, providing robust constraints on the stiffness of the EoS \cite{Demorest2010,Antoniadis2013,Cromartie2019}. From these, until recently, the most massive was PSR J0740+6620, with a mass of
$2.08\pm0.07$~$M_\odot$, and a $95.4\%$ lower bound of $1.95$~$M_\odot$ \cite{Fonseca2021}. Since then, however, several other observations have also raised notable interest \cite{Linares:2018ppq,Romani:2021xmb}. Observation of the black-widow pulsar PSR J0952-0607 have measured its mass to be $2.35\pm0.17$~$M_\odot$ \cite{Romani:2022jhd}.

A recent observation discovered a very light central compact object within the supernova remnant HESS J1731-347. It is interpreted as the lightest NS observed so far or a quark star with a mass of $0.77^{+0.20}_{-0.17}$~$M_\odot$ and radius $10.4^{+0.86}_{-0.78}$~km.

Unlike masses, the measurement of radii of NSs is extremely challenging, and so far, the most accurate X-ray measurements were able to achieve a precision of $\sim10\%$. Recent measurements of the NICER collaboration use the ingenious idea of examining the rotation-resolved X-ray spectrum of NSs with hot spots. This, for the first time enables the simultaneous measurement of the mass and radius of a single NS. Two NSs have been measured with this method so far, one is PSR J0030+0451 with a mass and equatorial radius of $1.44^{+0.15}_{-0.14}$~$M_\odot$ and $13.02^{+1.24}_{-1.06}$~km \cite{Miller2019}, or $1.34^{+0.15}_{-0.16}$~$M_\odot$ and $12.71^{+1.14}_{-1.19}$~km \cite{Riley2019}, according to different collaborations using slightly different methods. The other pulsar measured was the massive pulsar PSR J0740+6620, with mass $2.08\pm0.07$~$M_\odot$ and reported radii of $13.7^{+2.6}_{-1.5}$~km \cite{Miller2021} or $12.39^{+1.30}_{-0.98}$~km \cite{Riley2021} at the $68\%$ credible interval.

In addition to these measurements, we have also witnessed the first multi-messenger observation of a binary NS merger, with its GW signal being labeled GW170817. The first analysis of GW170817 performed by the LIGO-Virgo Collaboration (LVC) inferred a value of $\Lambda<800$ for $1.4$~$M_\odot$ NSs in the low-spin limit \cite{LIGOScientific2017}. A thorough investigation of this constraint performed by Ref.~\cite{Annala2017} using a generic family of EoSs found an upper radius limit of $13.6$~km for $1.4$~$M_\odot$ NSs, while Ref.~\cite{Most:2018hfd} arrived at a radius limit of $13.7$~km with higher statistics. A subsequent study was also performed by the LVC, in which a combined analysis of tidal deformabilities and NS radii was performed, utilizing various assumptions for the EoSs. Here the values of $\Lambda(1.4M_\odot)=190^{+390}_{-120}$ and $R(1.4M_\odot)=10.8^{+2.0}_{-1.7}$~km were found \cite{LIGOScientific:2018cki}. An additional assumption of this study was to use a single EoS to describe both objects, whereas in Ref.~\cite{LIGOScientific2017} the two EoSs were varied independently. A similar study, also using a single EoS ansatz, was performed by Ref.~\cite{De:2018uhw}, where the authors arrived at a slightly higher upper limit ($\Lambda < 642$, $\Lambda < 698$ or $\Lambda < 681$ depending on the prior assumption on the component masses). A companion study of Ref.~\cite{LIGOScientific:2018cki} was also published by the LVC at around the same time, where an EoS agnostic approach was applied \cite{LIGOScientific:2018hze}. In their study they investigated the effect of using various waveform templates, and under minimal assumptions they found for the upper limit of the tidal deformability $\Lambda(1.4M_\odot)<720$ \cite{LIGOScientific:2018hze}.

In this paper we chose to utilize the results of this analysis with the upper limit of $\Lambda(1.4M_\odot)<720$. The minimal prior assumptions of this study make it suitable to use it as a conservative upper limit for the tidal deformability. Other, recent studies also utilize this constraint (e.g. \cite{Annala:2021gom}. Ref.~\cite{Kastaun:2019bxo}) examine previous studies \cite{LIGOScientific:2018cki,LIGOScientific:2018hze,De:2018uhw,Landry:2018prl,Capano2019} and investigates the impact of prior assumptions and argues that upper and especially lower limits on $\Lambda$ can be misleading without a more detailed discussion. Another reanalysis has also been done by Dietrich et al., which found similar upper limits for $\Lambda$ (see Table~S2 of Ref.~\cite{Dietrich:2020efo}).

The electromagnetic properties of the source of GW170817 were also used to put constraints on NSs. A lower radius constraint was inferred by Ref.~\cite{Bauswein2017} from the absence of prompt collapse during this event, while an upper mass limit of $2.16^{+0.17}_{-0.15}$~$M_\odot$ was proposed by Ref.~\cite{Rezzolla2017} using a quasi-universal relation between the maximum mass of static and the maximum mass of uniformly rotating NSs. This conclusion rests upon the assumption that the merging NSs first formed a differentially rotating hypermassive NS and not a uniformly rotating supermassive one. This hypothesis is supported by simulations of the dynamical ejecta and kilonova modeling (e.g. \cite{Shibata2017}), albeit other scenarios are not completely ruled out either.

The GW signal of another binary NS merger GW190425 was also observed by the LVC, however, no clear tidal signature or electromagnetic signal was measured. Due to this, the binary NS classification only rests on the estimated masses of the binary components. Yet another notable GW event was GW190814, where one of the binary components resided in the so-called 'mass-gap', with a mass of $2.5-2.67$~$M_\odot$ \cite{LIGOScientific:2020zkf}, which could either mean it was the lightest black hole (BH), or the heaviest NS observed. Although the NS scenario seems unlikely, it should not be ruled out until further evidence is found against it.

Several studies exist that combine all these astrophysical measurements with nuclear physics and heavy-ion data to give stringent constraints on the nuclear EoS and the $M-R$ relation of NSs  (e.g. \cite{Capano2019,Huth2021,Ghosh:2021bvw,Ghosh:2022lam,Pradhan:2022txg,Shirke:2022tta,OmanaKuttan:2022aml}). Bayesian investigations are also available in this field (e.g. \cite{Jiang:2022tps,Annala:2023cwx,Ayriyan:2021prr,Pfaff:2021kse}), while other studies use deep neural networks to constrain the EoS \cite{Soma:2022qnv,Soma:2022vbb}. In this paper we also apply a Bayesian approach. However, we concentrate on hybrid stars, where the properties of quark matter are calculated from an effective model of QCD, with the correct behavior at zero density and finite temperature. Among others we focus on the restriction of quark model parameters and the parameters of the concatenation, and on the conditions for the existence of a pure quark core.

\subsection{Bayesian inference}
\label{ssec:BayesIntro}

Suppose our EoS, and hence the properties of NSs can be described by a set of parameters, $\boldsymbol{\vartheta}$. The probability of a specific data being measured, given a specific EoS is $p(\mathrm{data}|\boldsymbol{\vartheta})$. Then we can use Bayes' theorem to determine the probability of a specific parameter set, given data from a measurement:
\begin{equation}
    p(\boldsymbol{\vartheta}|\mathrm{data}) = \frac{p(\mathrm{data}|\boldsymbol{\vartheta}) \, p(\boldsymbol{\vartheta})}{p(\mathrm{data})} \: ,
\end{equation}
where $p(\boldsymbol{\vartheta})$ is our prior assumption about the parameter sets, and $p(\mathrm{data})$ is just a normalization constant. 

Our parameter space consists of the four parameters: $m_\sigma$, $g_V$, $\bar{n}$ and $\Gamma$. We vary $\bar{n}$ between $2n_0$ and $5n_0$, $\Gamma$ between $n_0$ and $4n_0$, and $g_V$ between $0$ and $10$. In order to vary $m_\sigma$, we need to reparameterize our model, which is computationally expensive and even conceptually different since we end up with a reparameterized model, which is not needed if we change other parameters of the model. Therefore, we do not vary the sigma meson mass continuously, but rather choose 5 different values for $m_\sigma$: $290$~MeV, from the original parameterization of the eLSM \cite{Kovacs2021}, as well as $400$~MeV, $500$~MeV, $600$~MeV, and $700$~MeV. We sample the other three parameters on a grid. Many Bayesian studies apply a Markov chain Monte Carlo approach with the appropriate sampler to obtain the posterior distributions of the model parameters \cite{Landry:2018prl}. However, due to the low dimensionality of our parameter space, sampling the parameters on a grid in advance is justified. Many recent Bayesian studies also utilize such a pre-sampled set of EoSs \cite{Dietrich:2020efo,Huth2021,Blaschke:2020qqj}. Sampling the model parameters on a grid is also not unprecedented (see e.g. \cite{Blaschke:2020qqj,Ayriyan:2021prr}), and is justified by the discretized values we take for $m_\sigma$. We choose the a priori probability for all parameter sets to be equal, hence, in this sense, our prior can be considered nearly uniform in our four-dimensional parameter space.

The EoSs generated using these parameters still need to comply with some basic requirements. We ensure that the low density EoS is described by the hadronic EoS by discarding parameter sets with $\bar{n}-\Gamma<n_0$. Additionally, we require stability and causality for all EoSs. Ultimately, we end up with a set of $\sim18\:000$ EoSs, for which the posterior probabilities are calculated.

In addition to our choice for prior in the parameter space, when calculating posterior probabilities for different astrophysical observations, the prior for the NS mass distribution is assumed to be uniform. We note here that a choice for the NS mass prior that does not match the observed distribution can lead to large biases in the Bayesian inference after $\mathcal{O}(25)$ observational events (see e.g. Refs.~\cite{Agathos:2015uaa,Wysocki:2020myz}). For the time being, we can safely assume a uniform prior without having to worry about these biases. Eventually, however, a self-consistent hierarchical framework that simultaneously models EoSs and NS populations will be necessary \cite{Chatziioannou:2020pqz}. For further discussion about the uniform population prior we refer the reader to Refs.~\cite{Miller:2019nzo,Landry:2020vaw,Chatziioannou:2020pqz}.

The conditional probability $p(\mathrm{data}|\boldsymbol{\vartheta})$ can be obtained as a product of several independent astrophysical observations:
\begin{equation}
    p(\mathrm{data}|\boldsymbol{\vartheta}) = p(M_\mathrm{max}|\boldsymbol{\vartheta}) \, p(\mathrm{NICER}|\boldsymbol{\vartheta}) \, p(\tilde{\Lambda}|\boldsymbol{\vartheta}) \: ,
\end{equation}
where we detail the specific observational constraints below. Also note that since only the proportions of the probabilities for different parameter sets are meaningful, we can neglect constant normalization factors in front of our conditional probabilities.

\subsubsection{Compatibility with perturbative QCD}

Even without any astrophysical constraints, our EoS should comply with some basic physical requirements. First and foremost, our EoS should be causal, meaning
\begin{equation}
    c_s^2 = \frac{\mathrm{d}p}{\mathrm{d}\varepsilon} \leq 1 \: .
\end{equation}

In addition, however, we can also use input from perturbative QCD calculations, similarly to Refs.~\cite{Komoltsev:2021jzg,Gorda:2022jvk}. We know that at some density $n_\mathrm{QCD}$ strongly interacting matter should have a baryon chemical potential $\mu_\mathrm{QCD}$ and a pressure $p_\mathrm{QCD}$. On the other hand, we require our hybrid EoS to be valid up to the density present in the center of the most massive NS described by that specific EoS. This point is described by $n_\mathrm{NS}$, $\mu_\mathrm{NS}$ and $p_\mathrm{NS}$. However, these EoSs should be in accord with each other and therefore there should exist a thermodynamically allowed connection of the two. Therefore we assume in the core of NS that $\mu_\mathrm{NS} \leq \mu_\mathrm{QCD}$ and require stability and causality:
\begin{equation}
    n_{\text{NS}} \leq n_{\text{QCD}}\: , \quad
    p_\mathrm{NS} \leq p_\mathrm{QCD} \: ,
\end{equation}
\begin{equation}
    \frac{n_\mathrm{NS}}{\mu_\mathrm{NS}} \leq \frac{n_\mathrm{QCD}}{\mu_\mathrm{QCD}} \: ,
\end{equation}
where we used the fact that a causal EoS crossing the point $(\mu,n)$ has a slope $\mathrm{d}n/\mathrm{d}\mu \geq n/\mu$.

We obtain an additional integral constraint from the definition of the difference in pressure:
\begin{equation}
    \Delta p = p_\mathrm{QCD} - p_\mathrm{NS} = \int\limits_{\mu_\mathrm{NS}}^{\mu_\mathrm{QCD}} n(\mu) \mathrm{d}\mu \: .
\end{equation}
The integral here depends on the specific way we connect the two points, however it can be easily shown that it falls between the two limiting cases (see e.g. Ref.~\cite{Komoltsev:2021jzg}):
\begin{equation}
    \Delta p_\mathrm{min} \leq \Delta p \leq \Delta p_\mathrm{max} \: ,
\end{equation}
with
\begin{equation}
    \Delta p_\mathrm{min} = \frac{\mu_\mathrm{QCD}^2-\mu_\mathrm{NS}^2}{2}\frac{n_\mathrm{NS}}{\mu_\mathrm{NS}}
\end{equation}
\begin{equation}
    \Delta p_\mathrm{max} = \frac{\mu_\mathrm{QCD}^2-\mu_\mathrm{NS}^2}{2}\frac{n_\mathrm{QCD}}{\mu_\mathrm{QCD}} \: .
\end{equation}

For the perturbative QCD EoS we use the values calculated by Ref.~\cite{Gorda:2021znl} and utilized in Ref.~\cite{Komoltsev:2021jzg} with a renormalization scale parameter $X=2$, hence, $\mu_\mathrm{QCD}=2.6$~GeV, $n_\mathrm{QCD}=6.47$~1/fm$^3$ and $p_\mathrm{QCD}=3823$~MeV/fm$^3$.

\subsubsection{Mass constraints}

We use PSR J0348+0432 with a mass $2.01\pm0.04$~$M_\odot$ and PSR J1614-2230 with a mass $1.908\pm0.016$~$M_\odot$ to put a lower limit on the maximum mass of NS mass--radius relations. In order to avoid double counting, we do not include here the mass measurement of PSR J0740+6620, since it is included as a NICER measurement. We also similarly include the upper mass bound from Ref.~\cite{Rezzolla2017}. We then approximate the likelihood functions by error functions:
\begin{align}
    p(M_\mathrm{max}|\boldsymbol{\vartheta}) &\propto \prod_{i=1,2} \frac{1}{2} \left[ 1 + \mathrm{erf}\left( \frac{M_\mathrm{max}(\boldsymbol{\vartheta}) - M_i}{\sqrt{2} \sigma_i} \right) \right]\nonumber \\
    &\times \frac{1}{2} \left[ 1 - \mathrm{erf}\left( \frac{M_\mathrm{max}(\boldsymbol{\vartheta}) - M_U}{\sqrt{2} \sigma_U} \right) \right] \: ,
\end{align}

where erf is the error function. For the upper mass limit from the hypermassive NS scenario we use $M_U = 2.16$~$M_\odot$ and set the standard deviation conservatively to $\sigma_U = 0.17$~$M_\odot$. While $M_1 = 2.01$~$M_\odot$, $\sigma_1 = 0.04$~$M_\odot$ and $M_2 = 1.908$~$M_\odot$, $\sigma_2 = 0.016$~$M_\odot$.

\subsubsection{NICER measurements}

For the two NICER measurements we use the kernel density estimated probability density $p_\mathrm{N}(M,R)$, utilizing the data provided by Refs.~\cite{Miller2019,Miller2021}. The likelihood for a single measurement is then given by
\begin{align}
    p(\mathrm{NICER}|\boldsymbol{\vartheta}) &\propto \int \mathrm{d}M \mathrm{d}R \, p_\mathrm{N}(M,R) \delta(R-R(M,\boldsymbol{\vartheta})) \nonumber \\
    &= \int \mathrm{d}M \, p_\mathrm{N}(M,R=R(M,\boldsymbol{\vartheta})) \: .
\end{align}
Note that the uniform mass population prior is already included in this formula.

\subsubsection{Tidal deformability measurement}

The chirp mass of the source of GW170817 was measured very precisely by the LVC to be
\begin{equation}
    \mathcal{M} = \frac{(M_1 M_2)^{3/5}}{(M_1 + M_2)^{1/5}} = (1.186 \pm 0.001) \: M_\odot \: ,
\end{equation}
where $M_1$ is conventionally considered to be the mass of NS with the larger mass. We then use the joint posterior probability density $p_\mathrm{GW}(\tilde{\Lambda},q)$, provided by Ref.~\cite{LIGOScientific:2020zkf}, where $q=M_1/M_2$ is the mass ratio. The accurate measurement essentially determines the secondary mass $M_2$ for a specific primary mass $M_1$. Then, utilizing the EoS, $\Lambda_1$ and $\Lambda_2$ can be determined, and therefore $\tilde{\Lambda}$ as well. We then calculate the conditional probability as
\begin{equation}
    p(\tilde{\Lambda}|\boldsymbol{\vartheta}) \propto \int\limits_{M_\mathrm{eq}}^{M_\mathrm{max}} \mathrm{d}M_1 \, p_\mathrm{GW}(\tilde{\Lambda}(M_1,\mathcal{M},\boldsymbol{\vartheta}),q(M_1,\mathcal{M})) \: ,
    \label{eq:pGW}
\end{equation}
where $M_\mathrm{eq}=1.362$~$M_\odot$ corresponds to a mass ratio of $q=1$, and $M_\mathrm{max}$ is the mass of the maximally stable NS.

\subsubsection{BH hypothesis}

Based on some properties of the electromagnetic counterpart of GW170817 some previous works have suggested that the remnant collapsed to a BH (e.g. \cite{Margalit:2017dij,Shibata2017,Rezzolla2017}). We refer to this as the BH hypothesis. In order to incorporate this assumption in our analysis we utilize baryon number conservation during the merger event:
\begin{equation}
    N_1 + N_2 = N_\mathrm{remn} + N_\mathrm{ej}
\end{equation}
where $N_1$ and $N_2$ are the baryon numbers of the two component NSs, while $N_\mathrm{remn}$ and $N_\mathrm{ej}$ are the baryon numbers corresponding to the remnant and the ejecta, respectively. Similarly to Refs.~\cite{Annala:2021gom,Gorda:2022jvk} we use the assumption $N_\mathrm{ej}\approx0$. Hence, in order for the remnant to collapse to a BH we must have $N_1+N_2>N_\mathrm{max}$, where $N_\mathrm{max}$ is the baryon number of the maximally massive stable NS. To add this assumption to our analysis we discard every pair of NS during the integral in Eq.~\ref{eq:pGW}, for which $N_1+N_2\leq N_\mathrm{max}$. Since values that $N_1+N_2$ can take are primarily determined by experiment, and higher values become more and more improbable, this gives an upper bound for $N_1+N_2$, which in turn gives an upper bound on the maximum mass of NSs, $M_\mathrm{max}\lesssim 2.53~M_\odot$.

Somewhat more speculatively one can assume that the remnant for a brief time remained a hypermassive NS, after which it quickly collapsed to a BH (e.g. \cite{Rezzolla2017}). We implement this assumption using the upper mass bound $M_U$ mentioned earlier in this section. We refer to this scenario as the hypermassive NS hypothesis.

In addition, we can include the assumption that the inspiral did not end in a prompt collapse to a BH. In this case we discard pairs of NSs, for which the total mass is above the threshold mass for prompt collapse to a BH. We use this assumption in all of our results that include the BH constraint in any form. Several approaches exist to calculate this threshold mass \cite{Bauswein2013,Koppel:2019pys}. Here we utilize the nonlinear relation given by Ref.~\cite{Koppel:2019pys}, calibrated by numerical relativity simulations:
\begin{equation}
    \frac{M_\mathrm{th}}{M_\mathrm{max}} = a - \frac{b}{1-c \cdot C_\mathrm{max}} \: ,
\end{equation}
where $C_\mathrm{max}=M_\mathrm{max}/R_\mathrm{max}$ is the compactness of the maximum mass configuration, and the parameters are $b=1.01$, $c=1.34$ and $a=2b/(2-c)$. Hence, we only perform the integration in Eq.~(\ref{eq:pGW}) for configurations where $M_\mathrm{tot}=M_1+M_2<M_\mathrm{th}$.

\subsubsection{Mass-gap compact object}

As we discussed in Sec.~\ref{ssec:NSobs}, an object in the mass gap was observed in the event GW190814 with a mass $M=2.59^{+0.08}_{-0.09}$~$M_\odot$ in the $90\%$ credible interval \cite{LIGOScientific:2020zkf}. In our analysis we also investigate what happens when we require this object to be described as a NS. We use a similar error function as for the other mass constraints with a mean $M_\mathrm{gap}=2.59$~$M_\odot$ and a standard deviation $\sigma_\mathrm{gap}=0.055$~$M_\odot$, assuming normal distribution.

\section{Results}
\label{sec:results}

In this section we discuss our results from our analyses. In Sec.~\ref{ssec:Mmax} we investigate how well the maximum mass of hybrid stars can be predicted by the parameters of the quark component. In Sec.~\ref{ssec:Bayes} we show our results from our Bayesian analysis with various constraints included.

\subsection{Dependence of $M_\mathrm{max}$ on constituent quark model parameters}
\label{ssec:Mmax}

In Ref.~\cite{Kovacs2021} we already showed in selected cases how the maximum mass of hybrid star sequences produced by using the eLSM for the quark component correlates with the parameters chosen for our quark model. Here we also investigate this correlation on whole span of the parameter space.

\begin{figure}[tb!]
  \centering
  \includegraphics[width=0.48\textwidth]{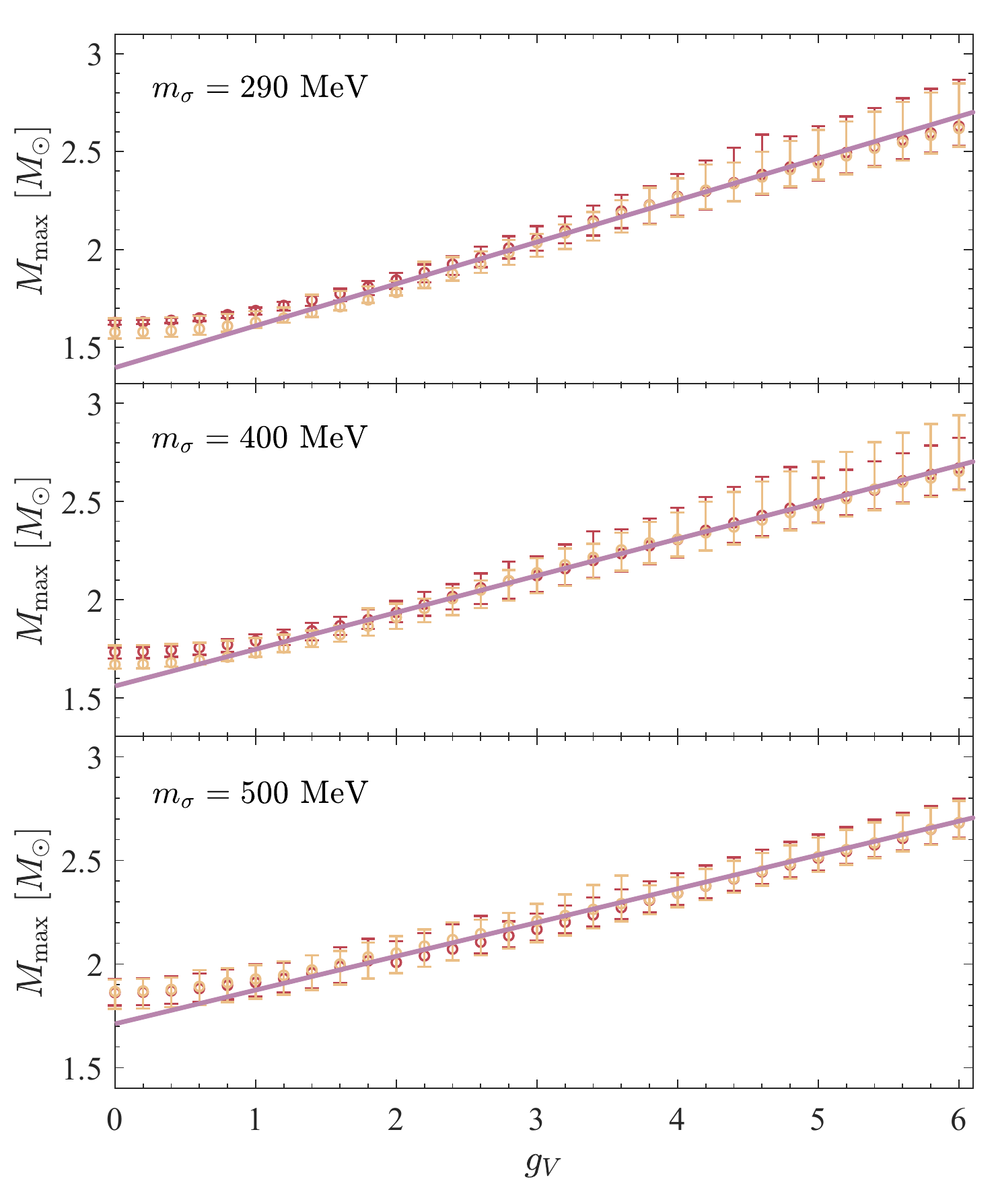}
  \caption{\label{fig:Mmax}The maximum mass of stable NSs as a function of the $g_V$ vector meson coupling, for different sigma masses. For specific constituent quark model parameters the circles denote the median, while the errorbars denote the 90\% confidence interval of maximum masses obtained by applying the complete ensemble of different concatenation parameters. The two different colors correspond to the SFHo (red) and the DD2 (yellow) hadronic EoS. The fitted relation is visualized by the purple lines.}
\end{figure}

The results are shown in Fig.~\ref{fig:Mmax}. The three different panels show results for three different values of the sigma meson mass, with $m_\sigma=290$~MeV being the one preferred by the parameterization. For a specific parameter set $\{m_\sigma,g_V\}$, we have gathered the maximum masses from all the different concatenations and plotted the median and the $90\%$ confidence intervals. The width of these intervals can be even lower than $\pm0.05$~$M_\odot$ for $M_\mathrm{max}\sim2$~$M_\odot$, while they moderately increase for higher median masses to $\pm0.15-0.2$~$M_\odot$. We can also observe that changing the hadronic EoS (red and yellow points) does not make any significant change in the maximum masses.

We try to quantify this correlation by making a linear fit to the points above $g_V=1$, since below that the dependence is clearly non-linear. The fitting function is then
\begin{equation}
    \frac{M_\mathrm{max}}{M_\odot} = \alpha \left( 1 + \gamma \cdot \overbar{m}_\sigma \right) + \beta \cdot g_V \left( 1 + \delta \cdot \overbar{m}_\sigma \right) \:,
\end{equation}
where
\begin{equation}
    \overbar{m}_\sigma = {m_\sigma \over 500\,\mathrm{MeV}} \: ,
\end{equation}
and where the cross-term with the coefficient $\delta$ is necessary since the slope of the linear fit changes for different sigma masses. 

The parameters obtained from the fit are
\begin{align}
    \alpha &= 0.962 \pm 0.010 \nonumber\\
    \beta &= 0.284 \pm 0.003 \nonumber\\
    \gamma &= 0.780 \pm 0.013 \nonumber\\
    \delta &= -0.426 \pm 0.014 \: ,
\end{align}
with a goodness-of-fit value $R^2=0.952$. The fitted function is shown in Fig.~\ref{fig:Mmax} by the purple lines. Due to $\delta$ being negative, we get the largest slope for $m_\sigma=290$~MeV.

\subsection{Bayesian analysis results}
\label{ssec:Bayes}

During our Bayesian analysis we incorporate constraints from astrophysical observations in a specific order. After establishing our prior we include the minimal constraints, namely, the requirement for consistency with pQCD calculations, and lower mass limits from the $2$~$M_\odot$ NSs. Then we apply the two NICER measurements, since these are the least constraining on our prior. After that, as another well-established constraint, we apply the tidal deformability measurement of GW170817, which, generally speaking, constrains the radii of $1.4$~$M_\odot$ NSs from above. These measurement constitute our canonical set of constraints.

On top of these, we also investigate the effect of other measurements as well. First, based on the hypermassive NS hypothesis, we put an upper limit on the maximum mass of NSs. As an alternative scenario, we incorporate a weaker constraint, the BH hypothesis, and explore the consequence of assuming the mass-gap object in GW190814 was a very massive NS. Finally, we also briefly review the effect of adding the recent measurement of the light compact object in HESS J1731-347 to our canonical set of constraints on top of the BH hypothesis. During these steps we try to monitor how the posterior probabilities evolve in the parameter space. We also investigate the radius distribution of $1.4$~$M_\odot$ and $2$~$M_\odot$ NSs during each step. The parameter set with the maximum posterior probability is also determined for each case.

Additionally, we also explore how sharp cut-offs at the $90\%$ credible intervals of measurable quantities from each astrophysical constraint would affect the allowed regions on the mass--radius diagram, and compare these results to the ones obtained from the Bayesian analysis. For the $2$~$M_\odot$ constraint the cut is achieved by requiring $M_\mathrm{max}>1.95$~$M_\odot$, which corresponds to the 2-sigma lower bound for the mass of PSR J0740+6620 \cite{Fonseca:2021wxt}. For the NICER measurements and the HESS object the requirement is that the $M-R$ curves should cross the 2-sigma contour lines of the given measurement. The cut for the tidal deformability measurement of GW170817 is established in the following way. For a given EoS we calculate all the possible $\tilde{\Lambda}$ values between $M_\mathrm{eq}<M_1<1.6~M_\odot$, and keep the EoS if $\tilde{\Lambda}<720$ for any configuration. Pairs of NSs with $M_\mathrm{tot}>M_\mathrm{th}$ are discarded, while pairs with $N_1+N_2\leq N_\mathrm{max}$ are also discarded when the BH hypothesis is included. The upper mass bound from the hypermassive NS hypothesis is taken to be $2.33$~$M_\odot$, while the lower mass bound from the mass-gap object is taken as $2.5$~$M_\odot$.

We divide our analysis into two separate parts. First, we restrict the sigma meson mass to $m_\sigma=290$~MeV, and the hadronic EoS to the relatively soft SFHo EoS. This keeps our discussion more transparent when investigating the evolution of probabilities in the parameter space. We also investigate how the predictions change when we use different combinations of the hadronic EoS and $m_\sigma$, and include the results in tables. For the second analysis we combine all EoSs with different hadronic EoSs and $m_\sigma$, and try to draw more general conclusions from the results.

\subsubsection{Results with the SFHo EoS and $m_\sigma=290$~MeV}

\begin{figure}[t]
  \centering
  \includegraphics[width=0.45\textwidth]{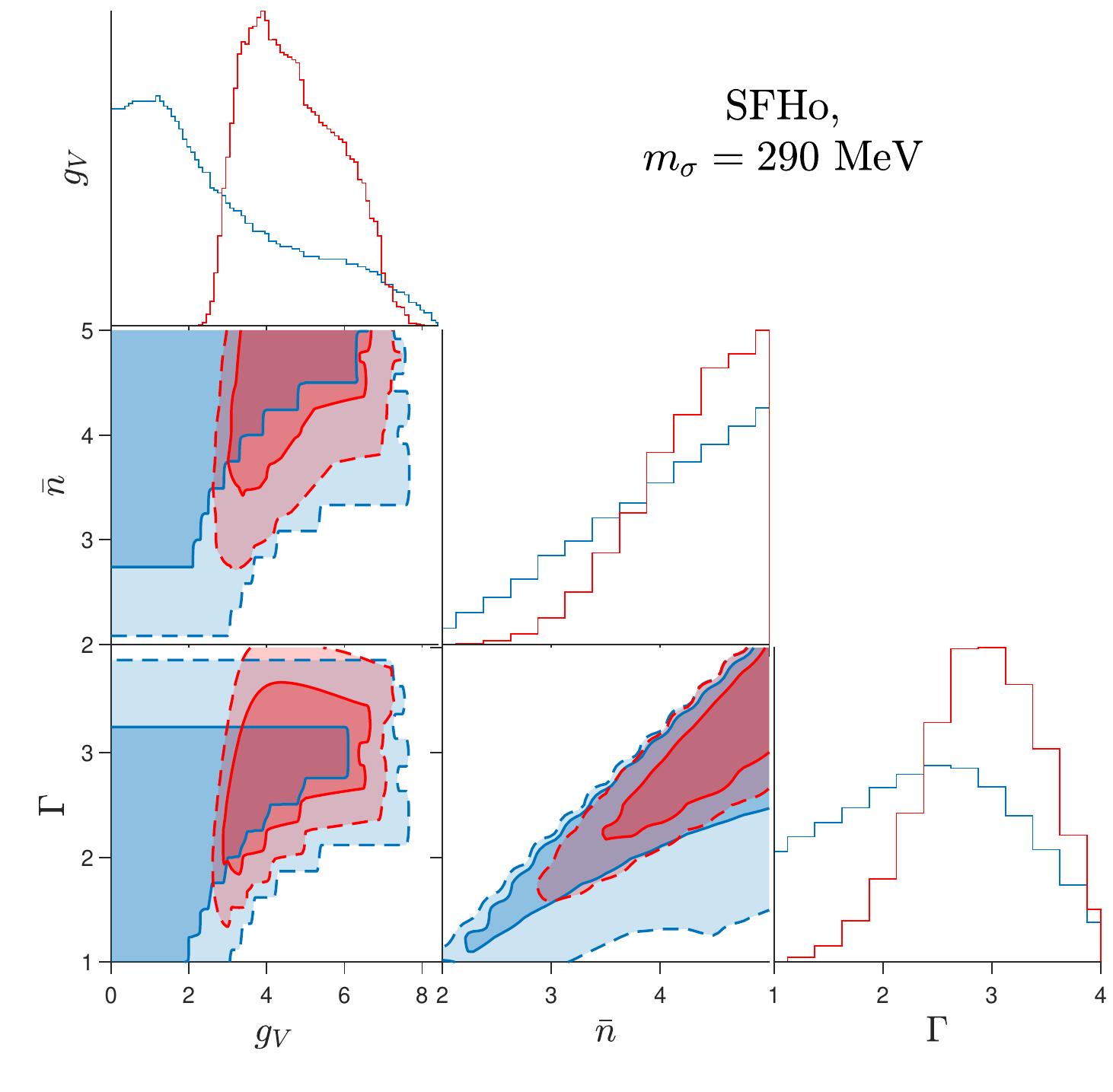}
  \caption{\label{fig:corner_SFHo}Marginalized priors (blue) and posteriors (red) for the three unfixed parameters characterizing the constituent quark model and the phase transition. The posterior corresponds to the canonical set of measurements (i.e. the pulsar mass measurements, the NICER measurements and the tidal deformability measurement of GW170817). For the two-dimensional marginals, the two contours correspond to the 68\% (solid) and 95\% (dashed) confidence intervals.}
\end{figure}

The marginalized priors and posteriors for the three unfixed parameters are shown in Fig.~\ref{fig:corner_SFHo}. The posteriors correspond to the canonical set of measurements. Even though we have a uniform grid in the parameter space for the prior, due to the requirements for stability and causality, large areas of the parameter space are excluded, and therefore, the marginals appear to have a structure. The posterior for the vector coupling $g_V$ is consistent with our expectations from Sec.~\ref{ssec:Mmax}, considering values below $g_V\approx3$ become increasingly improbable due to the $2M_\odot$ constraint. Lower values for $\bar{n}$ and $\Gamma$ also become disfavoured. Even though a figure such as Fig.~\ref{fig:corner_SFHo} contains a lot of information about the parameter distributions, due to the marginalized nature of the probability density functions (PDFs) in such a figure, some information is necessarily lost. Therefore, besides the marginalized PDFs shown in Fig.~\ref{fig:corner_SFHo}, during the following parts of our analysis we also show different slices of the PDFs in order to analyze the results from a different perspective. We can do this due to the low dimensionality of our parameter space.

\begin{figure*}[htbp]
  \centering
  \includegraphics[width=0.32\textwidth]{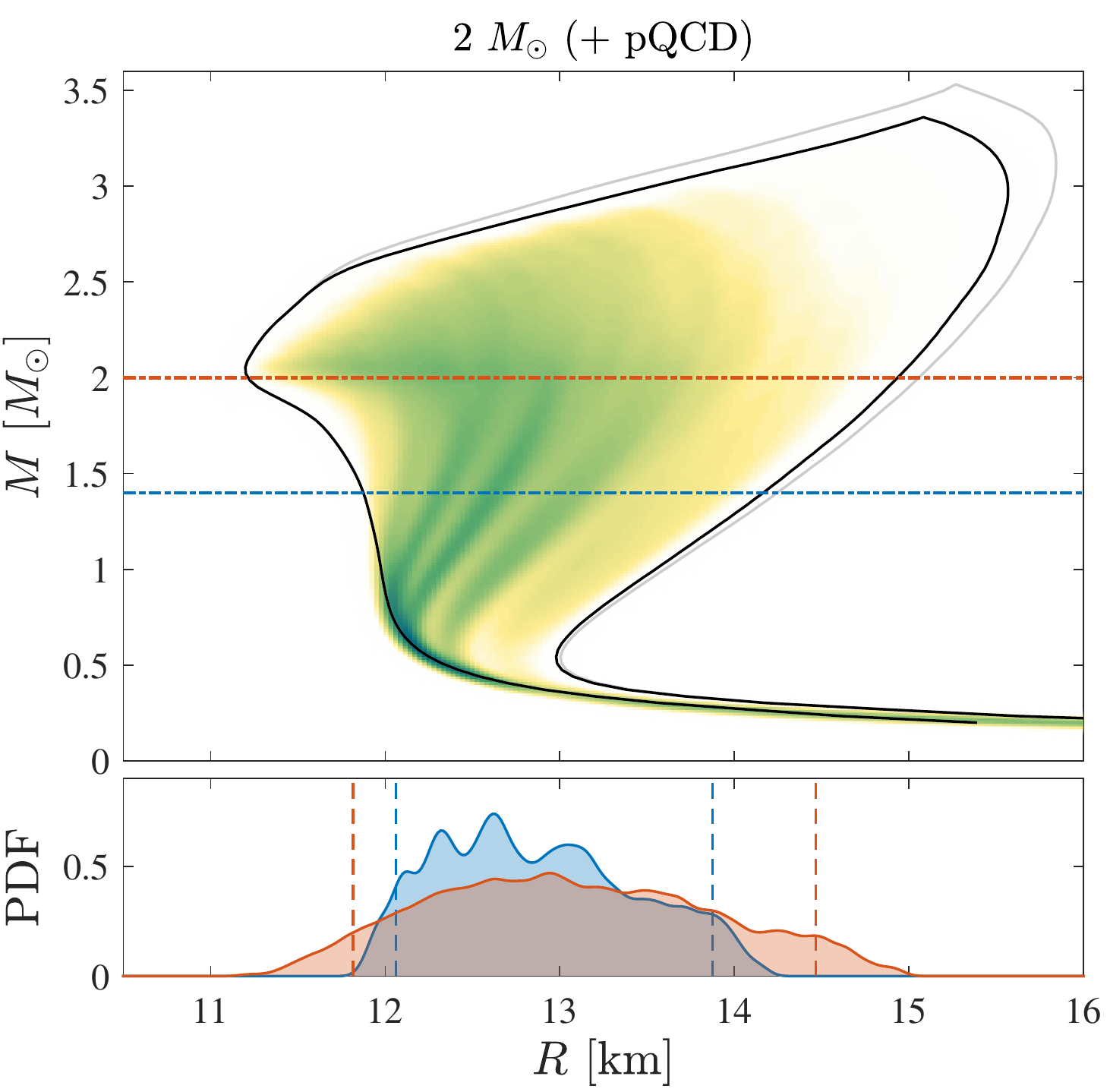}
  \includegraphics[width=0.32\textwidth]{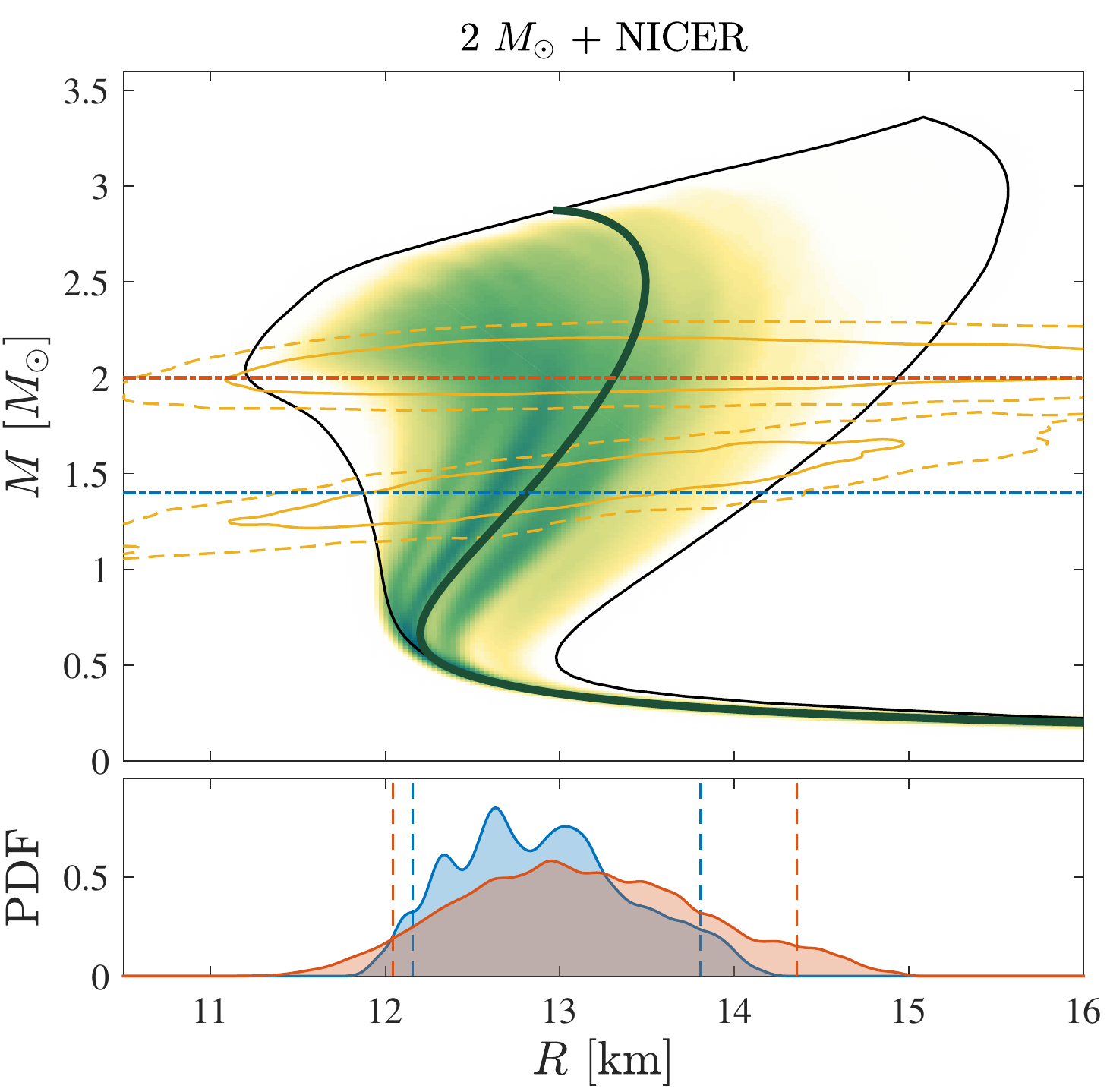}
  \includegraphics[width=0.32\textwidth]{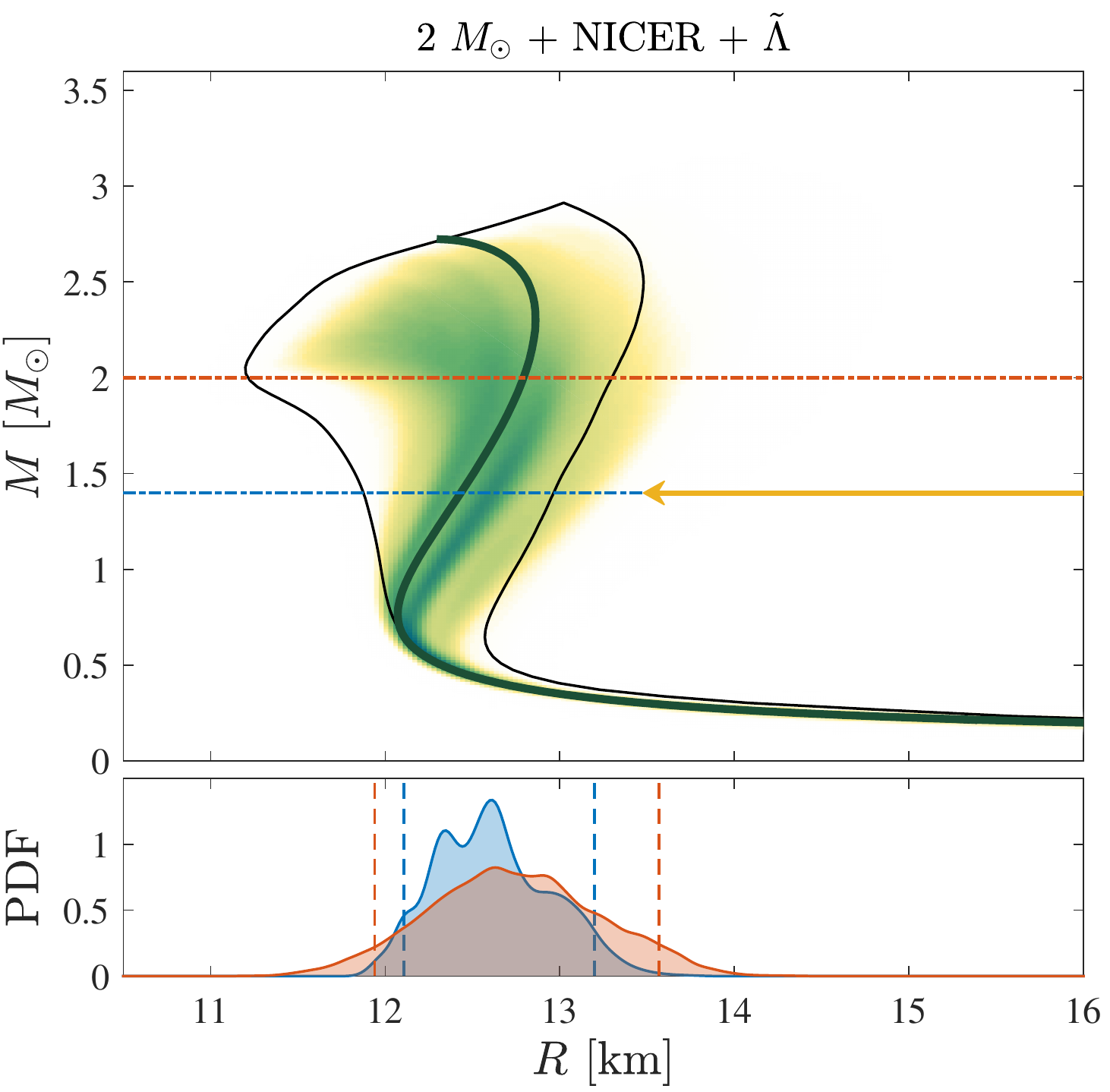}
  \includegraphics[width=0.32\textwidth]{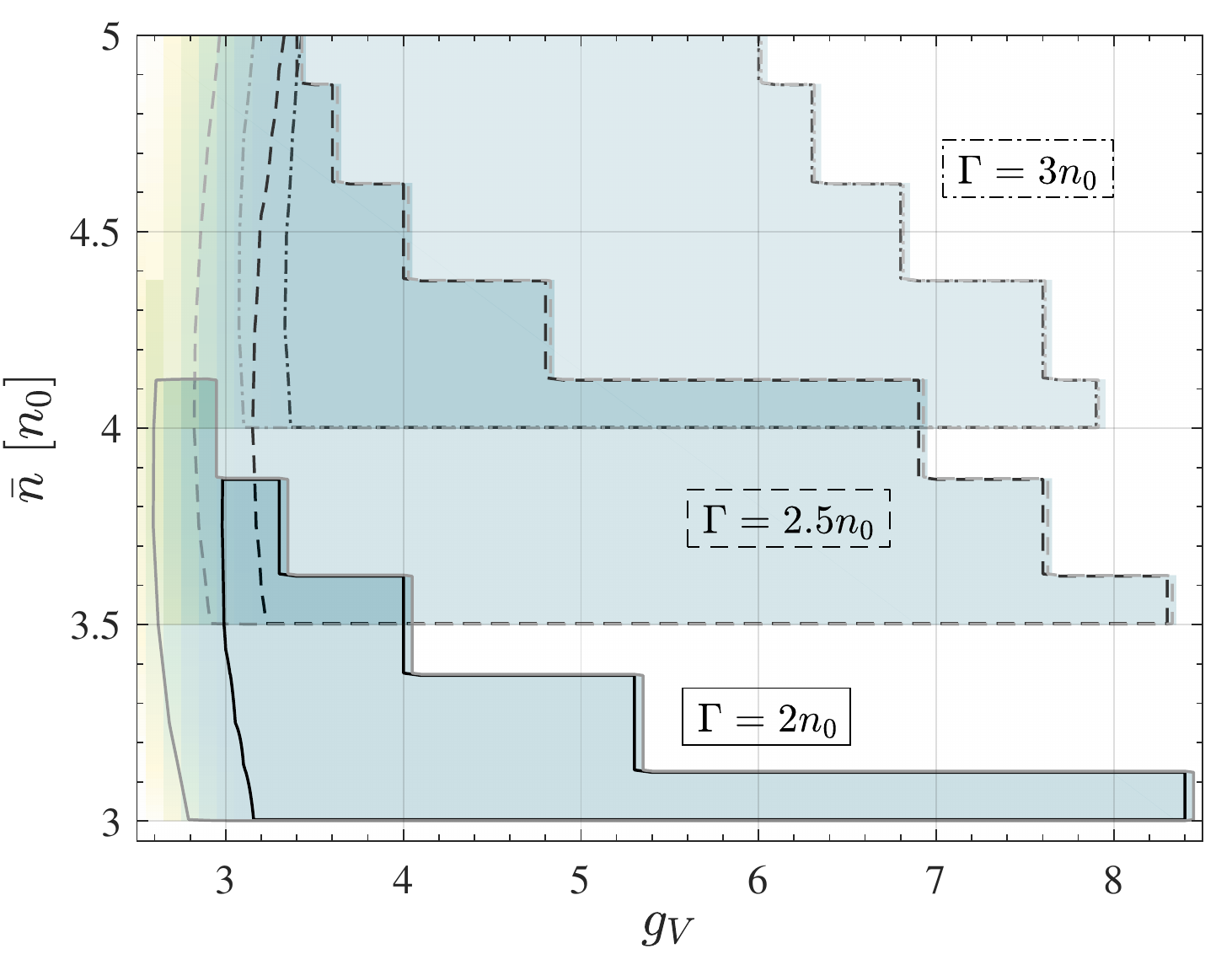}
  \includegraphics[width=0.32\textwidth]{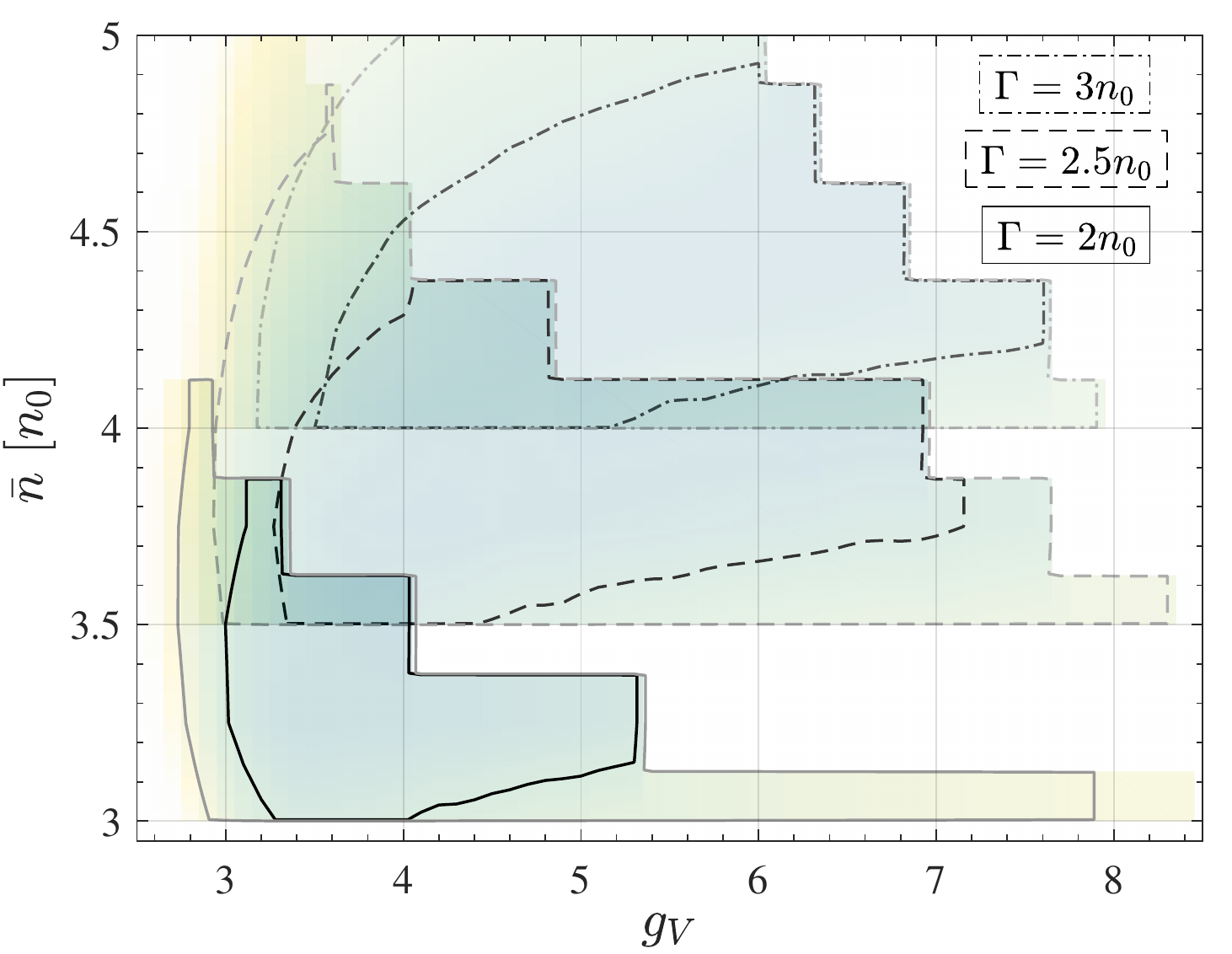}
  \includegraphics[width=0.32\textwidth]{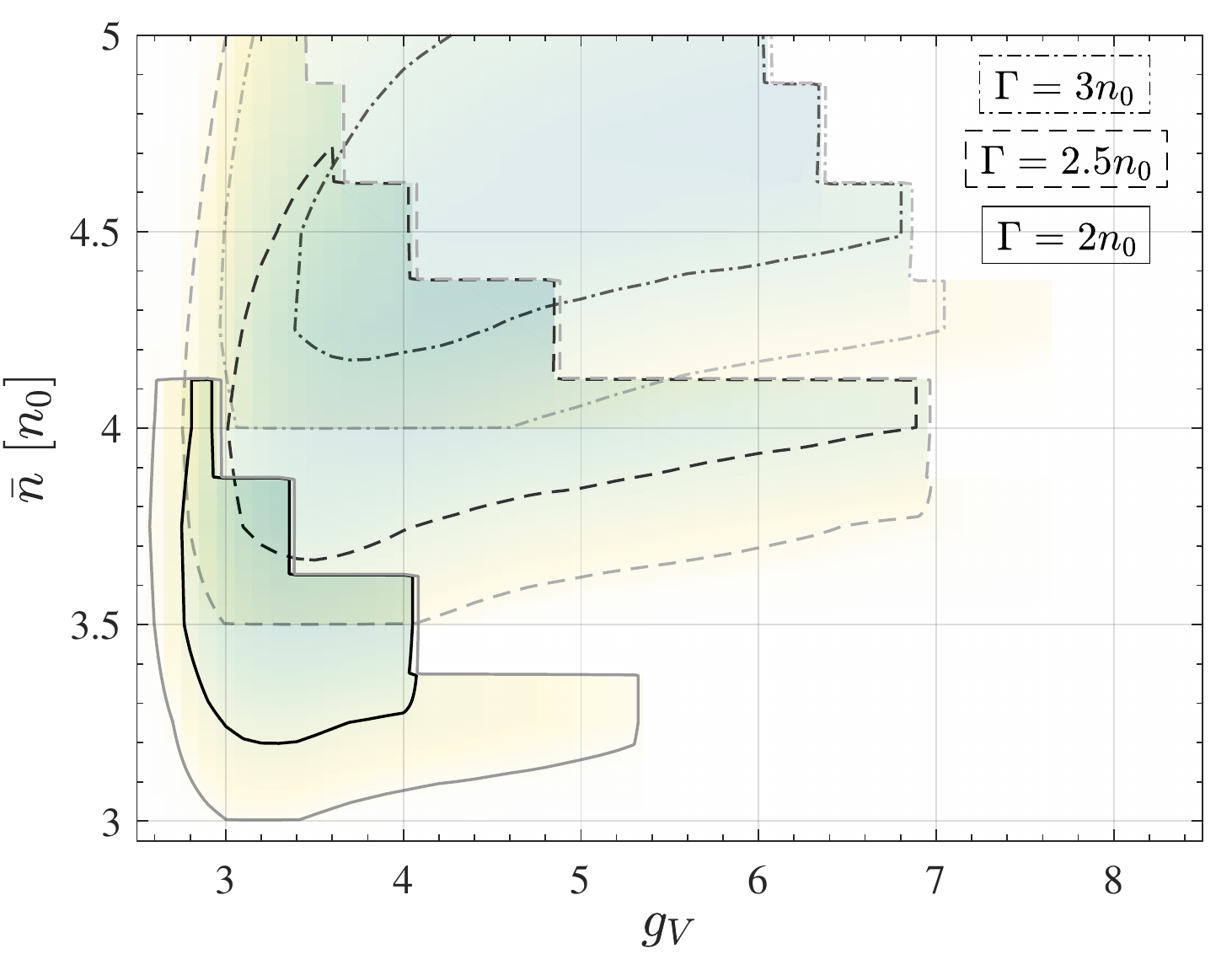}
  \caption{\label{fig:MR_Bayes}Posterior probabilities from our Bayesian analysis in the mass--radius plane (top), as well as in different slices of the parameter space (bottom). The probabilities displayed correspond to the SFHo hadronic EoS and $m_\sigma=290$~MeV, and darker colors indicate higher probabilities. On the mass--radius diagrams the outer black contours represent the boundaries of all the possible M--R curves using the given constraints. Below the mass-radius diagrams we show the radius distribution of $1.4$~$M_\odot$ (blue) and $2$~$M_\odot$ (red) NSs, with the 90\% confidence intervals indicated by the vertical dashed lines. In the bottom, the posterior probabilities with different measurements in the parameter space are shown in a contour plot, with the the two levels indicating the $68\%$ (black) and the $95\%$ (grey) credible intervals, while different contour styles represent different slices in $\Gamma$. The three panels side-by-side correspond to the posterior with the $2$~$M_\odot$ minimal constraint (left), the posterior with the two NICER measurements (middle), and the posterior with the NICER and tidal deformability measurements from GW170817 (right). All panels contain the pQCD minimal constraint as well. On the left, the grey contour represents all $M-R$ curves without the pQCD constraint applied (top), while on the middle and right panels} the dark green curves display the maximum posterior probability configurations.
\end{figure*}

We can also explore the effect of different measurements on the mass--radius diagram. The results for our canonical set of measurements are shown in Fig.~\ref{fig:MR_Bayes}. The left panel shows the results with the minimal constraints. As mentioned before, our prior is taken to be uniform in the parameter space, as it is clearly visible now in the bottom panel, where slices of the PDF for different values of $\Gamma$ are shown (see the uniform color of the PDFs, apart from the excluded regions). This does not apply to the leftmost region, where parameter sets with low $g_V$ are not preferred by the $2$~$M_\odot$ constraint. Other regions of zero probability correspond to exclusions. Values of $\bar{n}$ for which $\bar{n}-\Gamma<n_0$ are excluded since we require our EoS to be described by the hadronic EoS at least up until $n_0$. The upper right regions with high $\bar{n}$ and $g_V$ are mostly excluded by the instability or acausality of the intermediate interpolated region, and some of them are excluded due to the pQCD constraint. The sharp edges are the result of our finite grid on the parameter space.

Regions of the mass--radius diagram retained after the sharp cut-offs are encompassed by the solid contours. Examining the top left panel one can observe that the highest mass NSs with masses of $\sim3.5$~$M_\odot$ are excluded by the pQCD constraint (see the region between the black and grey contours). Our construction for the EoS results in a stiffening in the intermediate-density region, as we showed in Ref.~\cite{Kovacs2021}. A result of this is that even with the relatively soft SFHo model as the hadronic EoS we get radii $R(1.4~M_\odot)\gtrsim12$~km and hence NSs with $M_\mathrm{max}\gtrsim2~M_\odot$ and $R(1.4~M_\odot)\lesssim12$~km are absent from our prior.

Even though we take a uniform grid on the parameter space, the probability distribution in the $M-R$ plane might still exhibit irregularities, which is visible in the radius distributions as well. During the creation of the probability density plot in the $M-R$ plane we have to introduce a metric to be able to properly define densities of curves, since there is no natural connection between 'lengths' in masses and radii. We can sample points along each $M-R$ curve evenly in baryon number density, or explicitly define a metric connection between line elements in $M$ and $R$, etc. Since there is no unique way to introduce this metric, the distribution obtained will be somewhat arbitrary. Hence, the prior distribution in itself will not present definitive information. However, the change between the prior and posterior distributions is independent of the chosen metric and hence portrays faithful information about the posterior probabilities. In our analysis we choose a metric that suppresses the PDF at low densities and enhances it at high densities, in order to obtain an even distribution of points at all densities.

After taking into account the two NICER measurements (middle panel in Fig.~\ref{fig:MR_Bayes}) the probabilities are only slightly modified, since, as showed in the top panel, even the 1-sigma contours (solid yellow lines) of the two measurements completely overlap with the whole set of $M-R$ diagrams. The EoS parameters for the maximum posterior probability case are $g_V=6.9$, $\bar{n}=4n_0$ and $\Gamma=2.5n_0$. The $M-R$ curve for this parameter set is displayed in the middle panel of Fig.~\ref{fig:MR_Bayes}.

The change is more drastic when the tidal deformability measurement is taken into account as well. This measurement significantly constrains radii from above (see the indication by the yellow arrow) and consequently reduces the maximally possible NS mass as well to $\mathrm{max}(M_\mathrm{max})<2.8~M_\odot$. In the parameter space, this measurement constrains the value of the vector coupling from above, since large values of $g_V$ would correspond to stiff EoSs, which in turn would create NS sequences with large maximum masses and radii. One can observe that the probability density plot in the $M-R$ diagram extends over the black contour to the right. Examining the distribution of $R(1.4~M_\odot)$ one can also verify that while the black contour -- corresponding to the $90\%$ bound of $\tilde{\Lambda}<720$ -- crosses the $1.4~M_\odot$ line at $\sim13$~km, the $90\%$ bound of the radius distribution is $\sim13.2$~km. This phenomenon was also reported in e.g. Ref.~\cite{Jiang:2022tps} and reinforces the necessity of taking the complete data from a given measurement into account instead of simply bounds from some credible intervals. The parameter set corresponding to the maximum probability EoS in this case is $g_V=6.5$, $\bar{n}=4.5n_0$ and $\Gamma=2.75n_0$.

\begin{figure}[!tb]
  \centering
  \includegraphics[width=0.4\textwidth]{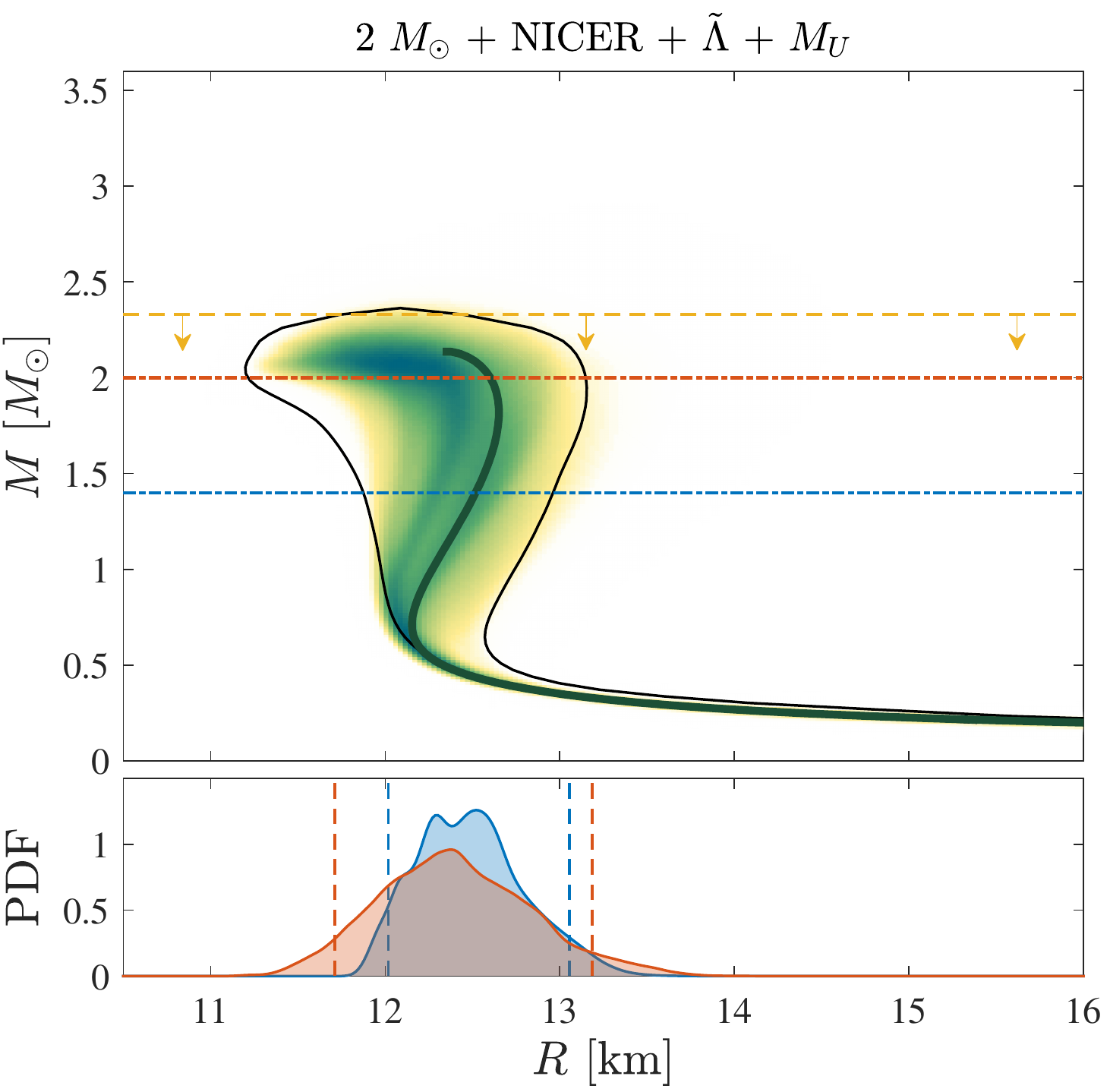}
  \includegraphics[width=0.4\textwidth]{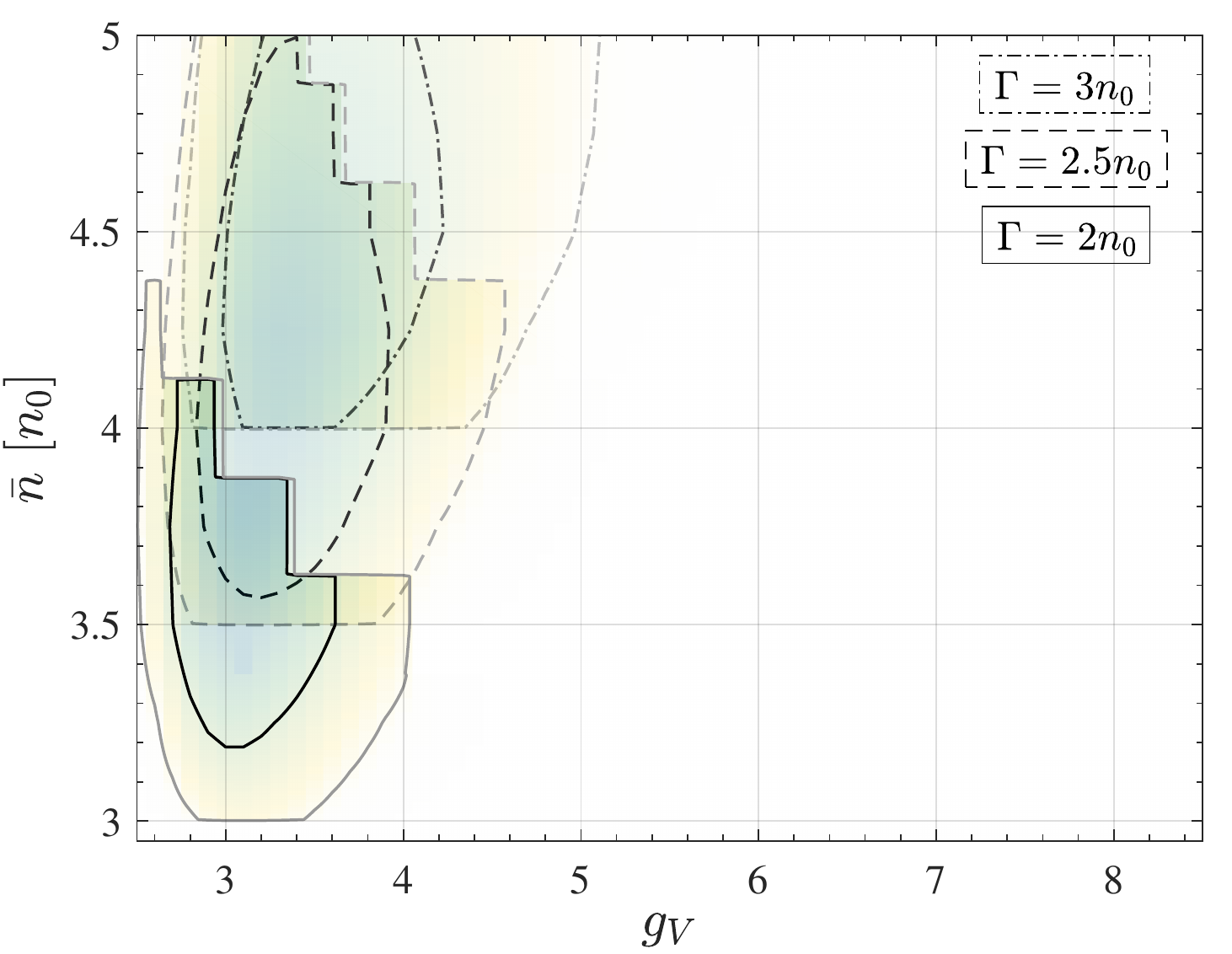}
  \caption{\label{fig:MR_Bayes_Mu}Same as in Fig.~\ref{fig:MR_Bayes} but with the upper mass constraint from the hypermassive NS hypothesis also applied, in addition to the NICER and tidal deformability measurements.}
\end{figure}

In addition to the canonical measurements we can investigate the constraint imposed by the hypermassive NS hypothesis. Specifically, we can use an upper mass bound based on Ref.~\cite{Rezzolla2017}. We did not include this measurement in our canonical set, since there is still some ambiguity around the modeling of the kilonova signal AT2017gfo, and therefore the ejected mass in the merger event. The results for this scenario are shown in Fig.~\ref{fig:MR_Bayes_Mu}. The black contour in the top panel that encompasses all the possible $M-R$ curves that meet all the requirements does not shrink at lower masses, which is to be expected for a sufficiently robust ensemble of EoSs. On the other hand, the $90\%$ credible interval for $R(1.4M_\odot)$ shrinks from a width of $1.10$~km to $1.04$~km and shifts to lower values from an upper bound of $13.20$~km to $13.06$~km. The main effect of this step on the parameter space, as expected from Sec.~\ref{ssec:Mmax}, is an upper bound on the vector coupling, as can be seen in the lower panel of Fig.~\ref{fig:MR_Bayes_Mu}, which shows the probability densities for $m_\sigma=290$~MeV. The maximum posterior probability corresponds to the parameter set $g_V=3.1$, $\bar{n}=3.5n_0$ and $\Gamma=2n_0$, which means $n_{BU}=5.5n_0$. Despite this moderate value, this NS does not develop a quark core.

\begin{figure}[t]
  \centering
  \includegraphics[width=0.4\textwidth]{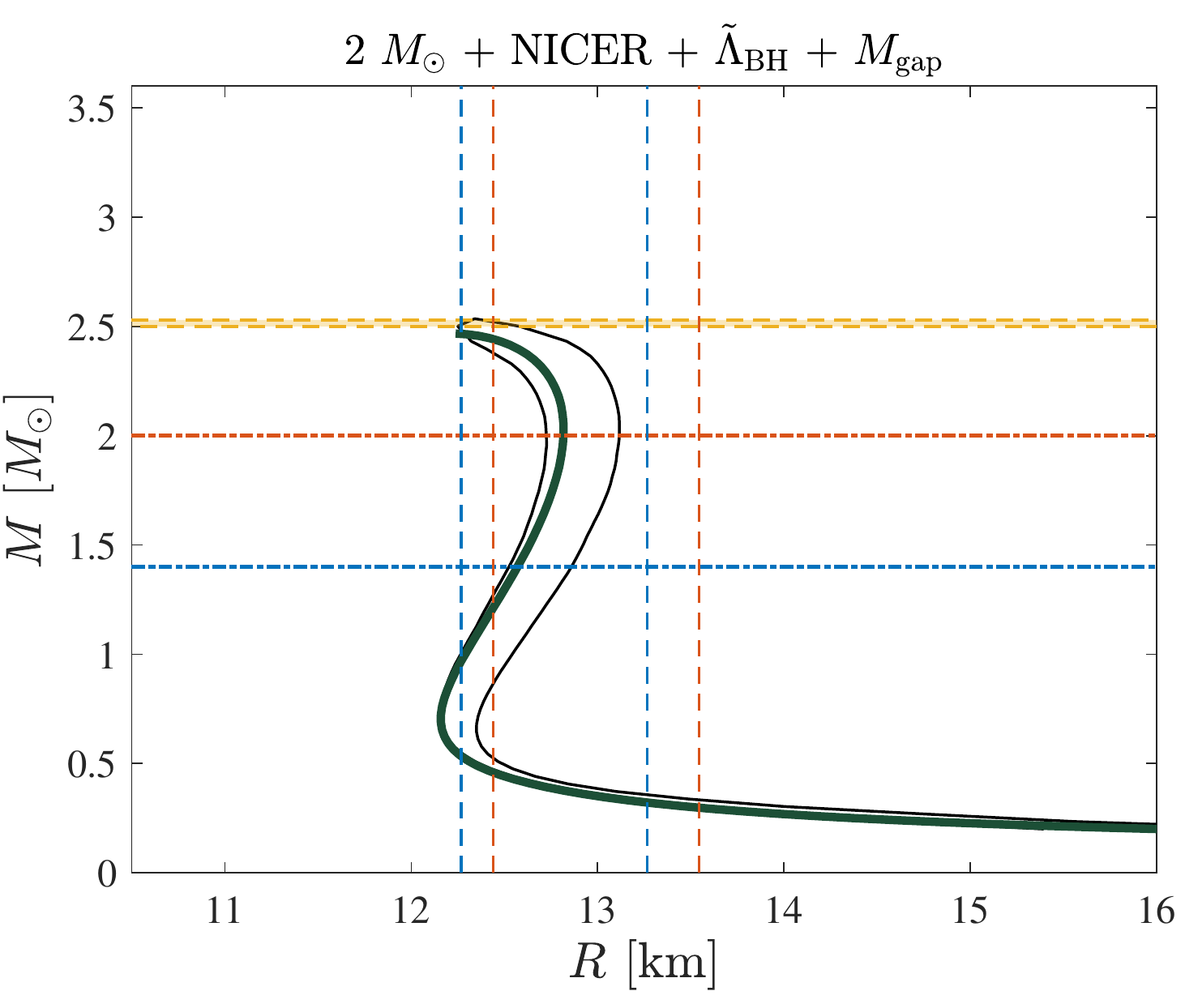}
  \includegraphics[width=0.4\textwidth]{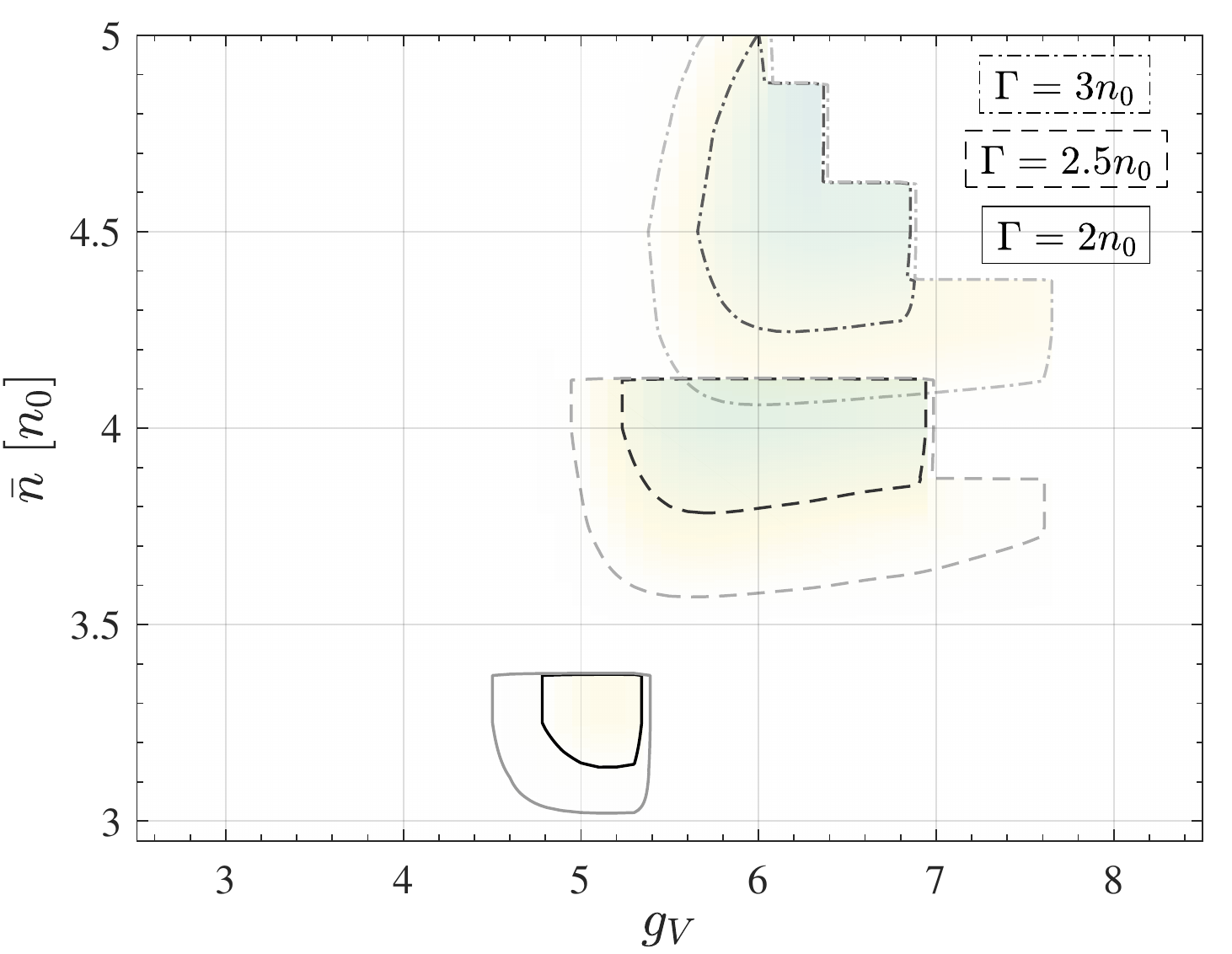}
  \caption{\label{fig:MR_Bayes_Mgap}Same as in Fig.~\ref{fig:MR_Bayes_Mu} but instead of taking into account the upper mass bound based on the hypermassive NS hypothesis we only include the constraint from the BH hypothesis, while identifying the mass-gap object in GW190814 as a NS. Due to the statistically low number of EoSs fulfilling these tight constraints, the shapes of the posterior PDFs become irregular, and hence we omit them, only showing the 90\% radius bounds.}
\end{figure}

Alternatively, resolving the hypermassive NS hypothesis, we can keep the BH hypothesis and assume, in addition, that the mass-gap object in GW190814 was an extremely massive NS. Even though by our current understanding of the nuclear EoS, such a massive NS seems unlikely, it is still allowed by astrophysical measurements, as we show in Fig.~\ref{fig:MR_Bayes_Mgap}, although together with the BH hypothesis they leave only a narrow region for the maximum mass of NS sequences (see the yellow band between $2.5~M_\odot$ and $2.53~M_\odot$). Due to this narrow permitted region, our statistical ensemble of permitted EoSs will be greatly reduced, leaving only $\sim100$ EoSs with sufficiently large posterior probabilities. Therefore, the posterior radius distributions will have an irregular shape and will be unreliable, although the 90\% credible intervals might still be used. Note that the difference here between the black contour in the $M-R$ diagram and the 90\% credible intervals of the radius distributions is even more pronounced than in the right panel of Fig.~\ref{fig:MR_Bayes}. The $90\%$ upper bound on $R(1.4M_\odot)$ here is $\sim13.3$~km, in contrast to the value of $\sim12.9$~km predicted by the black contour. The parameters that correspond to the maximum posterior probability are $g_V=4.9$, $\bar{n}=4.25n_0$ and $\Gamma=2.75n_0$.

\begin{figure}[htb!]
  \centering
  \includegraphics[width=0.4\textwidth]{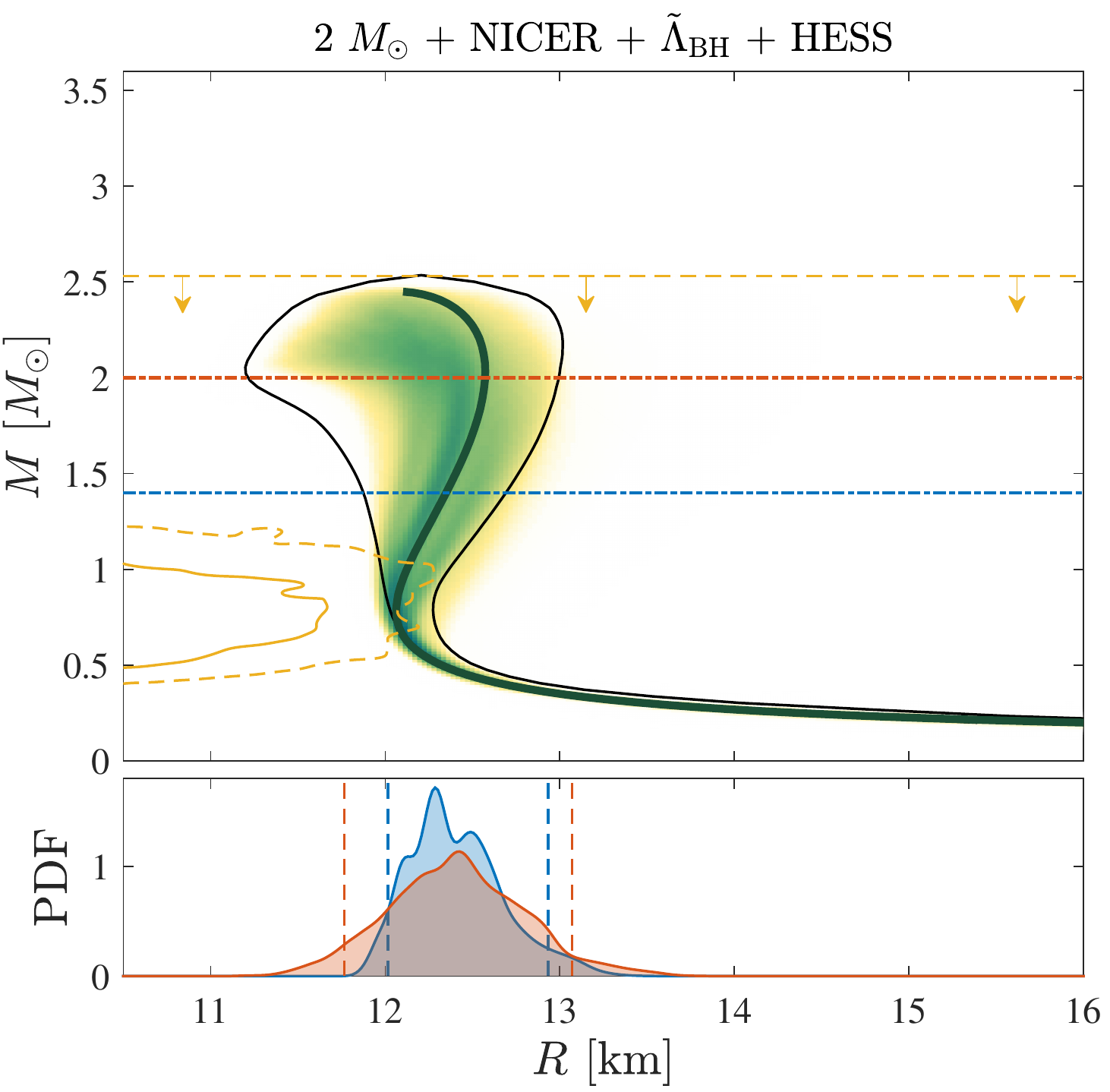}
  \includegraphics[width=0.4\textwidth]{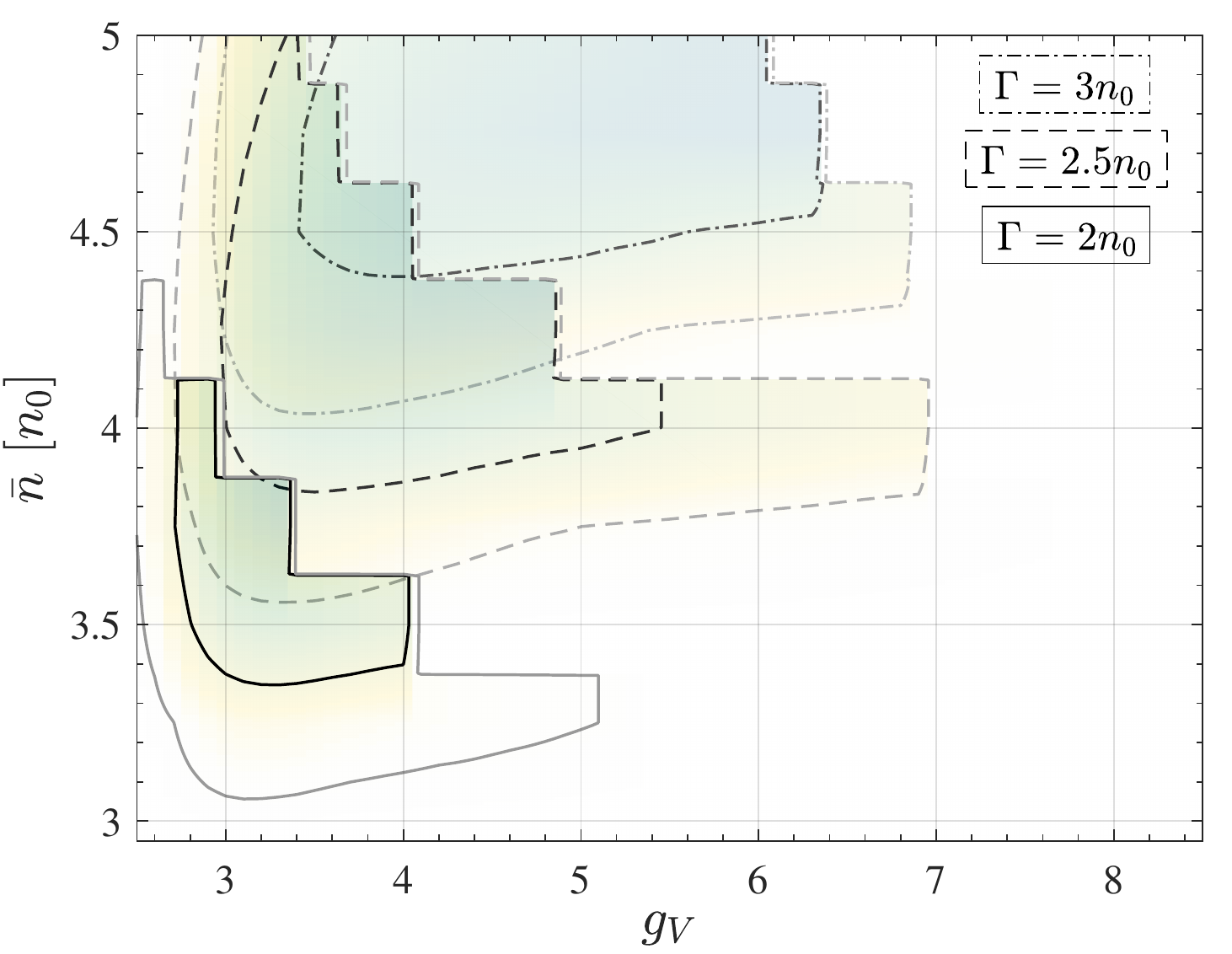}
  \caption{\label{fig:MR_Bayes_HESS}Same as in Fig.~\ref{fig:MR_Bayes} but with the constraint from the central compact object inside HESS J1731-347 also applied, in addition to the NICER and tidal deformability measurements, and the BH hypothesis.}
\end{figure}

Finally, Fig.~\ref{fig:MR_Bayes_HESS} shows the posteriors when in addition to the NICER and tidal deformability measurements, the BH hypothesis and the constraint from the central compact object inside HESS J1731-347 is also taken into account. The two-sigma credible interval of the measurement barely overlaps with our set of mass-radius curves, however, a considerable region is still allowed on the $M-R$ plane. Comparing the posterior probabilities on the parameter space to those in the bottom right panel in Fig.~\ref{fig:MR_Bayes}, we see that with this constraint included parameter sets with a low value of $\bar{n}$ are less probable, and the high-probability regions shift to higher values of $\bar{n}$. The maximum posterior probability corresponds to the parameter set $g_V=4.7$, $\bar{n}=4.25n_0$ and $\Gamma=2.5n_0$.

We also performed this analysis for different sigma meson masses and the DD2 hadronic EoS as well. Our results for the radius bounds and the highest posterior probability parameter sets can be found in Tables \ref{tab:RSFHo}, \ref{tab:RDD2}, \ref{tab:parSFHo} and \ref{tab:parDD2} of Appendix~\ref{App:tables_figs}. Radius bounds do not change drastically when using different sigma meson masses, although higher values of $m_\sigma$ correspond to larger radii. Using the stiffer, DD2 hadronic EoS on the other hand results in a significant increase in the radius bounds. In case we set the sigma meson mass to $600$~MeV or $700$~MeV, we are left with a low number of stable and causal EoSs, therefore we omitted the analysis for these parameters.

\begin{figure*}[htbp]
  \centering
  \includegraphics[width=0.24\textwidth]{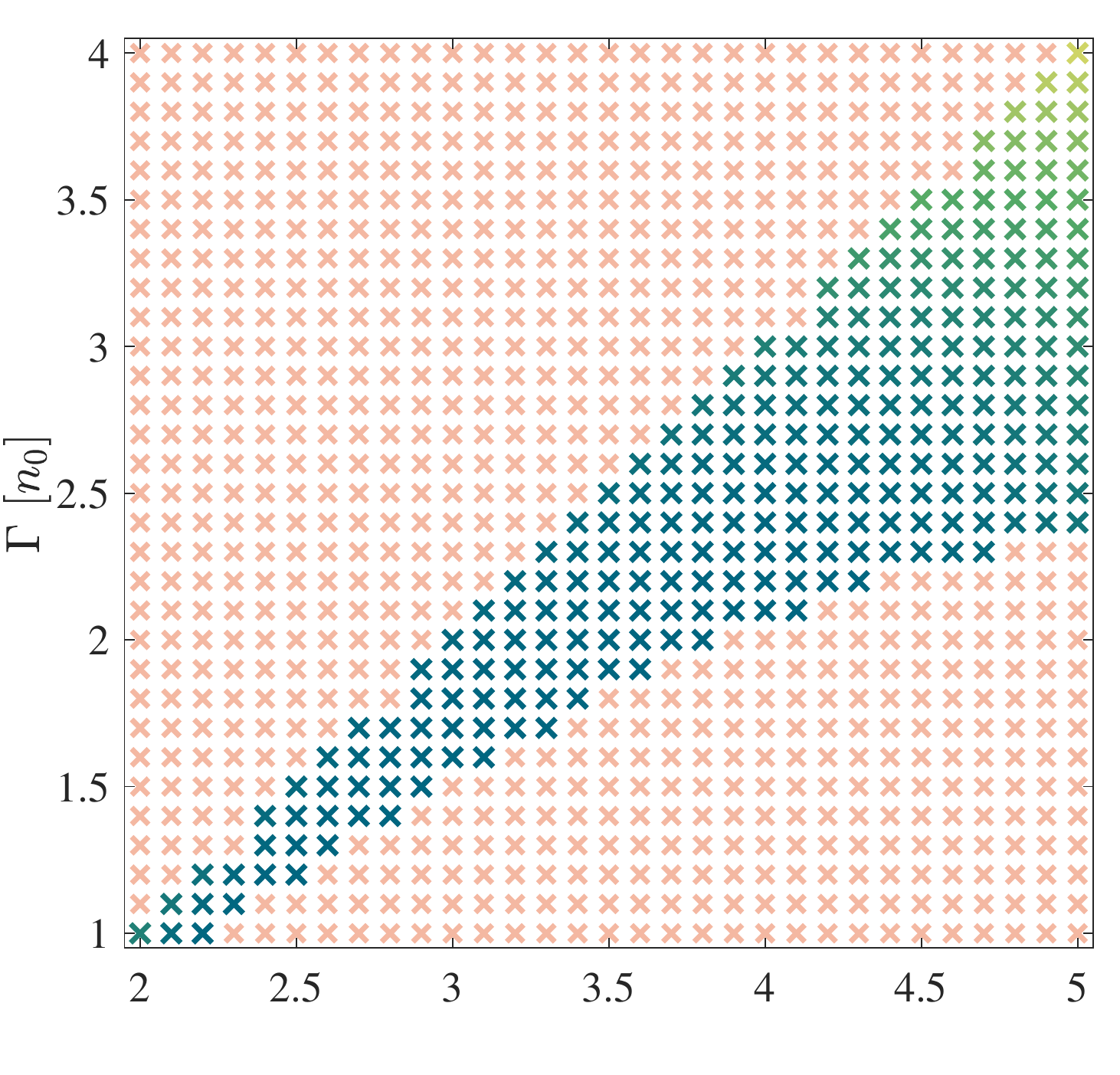}
  \includegraphics[width=0.24\textwidth]{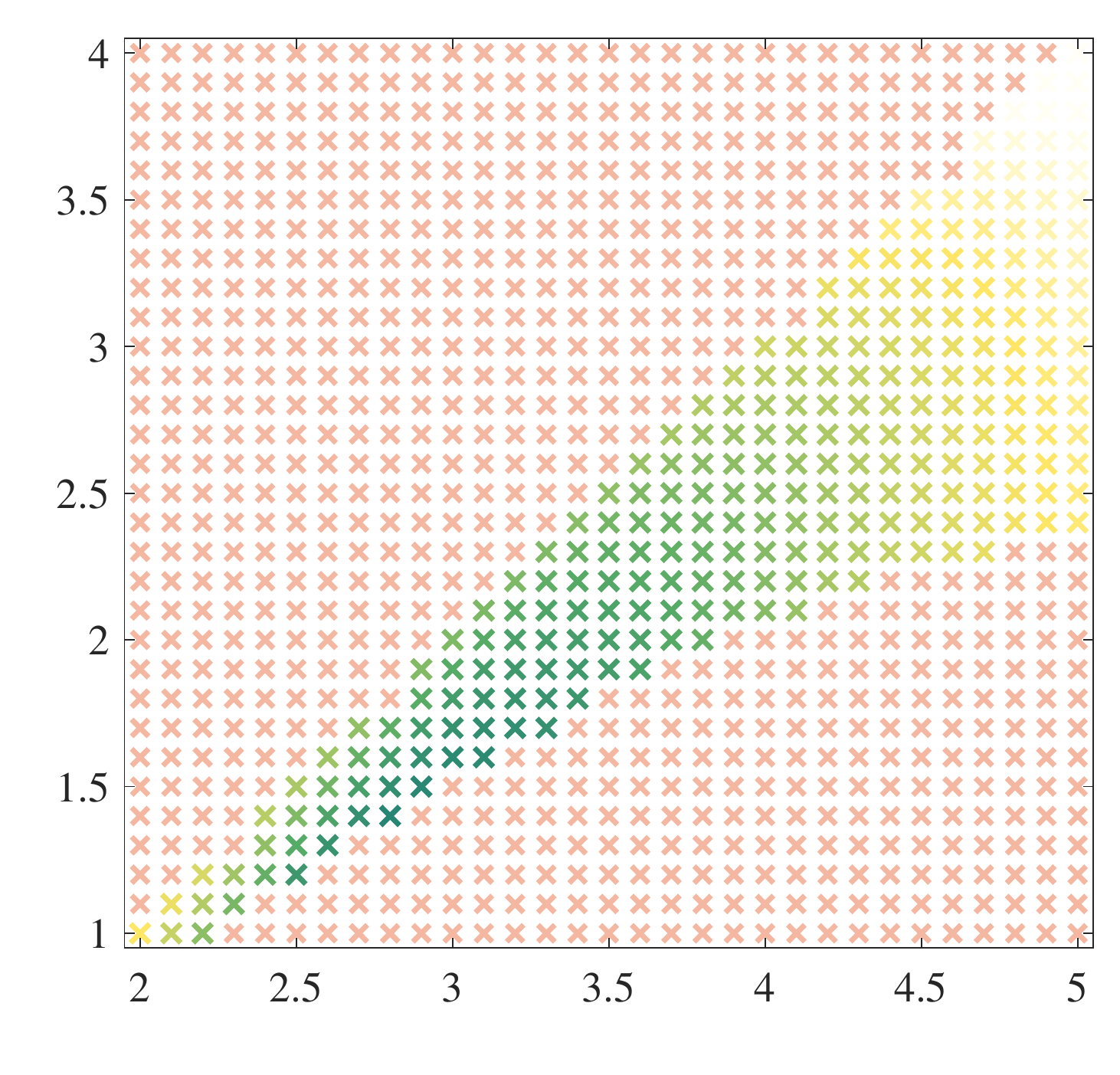}
  \includegraphics[width=0.24\textwidth]{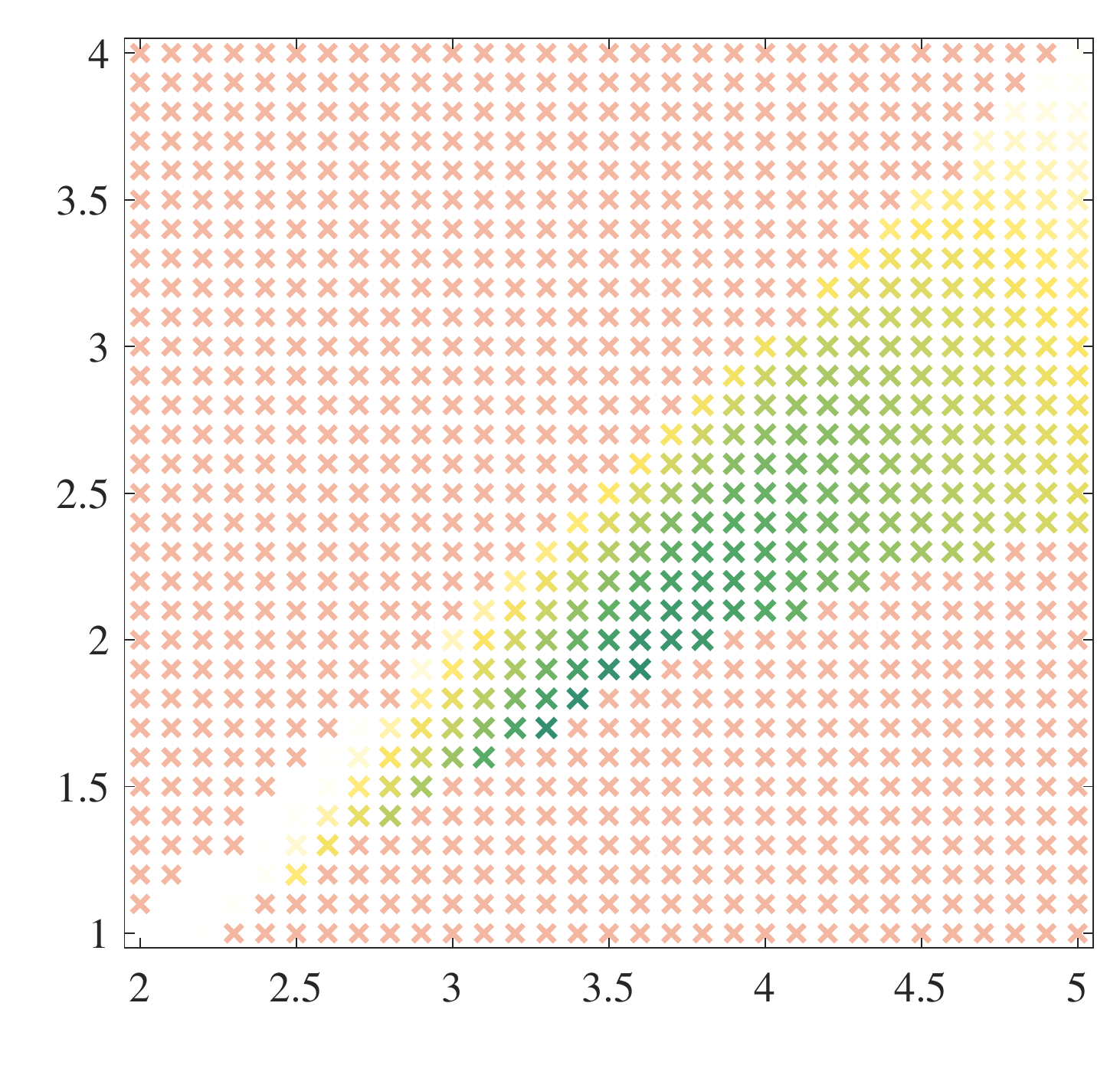}
  \includegraphics[width=0.24\textwidth]{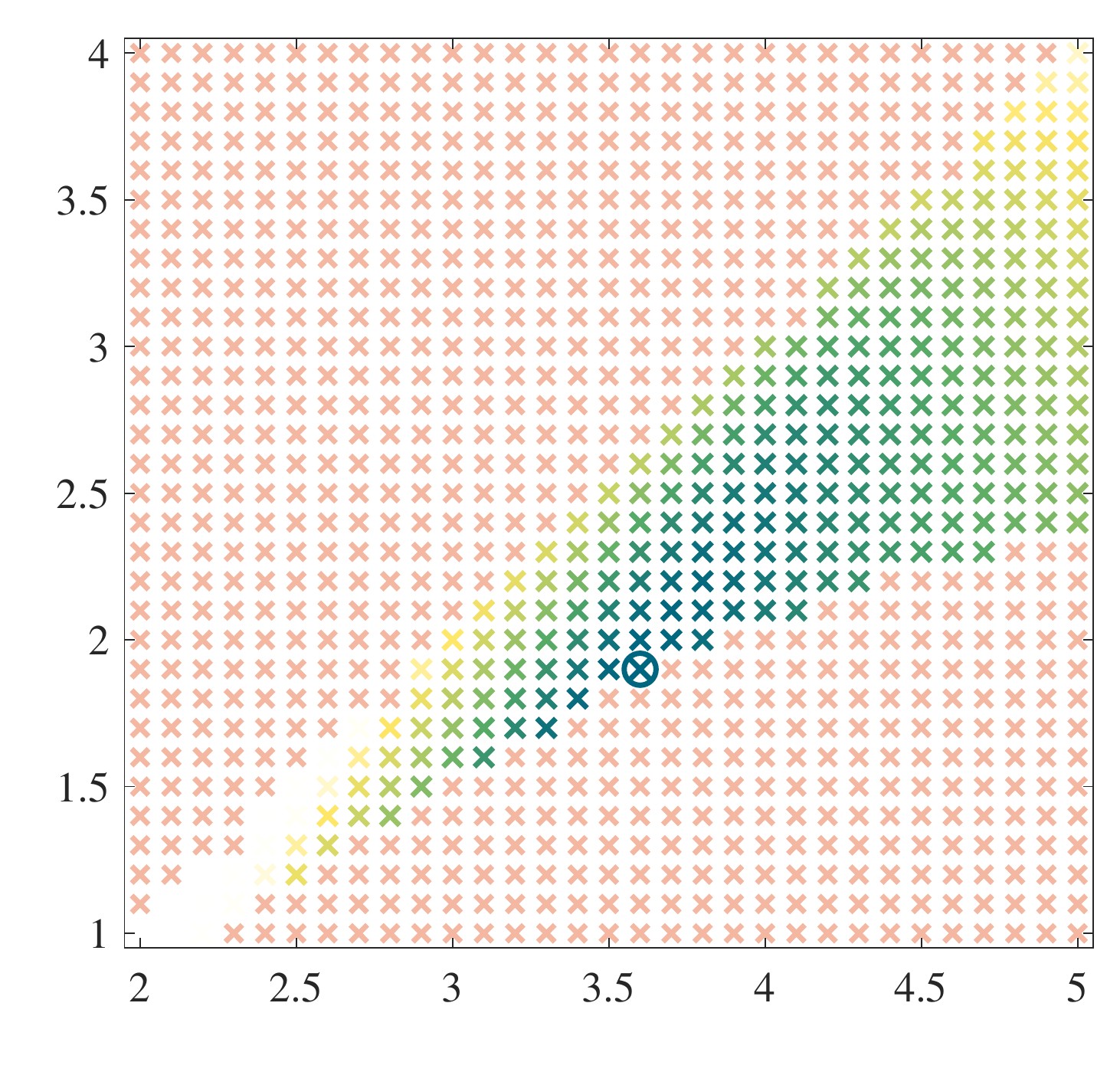}
  \includegraphics[width=0.24\textwidth]{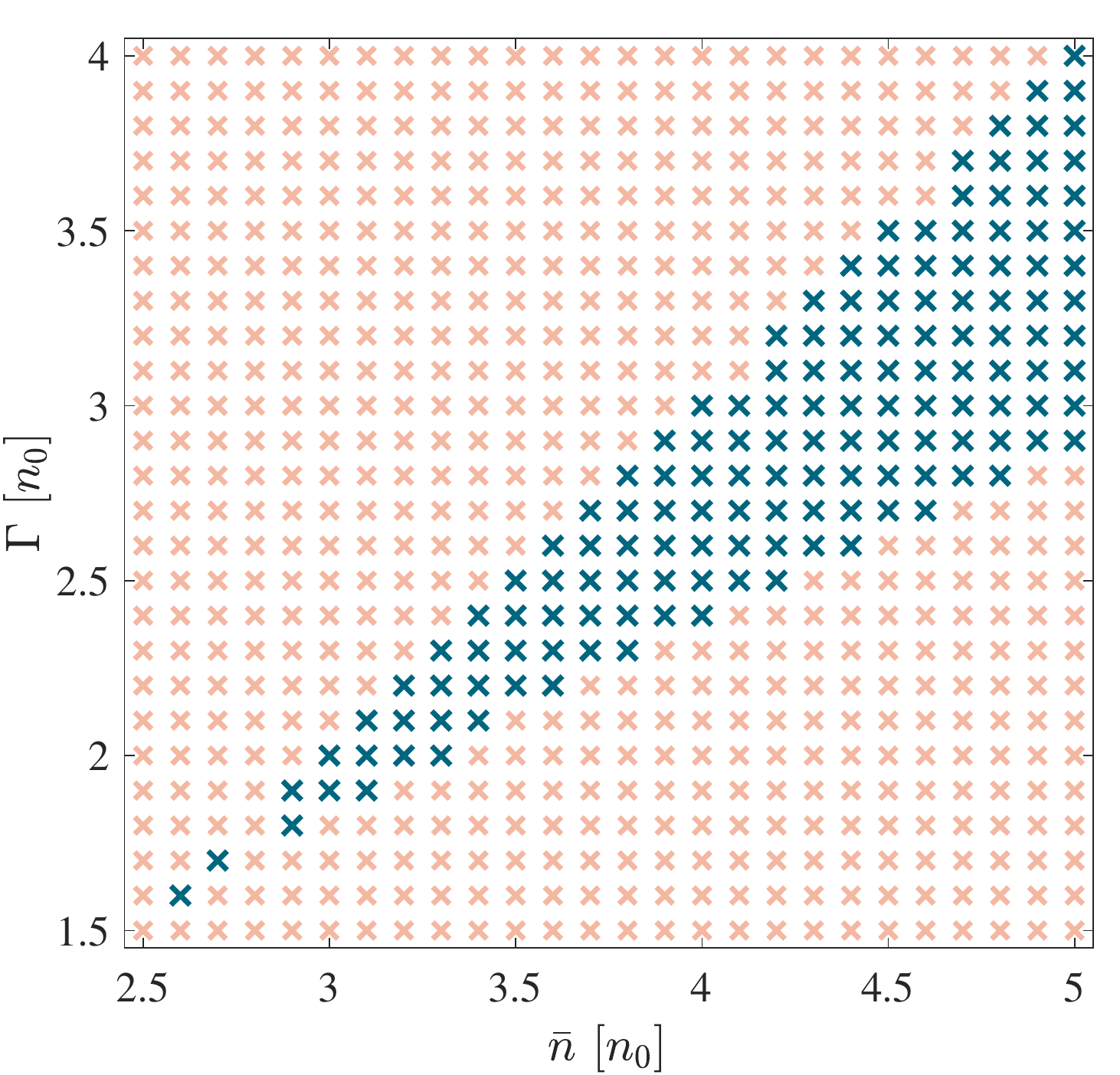}
  \includegraphics[width=0.24\textwidth]{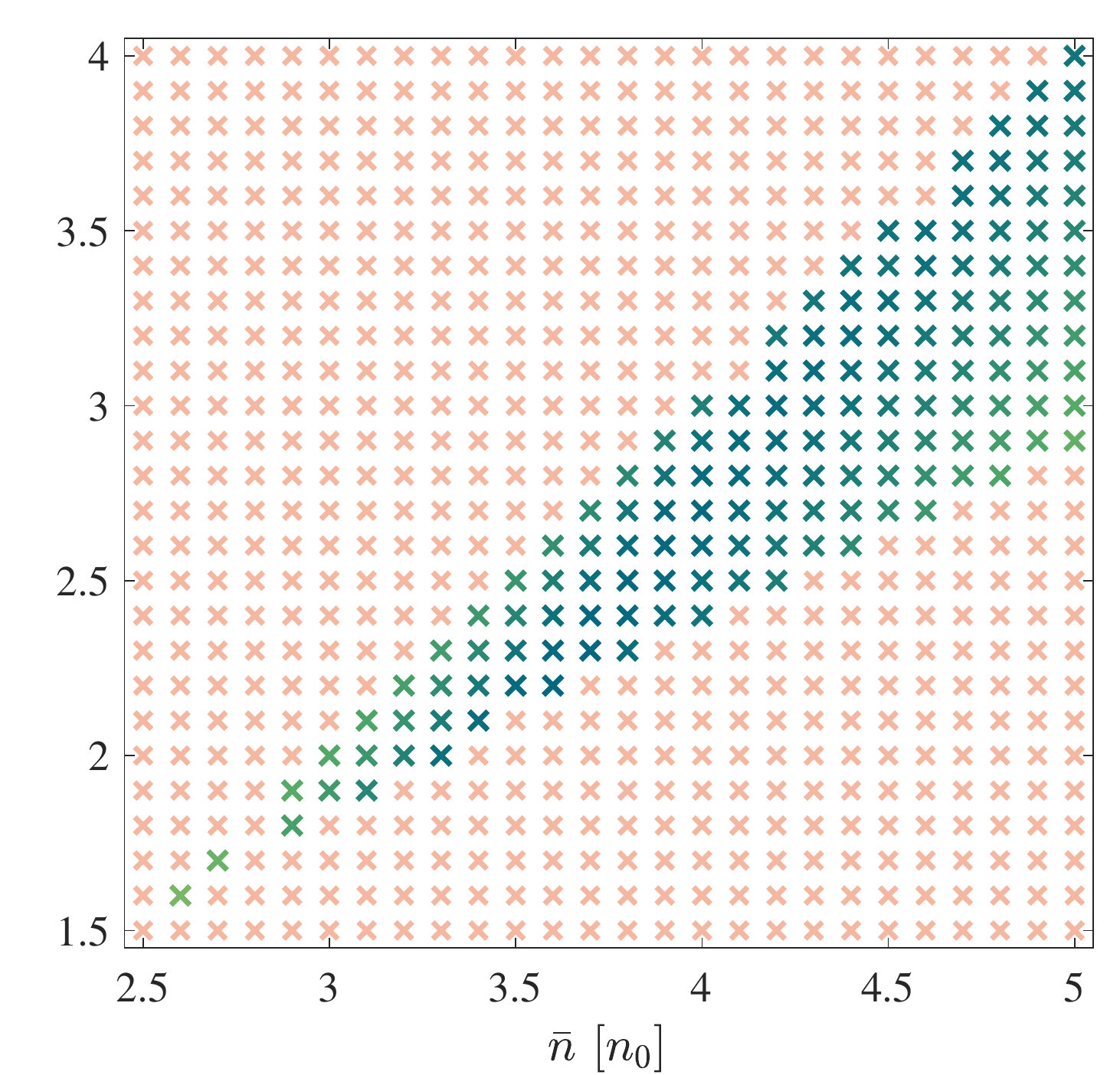}
  \includegraphics[width=0.24\textwidth]{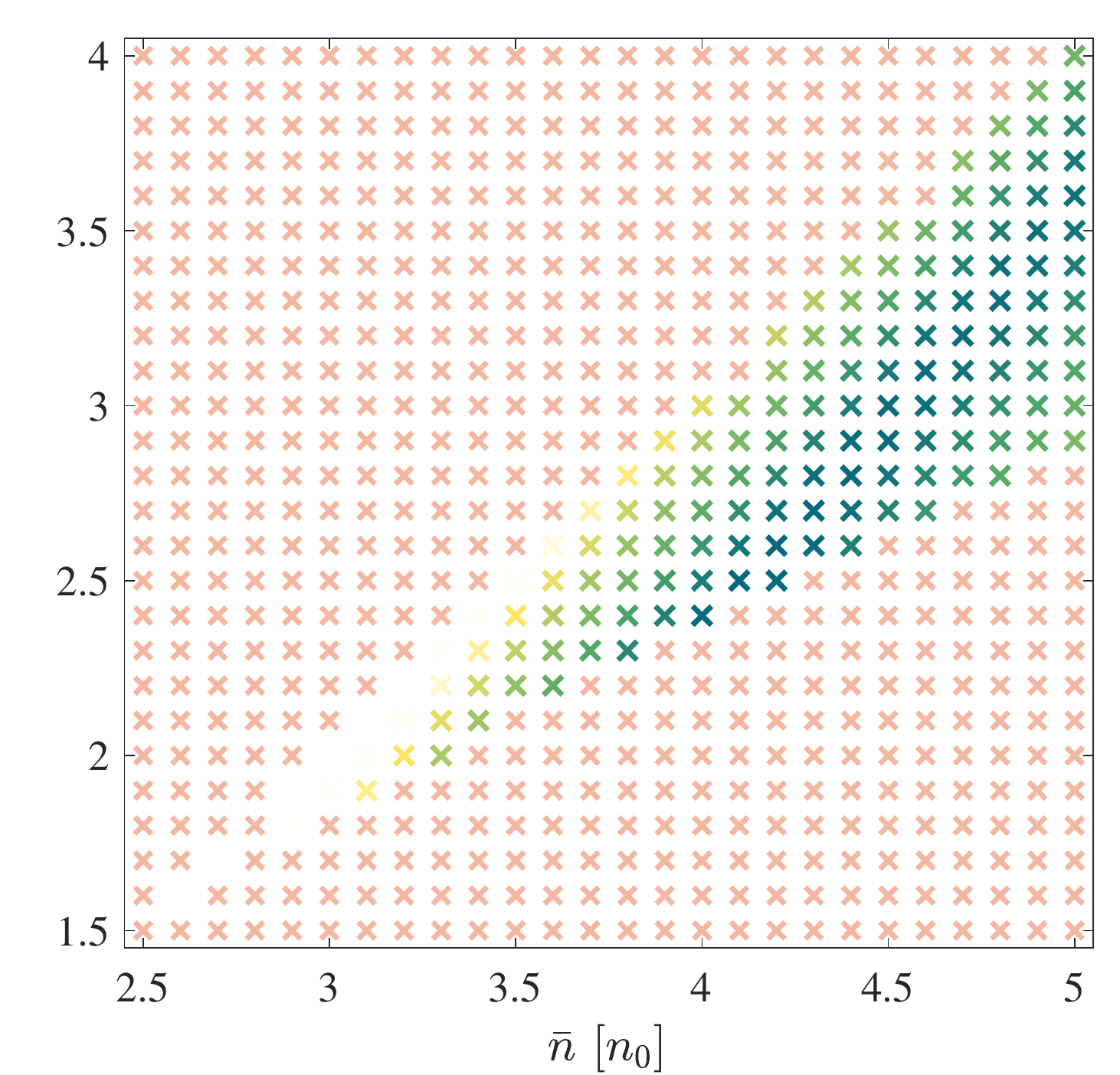}
  \includegraphics[width=0.24\textwidth]{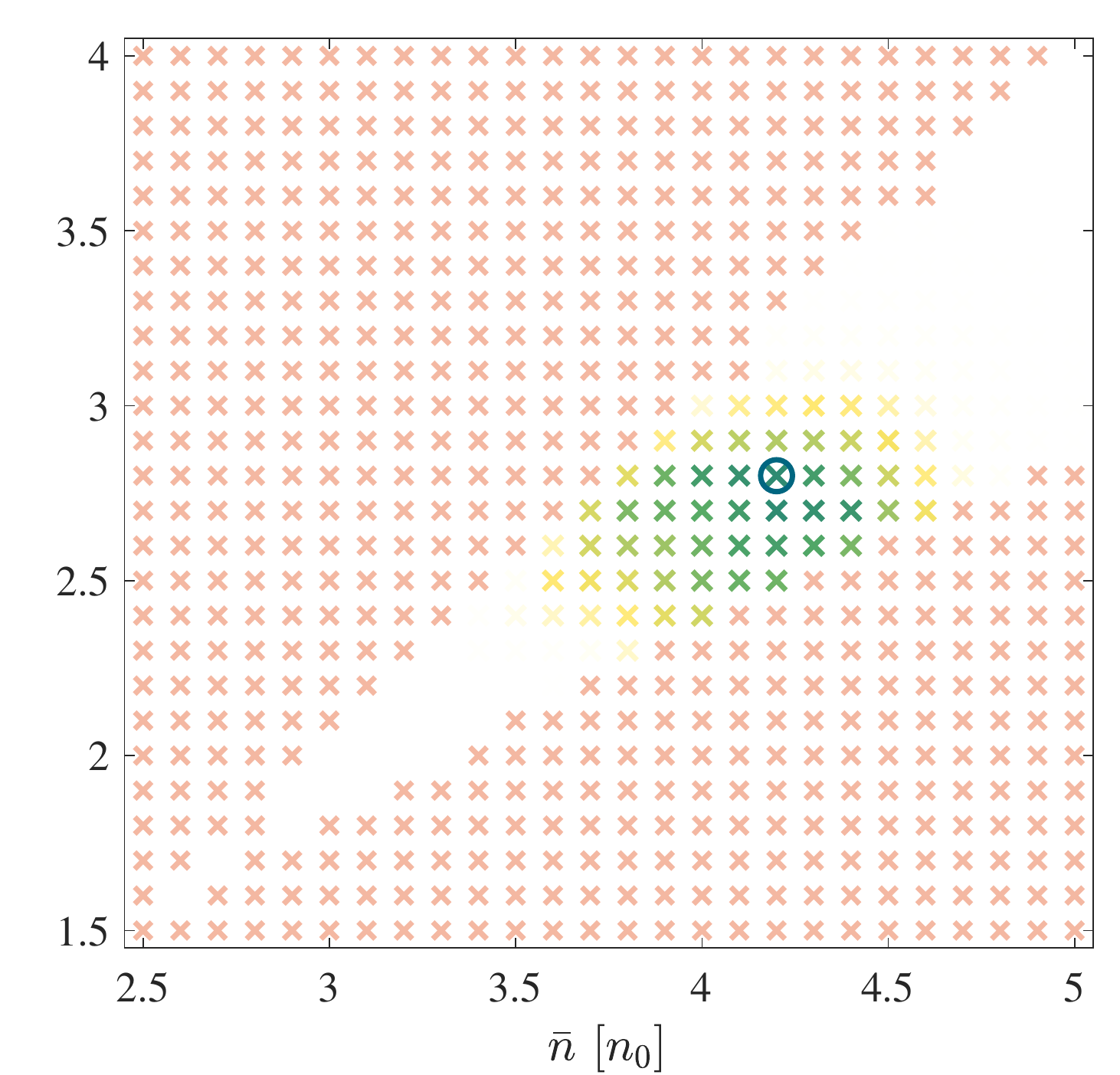}
  \caption{\label{fig:MR_Bayes_pt}Prior and posterior probabilities of different parameter sets in the plane of the phase transition parameters $\bar{n}$ and $\Gamma$, for two slices with fixed $g_V$. Darker colors indicate higher probabilities (white areas correspond to $\sim0$ probability), with probabilities having been normalized by the maximum probability in the complete parameter space (including all $g_V$). Parameter sets with an orange color are excluded by the requirement $\bar{n}-\Gamma\geq n_0$ or due to acausality and/or instability. The top panels correspond to slices in the parameter space with $g_V=3.1$, while the bottom panels have $g_V=4.7$. These values correspond to the EoSs with the maximum posterior probability for the hypermassive NS hypothesis and the mass-gap object scenarios, respectively. The first three panels from left to right both for the top and bottom panels show the posterior with the minimal constraints, the posterior with the NICER measurements, and the posterior with NICER and tidal deformability measurements, respectively. The rightmost panel at the top has the upper mass constraint from the hypermassive NS hypothesis included as well, while the one at the bottom contains instead the BH hypothesis together with the lower mass bound from the mass-gap object in GW190814. Parameter sets with the maximum probability are marked with a ring.}
\end{figure*}

So far we have only investigated the effect of various measurements on the mass--radius diagram and different $g_V-\bar{n}$ slices of the posterior PDFs. It is also instructive to look at different $\bar{n}-\Gamma$ slices and inspect what can be inferred about the parameters of the hadron-quark phase transition. This is shown in Fig.~\ref{fig:MR_Bayes_pt}, where we calculated the posteriors on a finer grid compared to the previous figures. The two rows correspond to two different slices with a fixed $g_V$. These two slices where chosen to match the parameters with the maximum posterior probability case in the hypermassive NS hypothesis scenario (top) and the mass-gap NS scenario (bottom). The first three panels in both rows from left to right show the posterior with the minimal constraints, the posterior with the NICER measurements, and the posterior with NICER and tidal deformability measurements, respectively. The rightmost panel at the top has the upper mass constraint from the hypermassive NS hypothesis included as well, while the one at the bottom contains the lower mass bound from the mass-gap object and exclusions from the BH hypothesis instead. Hence, these parameter planes can be viewed as an evolution of posterior probabilities with the inclusion of more and more constraints in these two scenarios. The upper left excluded region corresponds to the requirement $\bar{n}-\Gamma\geq n_0$, while the lower right part is excluded due to acausality or instability.

Looking at the top row, at first it might seem like the hypermassive NS hypothesis broadens the region of high-probability parameters from the third to the fourth panel. However, this illusion is due to the fact that these probabilities were normalized by the maximum posterior probability of the entire parameter space (including all $g_V$ values), and hence probabilities in the second and third panels at the top are suppressed, since the maximum posterior probabilities in these cases correspond to $g_V=6.9$ (NICER) and $g_V=6.5$ (NICER+$\tilde{\Lambda}$), respectively, which are outside the parameter space slices shown in Fig.~\ref{fig:MR_Bayes_pt}. Note however, that it is not the case for the bottom panels, where the maximum probability case is always closer in $g_V$ to the slice shown. Interestingly, in many cases, higher posterior probabilities are situated close to the edges of the allowed regions, adjacent to unstable or acausal EoSs. This means that the lowest possible value of $\Gamma$ is preferred for a given $\bar{n}$, which also means that astrophysical observations prefer a very stiff intermediate-density region. Such stiff intermediate regions are also predicted by the theory of the so-called quarkyonic matter (see e.g. Refs.~\cite{McLerran:2018hbz,Jeong:2019lhv}). The two scenarios depicted in Fig.~\ref{fig:MR_Bayes_pt} end up with different preferred values for $\bar{n}$ and $\Gamma$, and the preferred value of these parameters also varies from step to step, hence, no robust statement can be made about the values of the phase transition parameters. However, very low values of $\bar{n}$ and $\Gamma$ are disfavoured after taking into account the tidal deformability measurement, hence the existence of pure quark matter at densities below $\sim4n_0$ is also disfavoured.

\subsubsection{Results for the complete EoS ensemble}

\begin{figure*}[htbp]
  \centering
  \includegraphics[width=0.32\textwidth]{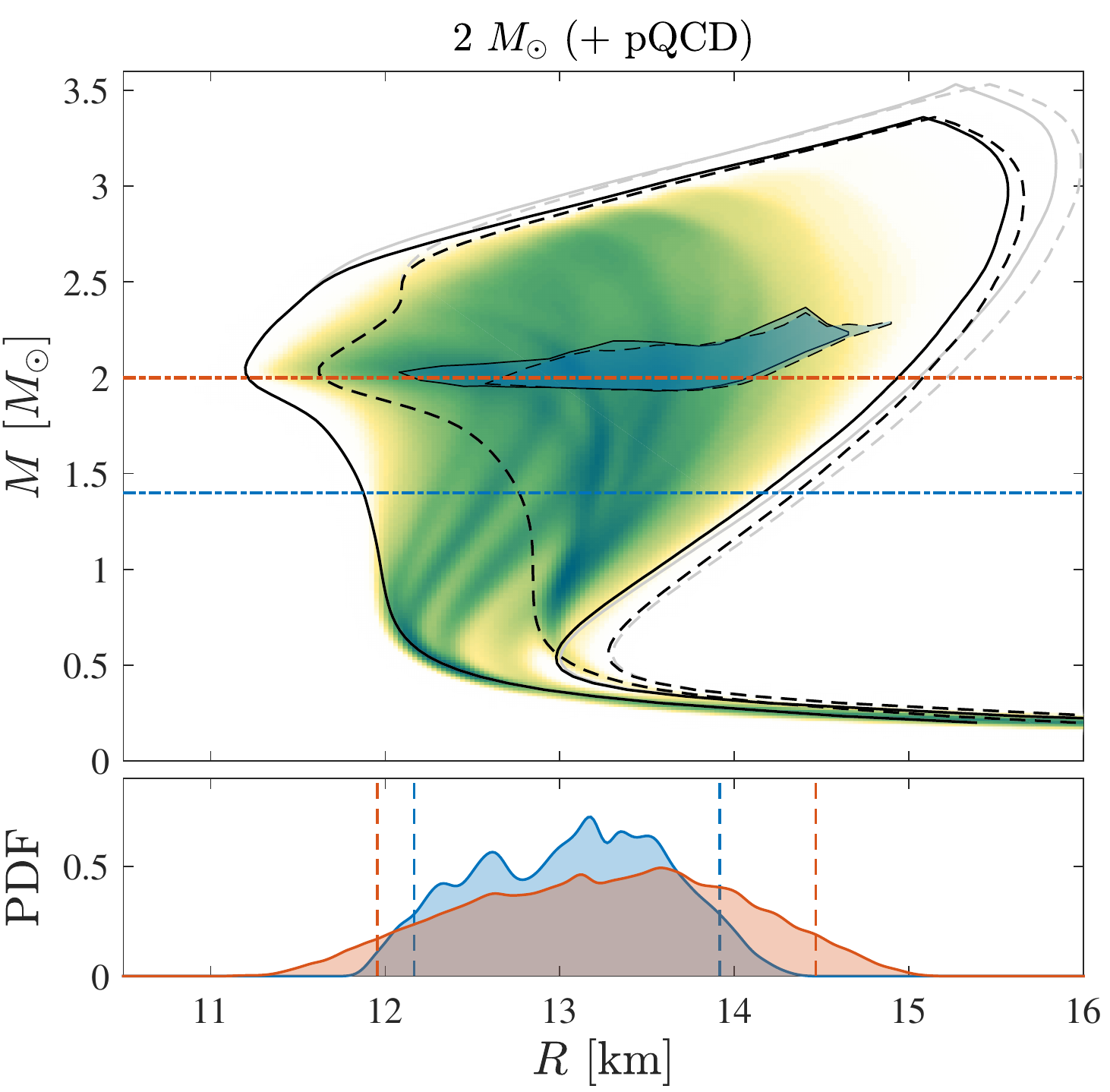}
  \includegraphics[width=0.32\textwidth]{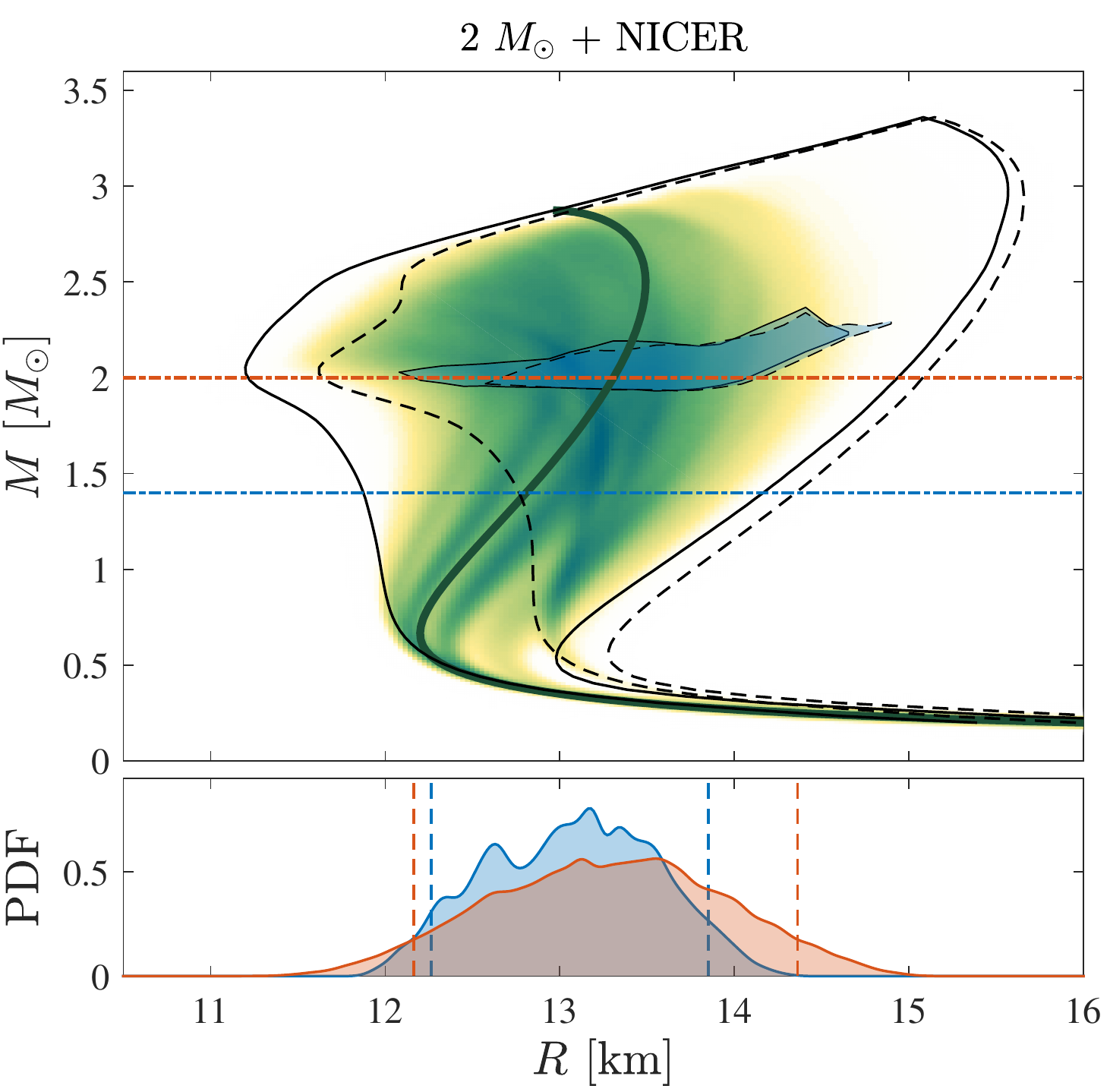}
  \includegraphics[width=0.32\textwidth]{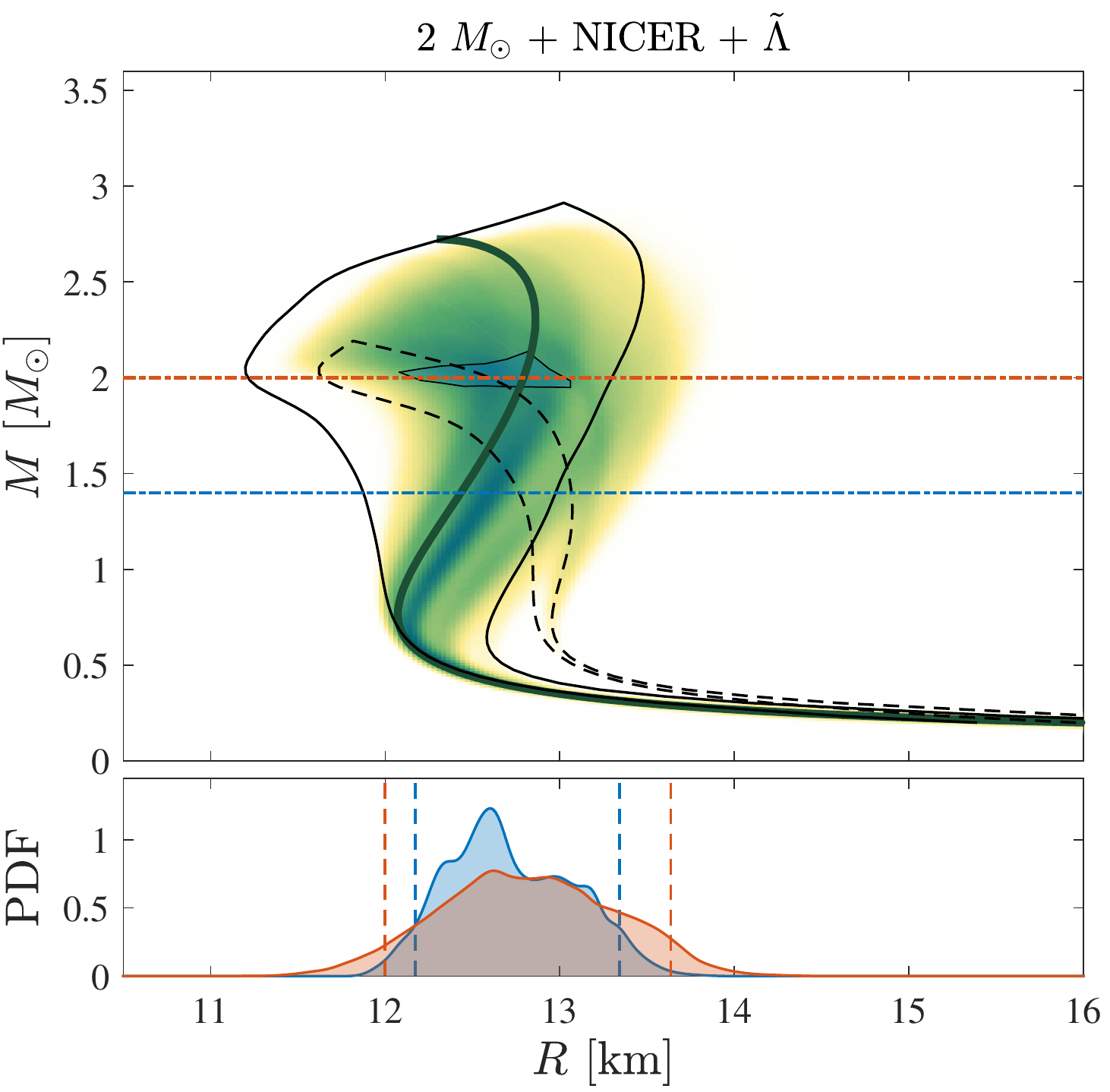}
  \includegraphics[width=0.32\textwidth]{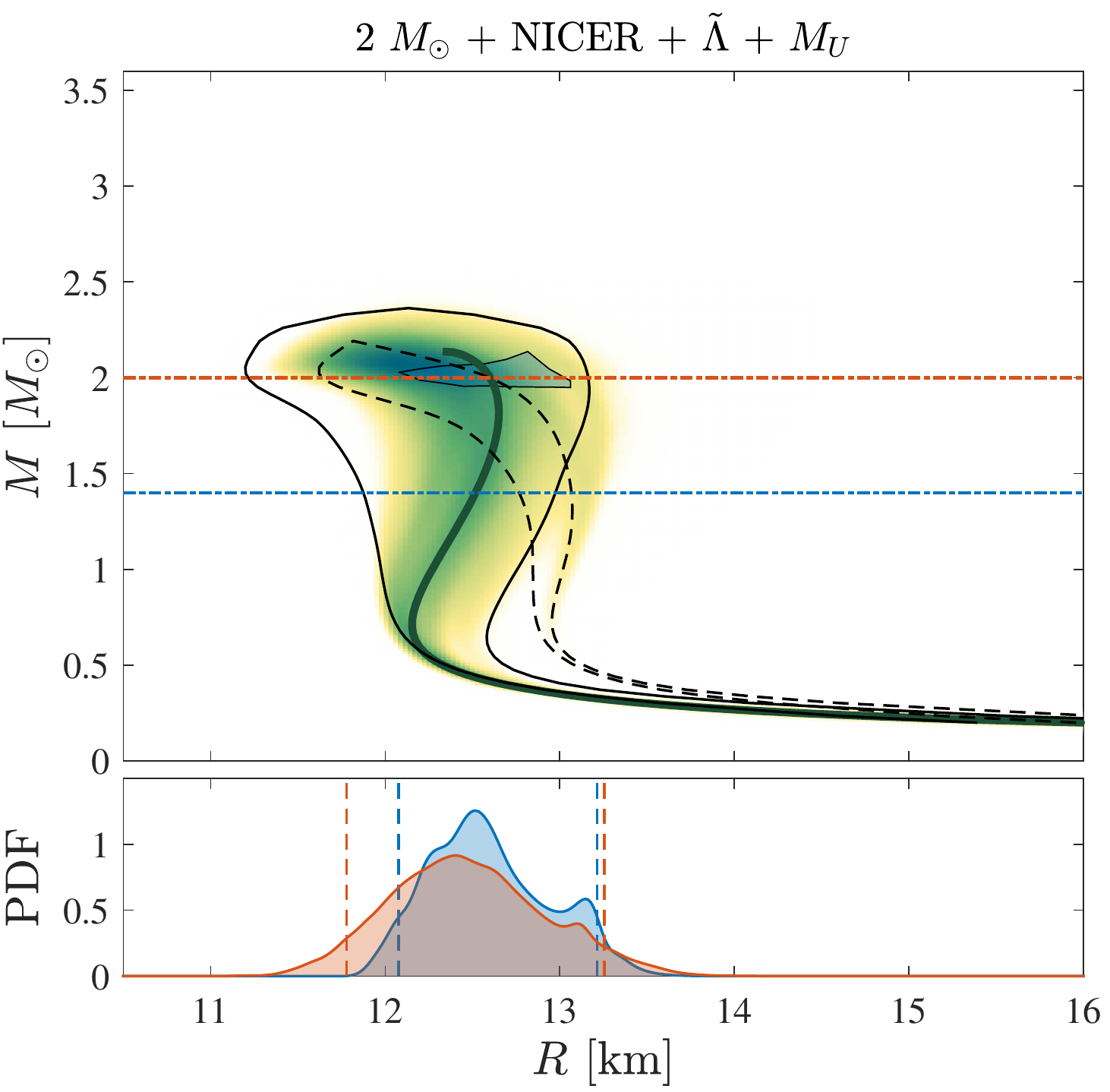}
  \includegraphics[width=0.32\textwidth]{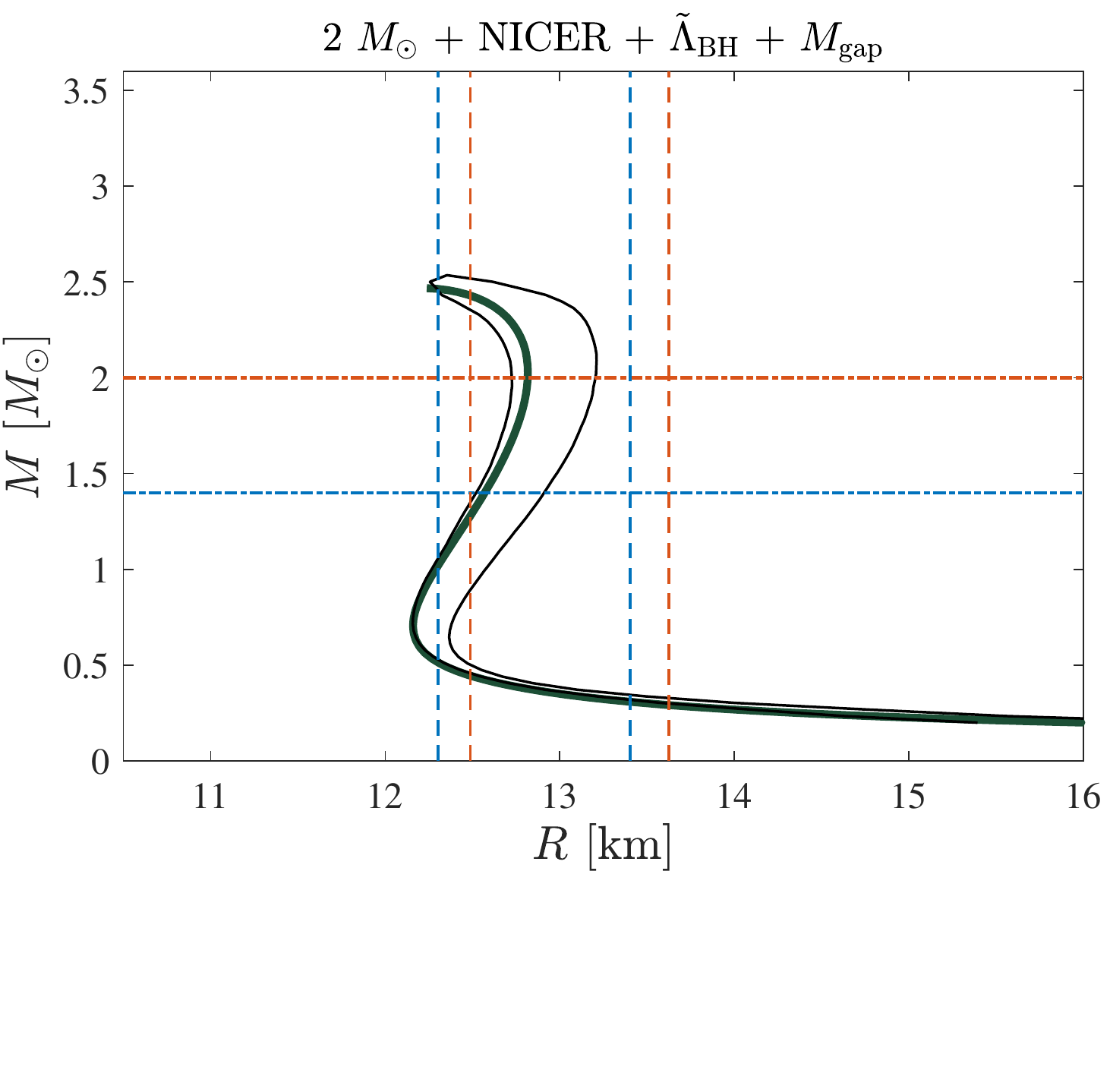}
  \includegraphics[width=0.32\textwidth]{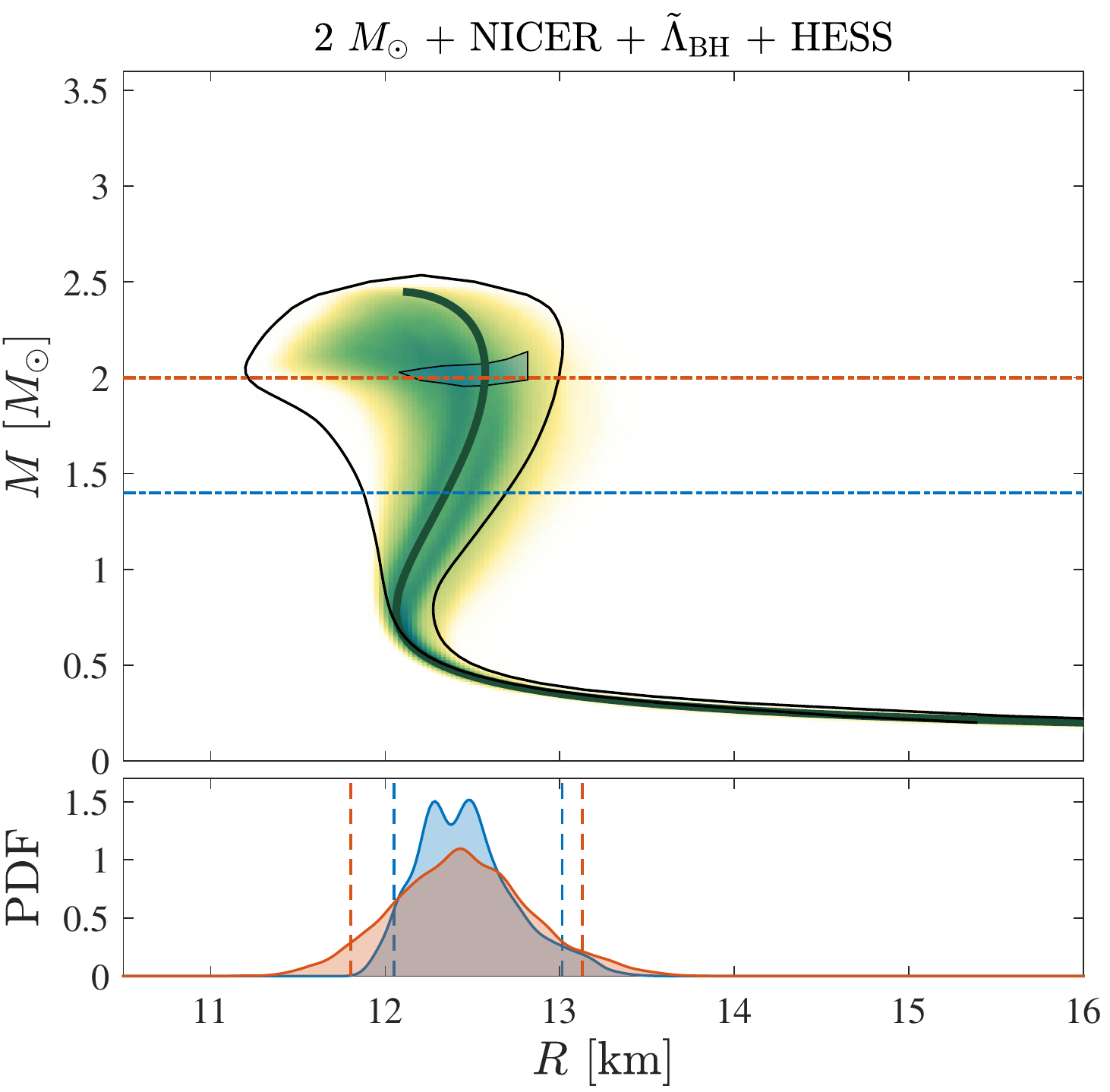}
  \caption{\label{fig:MR_Bayes_full}Posterior probabilities from our Bayesian analysis in the mass--radius plane, using the complete set of EoSs with different sigma meson masses, and the SFHo and DD2 hadronic EoSs combined. Darker colors indicate higher probabilities. The outer black contours represent the boundaries of all the possible M--R curves using the given constraints for the SFHo (solid) and the DD2 (dashed) hadronic EoSs. The radius distribution of $1.4$~$M_\odot$ (blue) and $2$~$M_\odot$ (red) NSs is also shown, with the 90\% confidence intervals indicated by the vertical dashed lines. The measurements taken into account for each panel are indicated above them. The pQCD minimal constraint is included in every panel. On the left, the grey contour represents all $M-R$ curves without the pQCD constraint applied (top), while on the middle and right panels the dark green curves display the maximum posterior probability configurations. The inner blue regions on the $M-R$ diagrams represent NSs with pure quark cores. In the mass-gap object scenario, due to the statistically low number of EoSs fulfilling the tight constraints, the shapes of the posterior PDFs become irregular, and hence we do not show them, only the 90\% radius bounds.}
\end{figure*}

After performing the analysis for the restricted case, we repeat it with our complete EoS ensemble in order to get more representative results for the radius bounds, the preferred parameters, and also generally for bounds on the EoS itself. Fig.~\ref{fig:MR_Bayes_full} yet again shows the posterior PDFs on the mass-radius plane and for the radii of $1.4$~$M_\odot$ and $2$~$M_\odot$ NSs. We also show the contours as a result of sharp cut-offs separately for the SFHo (solid contour) and the DD2 (dashed contour) hadronic EoSs, since at low densities the two hadronic EoSs produce very different radii. Additionally, we also show the regions where, given the specific constraints, hybrid stars with pure quark cores can exist. For this analysis, we define matter being in pure quark state when the density is above $n_B>n_{BU}=\bar{n}+\Gamma$, hence the EoS is characterized by our quark model.

An interesting feature of the region of NSs with a pure quark core is that there is an upper mass bound for these objects ($\sim2.35$~$M_\odot$), which might not seem obvious at first. However, one can understand this by examining the sound speed squared in the intermediate-density region (at $\sim3-5 n_0$, see e.g. Fig.~10. in Ref.~\cite{Kovacs2021}). First there is a stiff peak that is then followed by a valley, which, in some cases, can get close to $c_s^2\approx0$. After this valley, the sound speed increases and only then the quark EoS is reached. Therefore, EoSs that exhibit a large peak in the beginning and hence create high-mass NSs have a huge drop in the speed of sound, which makes these NS sequences prone to becoming unstable before they can develop a pure quark core. Interestingly, Ref.~\cite{Annala:2019puf} finds a similar upper mass bound for NSs with pure quark cores ($\sim2.25$~$M_\odot$) using a completely different definition for a pure quark core.

Due to the stiffness of the DD2 hadronic EoS, upper bounds on radii constrain mass-radius curves with the DD2 EoS even more. The tidal deformability constrain limits the maximum possible NS mass more significantly than in the case of the SFHo hadronic EoS, to $\mathrm{max}(M_\mathrm{max})<2.2~M_\odot$. The region of hybrid stars with a pure quark core also shrinks significantly for the SFHo EoS, and it even disappears for the DD2 EoS in this case. Adding the hypermassive NS hypothesis to this measurement does not constrain $M-R$ curves with the DD2 EoS any further, neither does it reduce the region of hybrid stars with a quark core for $M-R$ curves with the SFHo EoS. For the mass-gap object scenario, interestingly, EoSs created by using the relatively stiff DD2 EoS for the hadronic part can not produce any $M-R$ curves that can satisfy the conditions $\tilde{\Lambda}<720$ and $M_\mathrm{max}>2.5~M_\odot$ at the same time. This is due to the fact that the first of these two conditions limits radii of low mass NSs from above and since a stiff hadronic EoS generates larger radii in general, this condition will limit the maximum mass of NSs (as can be seen in the top right panel of Fig.~\ref{fig:MR_Bayes_full}). Therefore, the existence of very massive NSs is only possible in case the hadronic EoS is soft enough to produce sufficiently small radii. Another interesting consequence of the mass-gap object interpreted as a NS is that none of the NSs would have a core consisting of pure quark matter in this case. This can be understood by noting that the maximum mass of such hybrid stars is $\sim2.35~M_\odot$, while the minimum required mass for maximally stable NSs is higher in this case. Considering the measurement from the central object of HESS J1731-347, none of the NSs with the DD2 EoS is allowed, since they generally predict large radii for low mass NSs.

\begin{figure}[t!]
\centering
\begin{tabular}[c]{|c||c|c|}\hline
\label{tab:R}
Measurement & $R_{1.4}$ [km] & $R_{2.0}$ [km] \\\hline\hline
Prior (+pQCD) & $12.52^{+1.23}_{-0.71}$ & $13.25^{+1.21}_{-1.36}$ \\\hline
$2~M_\odot$ & $13.10^{+0.81}_{-0.94}$ & $13.28^{+1.18}_{-1.33}$ \\\hline
$2~M_\odot$ + NICER & $13.09^{+0.76}_{-0.82}$ & $13.29^{+1.08}_{-1.12}$ \\\hline
$2~M_\odot$ + NICER + $\tilde{\Lambda}$ & $12.67^{+0.67}_{-0.50}$ & $12.81^{+0.83}_{-0.81}$ \\\hline
$2~M_\odot$ + NICER + $\tilde{\Lambda}$ + $M_U$ & $12.56^{+0.66}_{-0.48}$ & $12.45^{+0.81}_{-0.67}$ \\\hline
$2~M_\odot$ + NICER + $\tilde{\Lambda}_\mathrm{BH}$ + $M_\mathrm{gap}$ & $12.82^{+0.58}_{-0.52}$ & $13.05^{+0.58}_{-0.56}$ \\\hline
$2~M_\odot$ + NICER + $\tilde{\Lambda}_\mathrm{BH}$ + HESS & $12.44^{+0.58}_{-0.39}$ & $12.43^{+0.69}_{-0.63}$ \\\hline
\end{tabular}
\captionof{table}{Median values of radii for $1.4~M_\odot$ and $2~M_\odot$ NSs for the different astrophysical constraints investigated in this paper. The errors represent the 90\% credible intervals. The pQCD minimal constraint is included in each case.}

\vspace{0.5cm}

\includegraphics[width=0.4\textwidth]{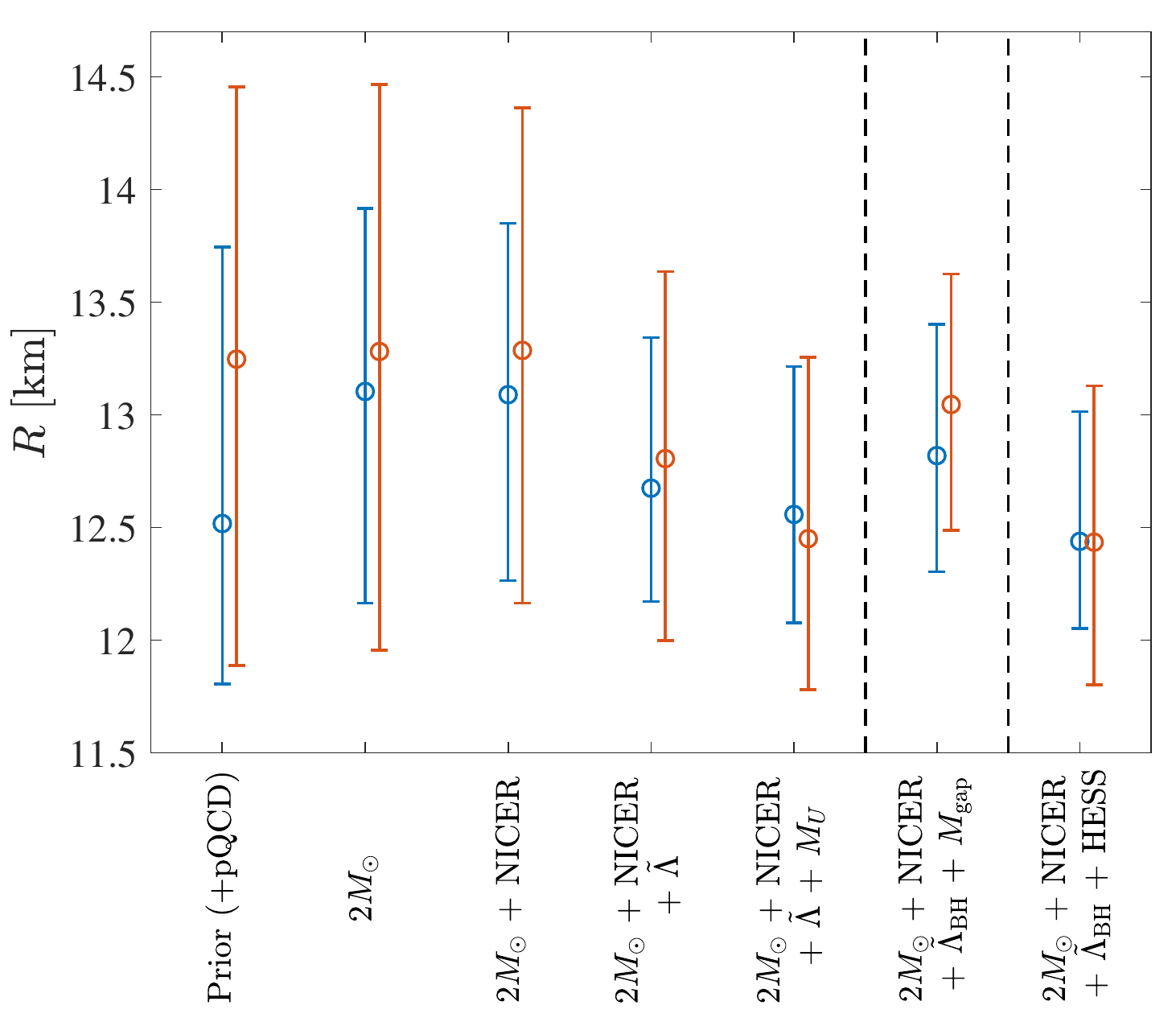}
\captionof{figure}{\label{fig:R_limits}Radius intervals for $1.4~M_\odot$ (blue) and $2~M_\odot$ (red) NSs. The circles represent the median values, while error bars correspond to the 90\% credible intervals. The vertical dashed lines separate alternative scenarios and additional, recent measurements. Data for these intervals can be found in Table~\ref{tab:R}.}

\end{figure}

We summarize our results for the calculated posterior radius distributions of $1.4~M_\odot$ and $2~M_\odot$ NSs for various astrophysical constraints in Table~\ref{tab:R} and Fig.~\ref{fig:R_limits}. Note, however, that these results should be taken with a grain of salt, since our prior was not preprocessed in order to acquire a uniform radius prior, which should be done in order to obtain meaningful results (see e.g. Ref.~\cite{Dietrich:2020efo}). Note also, that although it is not mentioned in the table and figure explicitly, our prior includes constraints from our constituent quark model implicitly, which restrict radii to values $R_{1.4}\gtrsim12$~km. The parameter sets corresponding to the maximum posterior probabilities for different sets of measurements are summarized in Table~\ref{tab:par}. Interestingly, all of the maximum posterior probability EoSs have $m_\sigma=290$~MeV and the SFHo EoS for the hadronic part.

\renewcommand*\arraystretch{1.3}

The prior and posterior marginalized one- and two-dimensional PDFs of the model parameters can be found in Fig.~\ref{fig:corner_all} in the Appendix. Looking at the prior we can see that there are very few stable and causal configurations for $m_\sigma=600$~MeV and there are almost none for $m_\sigma=700$~MeV. Even though the maximum posterior probabilities always correspond to $m_\sigma=290$~MeV, EoSs with higher sigma meson masses are not suppressed, and they are even enhanced for the hypermassive NS hypothesis scenario.

\begin{figure}[t]
  \centering
  \includegraphics[width=0.45\textwidth]{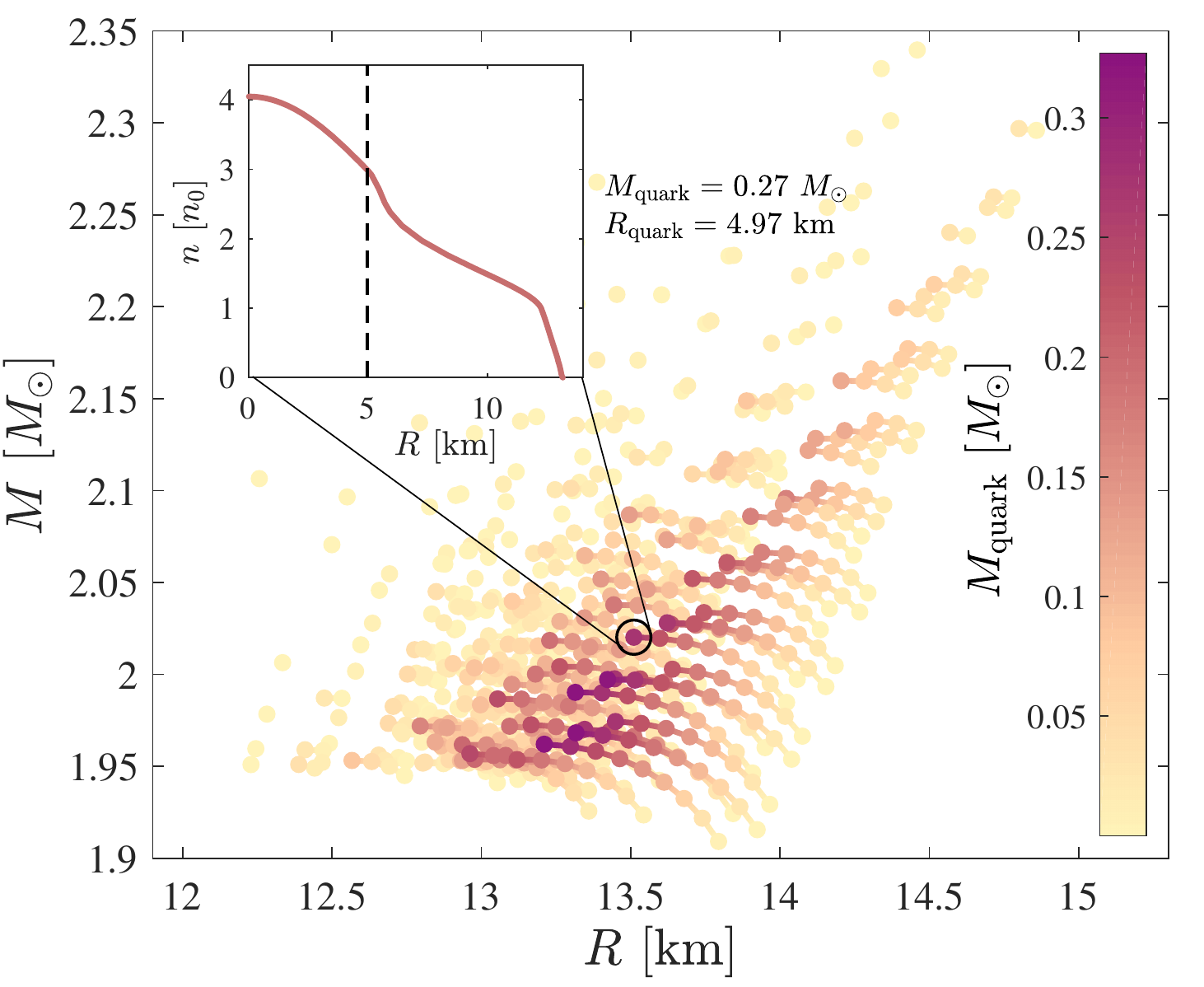}
  \caption{\label{fig:MR_quark}Masses of quark cores for hybrid stars that develop such a core. The inset shows the radial dependence of the baryon density inside one of the hybrid stars that have a sizeable quark core. The vertical dashed line represents the boundary between the quark core and the outer layers.}
\end{figure}

We also study the amount of quark matter contained in hybrid stars that have a quark core. We identify a NS core being made out of quark matter in case the baryon density rises above $n_{BU}=\bar{n}+\Gamma$. In addition to this condition, we also require the chiral phase transition to have occurred at that density. We define this by requiring the non-strange scalar condensate in our constituent quark model to drop below $10\%$ of its vacuum value. Even though this seems like an overly strict definition, this only excludes a few percent of NSs that would have a quark core by the first requirement only. In Fig.~\ref{fig:MR_quark} we show the amount of quark matter contained in hybrid stars that develop such a quark core. Here, no additional constraints were added on top of the minimal ones ($2~M_\odot$ and pQCD constraints). In many cases the quark core is light with masses of $M_\mathrm{quark}<0.05$~$M_\odot$. However, some hybrid stars can develop a sizeable quark core, with radii $R_\mathrm{quark}\gtrsim5$~km (see the inset in Fig.~\ref{fig:MR_quark}). More massive cores correspond to NS sequences with lower maximum masses with the most massive core having a mass $M_\mathrm{quark}\approx0.33$~$M_\odot$. This corresponds to a NS with a mass of $1.96~M_\odot$. A similar figure can be found in Ref.~\cite{Annala:2019puf} about the radii of quark cores in $2~M_\odot$ NSs.

\begin{table}[]
\begin{tabular}[c]{|c||c|}\cline{2-2}
\multicolumn{1}{c|}{} & $\vartheta_\mathrm{max}$ \\\hline
Measurement & EoS, $m_\sigma$, $g_V$, $\bar{n}$, $\Gamma$ \\\hline\hline
$2~M_\odot$ + NICER & SFHo, 290, 6.9, 4, 2.5 \\\hline
$2~M_\odot$ + NICER + $\tilde{\Lambda}$ & SFHo, 290, 6.5, 4.5, 2.75 \\\hline
$2~M_\odot$ + NICER + $\tilde{\Lambda}$ + $M_U$ & SFHo, 290, 3.1, 3.5, 2 \\\hline
$2~M_\odot$ + NICER + $\tilde{\Lambda}_\mathrm{BH}$ + $M_\mathrm{gap}$ & SFHo, 290, 4.9, 4.25, 2.75 \\\hline
$2~M_\odot$ + NICER + $\tilde{\Lambda}_\mathrm{BH}$ + HESS & SFHo, 290, 4.7, 4.25, 2.5 \\\hline
\end{tabular}
\caption{\label{tab:par}Parameter sets corresponding to EoSs with the maximum posterior probability for various measurements. $m_\sigma$ is given in MeV, while $\bar{n}$ and $\Gamma$ are given in units of the saturation density $n_0$.}
\end{table}

\begin{figure*}[tb!]
  \centering
  \includegraphics[width=0.32\textwidth]{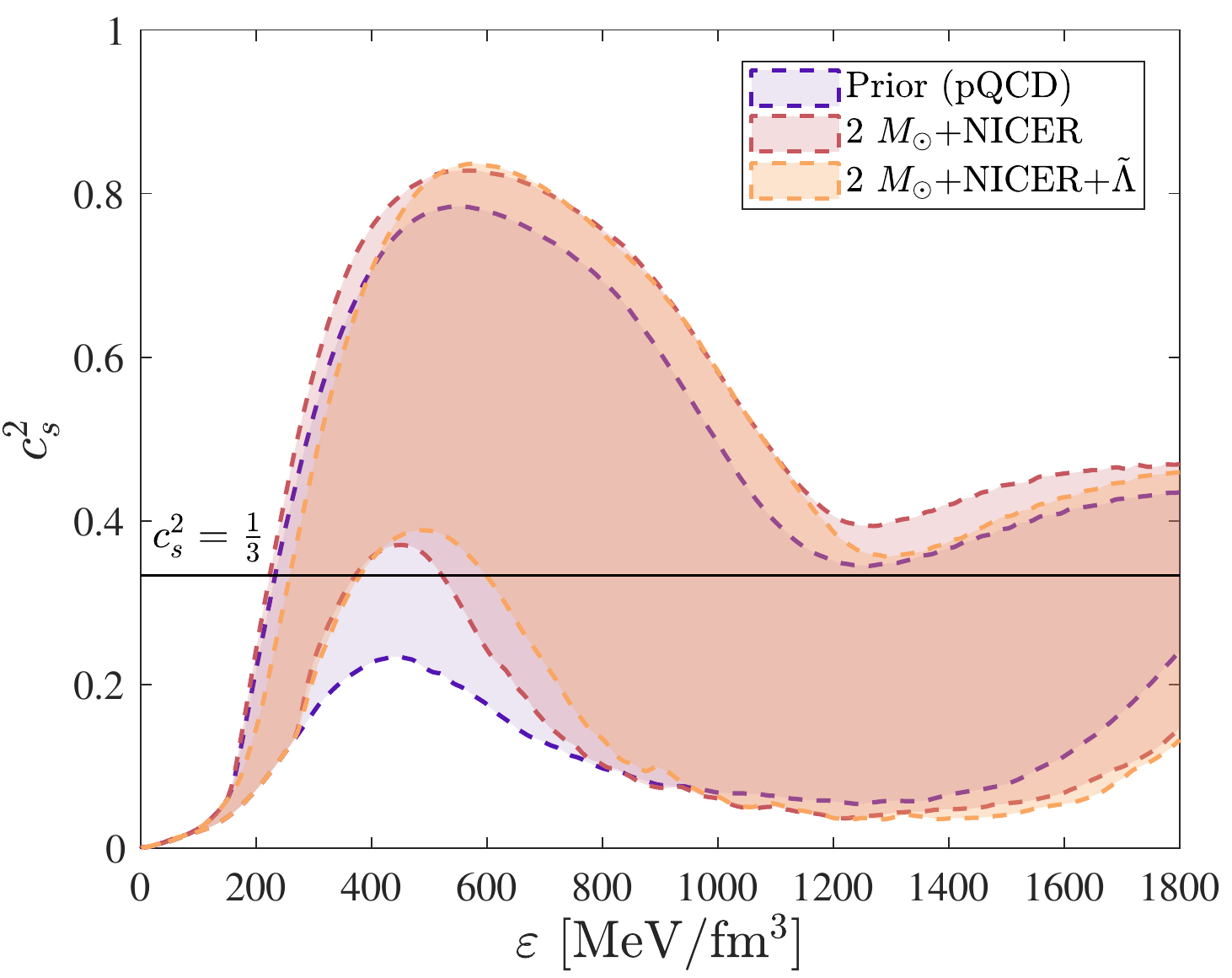}
  \includegraphics[width=0.32\textwidth]{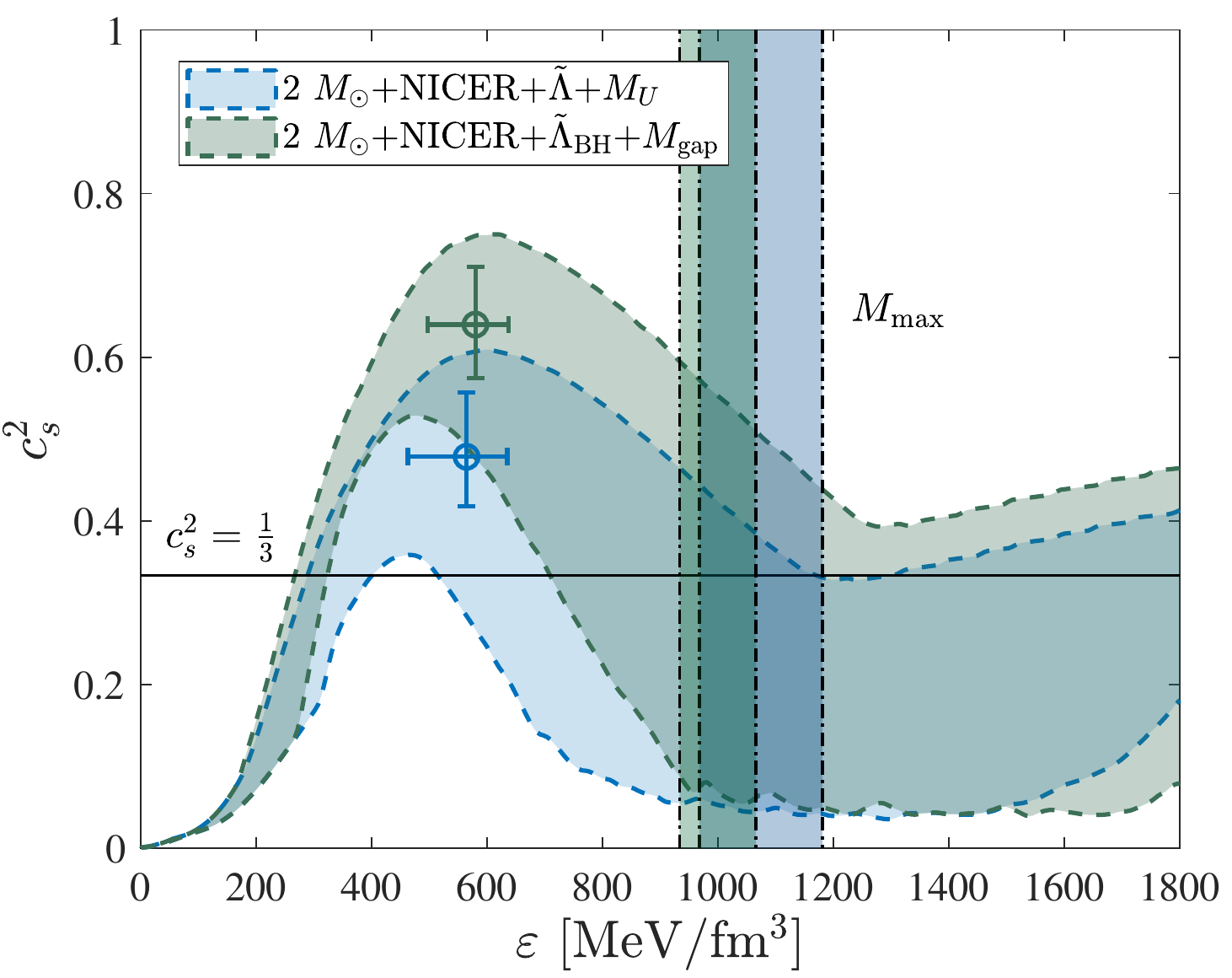}
  \includegraphics[width=0.32\textwidth]{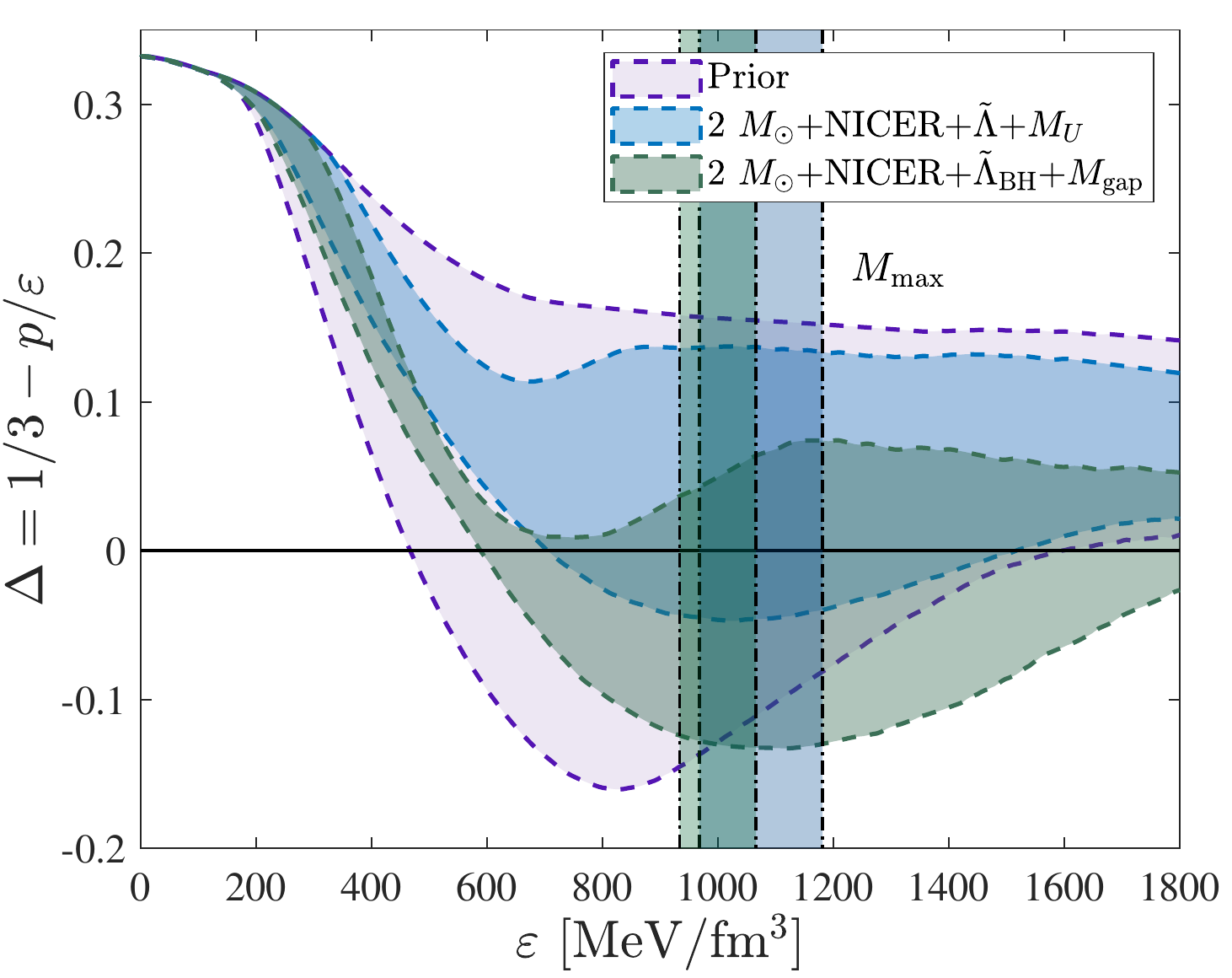}
  \caption{\label{fig:cs2}The $90\%$ credible intervals of the sound speed squared (left and middle) and the trace anomaly (right) as a function of energy density under various constraints, using the full ensemble of our EoSs. Different contours show results for the prior with the pQCD constraint (purple), the posterior with the NICER (red), and the NICER and tidal deformability measurements (yellow). In the middle the posteriors for the hypermassive NS hypothesis (blue) and for the alternative scenario where the mass-gap object in GW190814 is considered a NS (green) are shown. The position of the speed of sound peaks (circles in the middle panel) with the $68\%$ credible intervals denoted by the errorbars and the energy density at the center of maximally stable NSs (vertical lines in the middle and right panels, similarly with the $68\%$ credible intervals) is also displayed.}
\end{figure*}

Combining the complete ensemble of our EoSs we can investigate the effect of various constraints on the EoS itself. The left panel of Fig.~\ref{fig:cs2} shows the $90\%$ credible intervals of the sound speed squared as a function of energy density for various astrophysical constraints. The prior in itself exhibits a peak in the sound speed, which is located at $\varepsilon\approx400$~MeV/fm$^3$ for the lower bound. This can be attributed to the eLSM and the concatenation between the hadronic and quark EoSs. This peak translates to the lack of NSs with small radii in the $M-R$ plane in Fig.~\ref{fig:MR_Bayes}. However, this lower bound is slightly increased when we include the $2~M_\odot$, the NICER and tidal deformability measurements as well. This may be the result of the $2~M_\odot$ constraint, which requires a minimal stiffness for the EoSs, and also the NICER measurements, which disfavour small radii. The effect of the tidal deformability measurement of GW170817 is the reduction of the upper bound for energy densities $\varepsilon\lesssim500$~MeV/fm$^3$, which reduces the radii of $1.4~M_\odot$ NSs.

In the middle panel of Fig.~\ref{fig:cs2} we compare the two alternative scenarios with the hypermassive NS hypothesis and the mass-gap NS included, respectively. With the upper mass bound from the hypermassive NS hypothesis included in addition to the previous measurements, the upper bound of the sound speed squared is reduced from $\sim0.8$ to $\sim0.6$. The effect of astrophysical constraints on the upper bound above $\varepsilon\gtrsim1200$~MeV/fm$^3$ is minor, while it is negligible for the lower bound. In both cases an intermediate-density peak in the sound speed squared is preferred. The position of this peak is similar in the two cases, with $\varepsilon_\mathrm{p}=565^{+71}_{-102}$~MeV/fm$^3$ for the hypermassive NS and $\varepsilon_\mathrm{p}=580^{+57}_{-83}$~MeV/fm$^3$ for the mass-gap NS scenario. The values of the peaks are $0.48^{+0.08}_{-0.06}$ and $0.64^{+0.07}_{-0.07}$, respectively. These numbers correspond to medians and 68\% credible intervals. We note that Ref.~\cite{Marczenko:2022jhl} arrive at a similar result for the position of the peak in an independent analysis, however their median value of the peak is higher than ours. The energy density reached in the center of the maximally stable NSs in the two cases are $\varepsilon_\mathrm{max}=1089^{+93}_{-121}$~MeV/fm$^3$ and $\varepsilon_\mathrm{max}=1008^{+63}_{-71}$~MeV/fm$^3$, respectively, while for the baryon densities these values are $n_\mathrm{B,max}=0.89^{+0.07}_{-0.08}$~fm$^{-3}$ and $n_\mathrm{B,max}=0.80^{+0.04}_{-0.05}$~fm$^{-3}$. Note that the maximum energy density and baryon density is lower for the mass-gap NS case, in which the maximally massive NSs have larger masses. This follows from the fact that a larger peak in the speed of sound, which creates heavier NSs, will lead to an earlier destabilisation after the speed of sound drops to reach the pQCD limit at high densities.

One can interpret the peak in the speed of sound as a dominance of repulsive interactions, opposed to the finite temperature case, where it never exceeds the conformal limit \cite{HotQCD:2014kol,Kovacs:2016juc,Mykhaylova:2020pfk}. This might be interpreted as an indication of deconfinement, which might be linked to the percolation of hadrons. Using a simple model one can estimate the density at which percolation occurs (see e.g. Ref.~\cite{Marczenko:2022jhl}). In Ref.~\cite{Marczenko:2022jhl} the authors use an average mass radius of protons of $r_p=(0.80\pm0.05)$~fm (taken from Ref.~\cite{Wang:2022uch}) that is directly extracted from experimental data of $\phi$ meson photoproduction, yielding a density of $n_\mathrm{B,p}=0.57^{+0.12}_{-0.09}$~fm$^{-3}$. This density is obtained from percolation theory through the expression $n_\mathrm{B,p}=1.22/V_0$, where $V_0=4r_p^3\pi/3$ \cite{Castorina_2008,Braun-Munzinger:2014lba}. Ref.~\cite{Marczenko:2022jhl} arrive at a value $n_\mathrm{B,p}=0.56^{+0.09}_{-0.08}$~fm$^{-3}$, which is remarkably close to the estimated value of the percolation density. For the hypermassive NS and the mass-gap NS scenarios we calculate the density of the peak to be at $0.53^{+0.06}_{-0.08}$~fm$^{-3}$ and $0.53^{+0.04}_{-0.06}$~fm$^{-3}$, respectively, which are slightly lower but still consistent with the estimated density of percolation. It is worth noting that for the mass radius of proton there are several competing results on the market starting from as low as $r_p=(0.55\pm0.03)$~fm \cite{Kharzeev:2021qkd} up to $r_p=(0.86\pm0.08)$ \cite{Wang:2022vhr}. Moreover, beside the mass radius of protons there is also the charge radius of protons, which can be measured accurately by electron scattering experiments. Currently, there are two competing, non-overlapping values of $r_{\text{Ep}}=0.84$~fm and $0.88$~fm \cite{Tiesinga:2021myr,Mohr:2015ccw,Gao:2021sml}.
Thus, the size of the proton is still under debate, and can be as low as $0.55$~fm, which would give a much higher percolation density. Also note that the peak in the speed of sound might be a feature of the quark model itself, in case that the phase transition occurs at lower densities. Ref.~\cite{Ivanytskyi:2022bjc} finds that a low onset of the phase transition and a color superconducting quark model is consistent with the speed of sound peak from astrophysical data.

In the right panel of Fig.~\ref{fig:cs2} we also show the limits for the trace anomaly, defined as
\begin{equation}
    \Delta = \frac{1}{3}-\frac{p}{\varepsilon}.
\end{equation}
This was recently proposed as a measure of conformality \cite{Fujimoto:2022ohj}. As we approach the conformal limit the value of $\Delta$ will tend to zero. Similar to the results of Ref.~\cite{Marczenko:2022jhl} we find that the value of $\Delta$ approaches zero from above in the hypermassive NS scenario for large $\varepsilon$ values. This is to be compared to the mass-gap NS scenario, where the conformal limit can be reached both from above and below, in fact, quite remarkably, at $\varepsilon\approx800$~MeV/fm$^3$ a negative value for $\Delta$ becomes highly favoured.

\begin{figure}[t]
  \centering
  \includegraphics[width=0.45\textwidth]{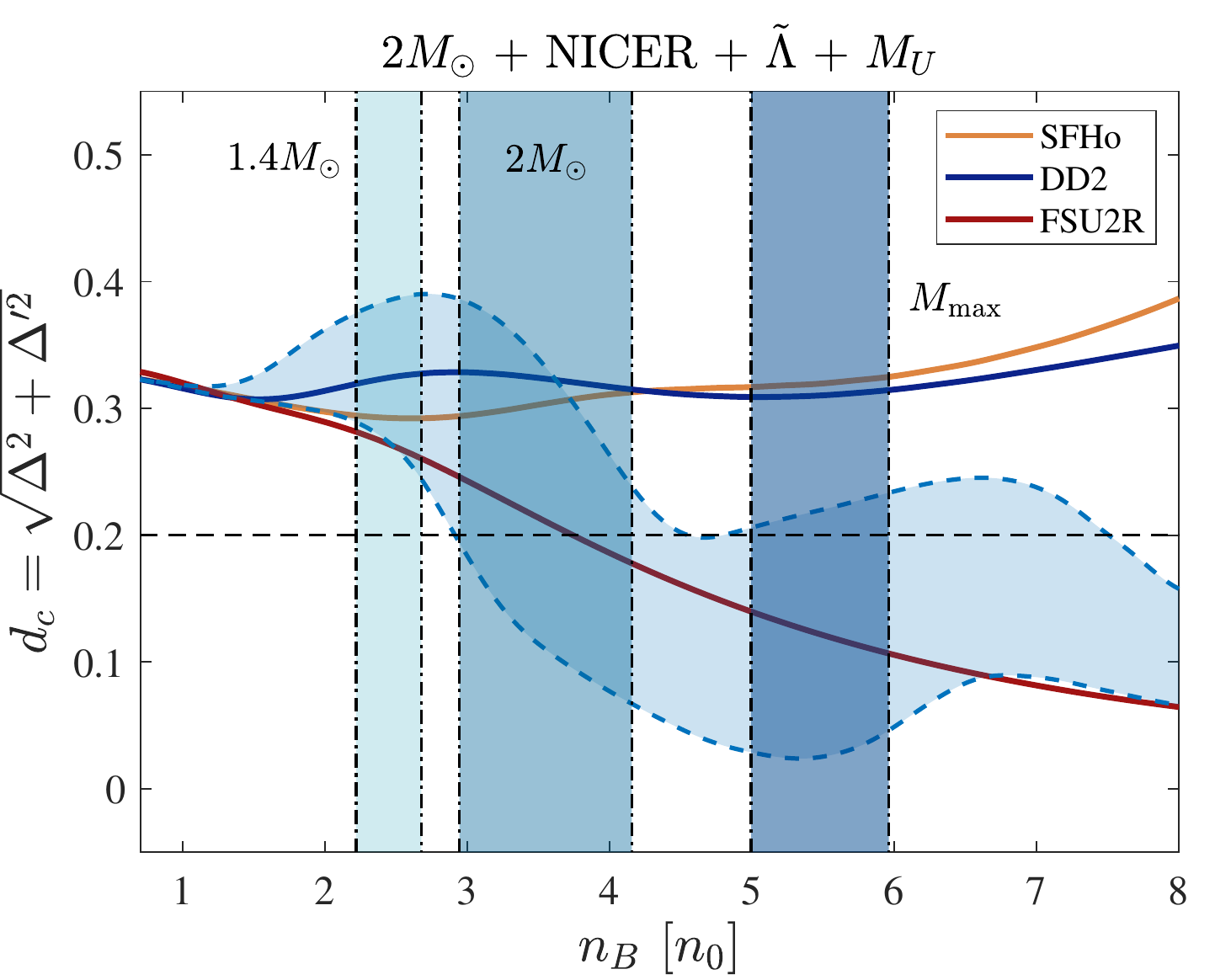}
  \includegraphics[width=0.45\textwidth]{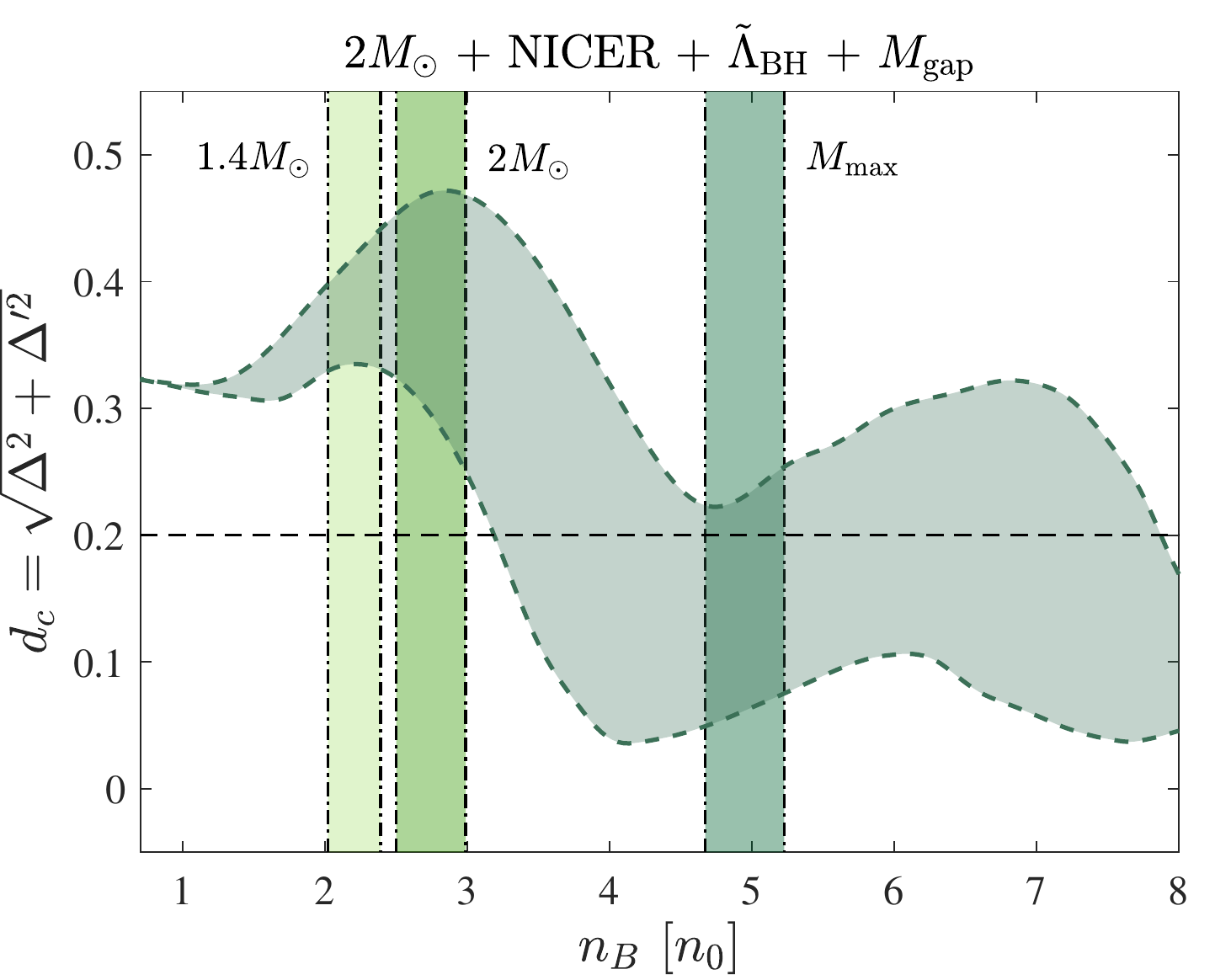}
  \caption{\label{fig:dc}$90\%$ credible intervals for the measure of conformality, $d_c$, as a function of baryon density, for the hypermassive NS scenario (top) and the mass-gap object scenario (bottom) using the full ensemble of our EoS. Vertical ares denote the $90\%$ intervals for the central densities of $1.4~M_\odot$,  $2~M_\odot$ and maximally stable NSs, with increasingly dark colors. Solid lines show the density dependence of $d_c$ for hadronic EoSs.}
\end{figure}

Yet another recent study \cite{Annala:2023cwx} proposes another quantity to measure conformality. They combine the trace anomaly with its logarithmic derivative, and define $d_c$ as
\begin{equation}
    d_c = \sqrt{\Delta^2 + \Delta'^2} \: ,
\end{equation}
where
\begin{equation}
    \Delta' = \frac{\mathrm{d}\Delta}{\mathrm{d}\:\mathrm{ln}\varepsilon} = c_s^2\left( \frac{1}{\gamma} - 1 \right) \: ,
\end{equation}
with $\gamma = \mathrm{d}\:\mathrm{ln}\:p/\mathrm{d}\:\mathrm{ln}\:\varepsilon$ being the polytropic index. Based on the observation that hadronic EoSs can be separated from conformal systems using this quantity and that for a first-order phase transition $d_c$ is bounded from below by $d_c\gtrsim0.236$ at the phase transition, they propose that conformal matter can be identified by the criterion $d_c<0.2$. We should add, however, that some hadronic EoSs do not comply with this conjecture. For example, in Fig.~\ref{fig:dc} we show the density dependence of $d_c$ for the FSU2R EoS \cite{Negreiros:2018cho} downloaded from the CompOSE database. This is a purely nucleonic EoS containing an $\omega^4$ interaction term, and yet it behaves similarly to hybrid EoSs, with $d_c<0.2$ for $n_B>4n_0$. Therefore, this criterion of conformality should be further studied.

We investigate this measure in Fig.~\ref{fig:dc} for the two alternative scenarios with the hypermassive NS hypothesis and the mass-gap object identified as a NS. In both cases $d_c$ drops below $0.2$ before reaching the density at the center of maximally stable NSs. However, for the hypermassive NS scenario, this happens at lower densities, at around $3-4n_0$, while for the mass-gap object scenario, this interval is at $\sim3.5-4.5n_0$. Interestingly, for the hypermassive NS scenario, NSs with $\sim2~M_\odot$ might already possess conformal matter, while for the other case, matter inside $2~M_\odot$ NSs is far from being conformal.

\section{Conclusion}
\label{sec:conclusion}

In this paper we have investigated what can be inferred from astrophysical observations about the properties of quark matter inside NSs and the phase transition between the hadronic and quark phases. For this purpose we have utilized the (axial)vector meson extended linear sigma-model to describe quark matter at high densities, the SFHo and DD2 models as hadronic EoSs representing soft and stiff hadronic models, respectively. To transition from the hadronic to the quark model we have used a general polynomial concatenation, the parameters of which can be varied to create phase transitions at different densities. The whole parameter space with 4 parameters (2 from the concatenation, and 2 from the constituent quark model) was studied by applying the full PDFs from recent astrophysical observations.

First of all, we have shown that there is a tight correlation between the parameters of the constituent quark model and the maximum mass attainable by heavy NSs described by that model, even though many of the maximum mass NSs do not have pure quark matter in their cores. Hence, some properties of quark matter at high densities might be inferred by only gaining information about the intermediate-density region, and therefore determining the maximum mass of NSs might be used to deduce information about the properties of strongly interacting matter at high densities.

In our Bayesian analysis, we have investigated the effect of different astrophysical measurements on mass--radius curves, the radius distribution of NSs with specific masses, and on the posterior probabilities of different parameter sets. In addition to the lower mass limit from $2~M_\odot$ stars and constraints from pQCD, we have also considered the two NICER measurements, and the tidal deformability data obtained from GW170817. Moreover, we have studied the effect of additional constraints, such as the upper mass bound inferred from the hypermassive NS hypothesis, interpreting the mass-gap object in GW190814 as a very massive NS, or the mass--radius data obtained from the light compact object in HESS J1731-34. We divided our analysis into two parts. First, we investigated the results in a reduced parameter space with $m_\sigma=290$~MeV and the SFHo hadronic EoS. We also calculated the results for reduced parameter spaces with other parameters. Then, we combined the full ensemble of $\sim18\:000$ EoSs to deduce more general conclusions about the EoS of strongly interacting matter.

We have shown that the $90\%$ credible regions on the mass--radius diagram obtained by using the complete observational data of GW170817 differ slightly from those originating from a sharp cut-off using the $90\%$ bound on the parameters of the binary corresponding to GW170817. This was also discussed in Ref.~\cite{Jiang:2022tps} and suggests the use of the full PDFs from astrophysical measurements for more precise predictions.

We have also found that the maximum mass of hybrid stars with a pure quark core is below $<2.35~M_\odot$. This is caused by the fact that there is a successive stiffening and softening in the intermediate-density region of our EoSs and in order to reach the density of our quark model, the NS sequence needs to go through the soft region without becoming unstable. Interestingly, Ref.~\cite{Annala:2019puf} find a similar upper bound for such hybrid stars in a completely independent analysis. Further constraints narrow down this region even more, leaving only a small space for hybrid stars with a pure quark core for EoSs with a soft hadronic part and none for ones with a stiff hadronic part. This is also in line with the findings of some other studies \cite{Weissenborn:2011qu,Nandi:2017rhy,Nandi:2020luz,Shirke:2022tta}, which also suggest the possible existence of pure quark matter inside massive NSs, although in a restricted parameter region, highly dependent on the hadronic EoS.

Additionally, we have also shown how the parameters of the hadron--quark concatenation are affected by the various astrophysical constraints. For the two main scenarios considered in this paper, we have found that the parameter encoding the central density of the phase is above $\bar{n}>3n_0$, and that the appearance of pure quark matter at densities below $\sim4n_0$ is disfavoured in both scenarios, using our parameterization.

Even though the presence of pure quark matter is restricted to a limited region in the mass--radius diagram, we find that a peak in the speed of sound is preferred by astrophysical observations, which might be interpreted as a consequence of reaching percolation densities. We have found that this peak is at $\varepsilon_\mathrm{p}=565^{+71}_{-102}$~MeV/fm$^3$ and $580^{+57}_{-83}$~MeV/fm$^3$ for the hypermassive NS and the mass-gap NS cases, respectively, or at $n_\mathrm{B,p}=0.53^{+0.06}_{-0.08}$~fm$^{-3}$ and $0.53^{+0.04}_{-0.06}$~fm$^{-3}$, regarding baryon densities. This is consistent with the findings of Ref.~\cite{Marczenko:2022jhl} and the prediction of simple percolation models. Ref.~\cite{Han:2022rug} also reports a value of $\sim3.3n_0$ for the location of the peak. We have also shown the dependence of the $\Delta$ trace anomaly on the energy density and found that in the mass-gap NS scenario a negative $\Delta$ is preferred at some energy density region. Additionally, we have investigated the baryon density dependence of $d_c$, proposed as another measure of conformality by Ref.~\cite{Annala:2023cwx}, and have found that the conformal region is reached in the centers of massive NSs for both scenarios. We have also found that in the hypermassive NS scenario, the conformal region can be alread reached by $2~M_\odot$ NSs.

\begin{acknowledgements}
The authors would like to thank Micha\l{} Marczenko for their useful advices on Fig.~\ref{fig:cs2} and the discussion about the speed of sound peak. They would also like to thank the Referee for their help in improving the manuscript considerably, and especially for their remarks on the hadronic EoSs that do not fulfill the $d_c>0.2$ conjecture. J. T.,  P. K. and G. W. acknowledge support by the National Research, Development and Innovation (NRDI) fund of Hungary, financed under the FK\_19 funding scheme, Project No. FK 131982 and under the K\_21 funding scheme, Project No. K 138277. J. T. was supported by the ÚNKP-22-3 New National Excellence Program of the Ministry for Culture and Innovation from the source of the NRDI fund. J. S. acknowledges support by the Deutsche Forschungsgemeinschaft (DFG, German Research Foundation) through the CRC-TR 211 'Strong-interaction matter under extreme conditions'– project number 315477589 – TRR 211. We also acknowledge support for the computational resources provided by the Wigner Scientific Computational Laboratory (WSCLAB).
\end{acknowledgements}

\bibliography{eLSM_bayes}{}

\appendix

\section{Tables and figures}
\label{App:tables_figs}

This section contains tables for the radius bounds of $1.4~M_\odot$ and $2~M_\odot$ NSs, as well as for the parameter sets corresponding to the maximum posterior probability cases, using different sets of EoSs with specific values for the sigma meson mass and the hadronic EoS. We also included figures about marginalized prior and posterior PDFs in the parameter space of our model, for different astrophysical measurements.

\begin{table*}
\begin{NiceTabular}{|c|c|c|c|c|c|c|}[hvlines,corners=NW]
 & \multicolumn{2}{c|}{$m_\sigma=290$ MeV} & \multicolumn{2}{c|}{$m_\sigma=400$ MeV} & \multicolumn{2}{c|}{$m_\sigma=500$ MeV}\\\hline
Measurement & $R_{1.4}$ [km] & $R_{2.0}$ [km] & $R_{1.4}$ [km] & $R_{2.0}$ [km] & $R_{1.4}$ [km] & $R_{2.0}$ [km] \\\hline\hline
Prior (+pQCD) & $12.13^{+1.49}_{-0.40}$ & $13.00^{+1.46}_{-1.25}$ & $12.35^{+1.29}_{-0.52}$ & $13.06^{+1.30}_{-1.22}$ & $12.69^{+0.94}_{-0.70}$ & $13.16^{+1.09}_{-1.22}$\\\hline
$2~M_\odot$ & $12.82^{+1.05}_{-0.76}$ & $13.02^{+1.44}_{-1.21}$ & $12.91^{+0.90}_{-0.76}$ & $13.09^{+1.28}_{-1.19}$ & $13.01^{+0.71}_{-0.68}$ & $13.20^{+1.07}_{-1.27}$ \\\hline
$2~M_\odot$ + NICER & $12.88^{+0.93}_{-0.72}$ & $13.09^{+1.26}_{-1.05}$ & $12.94^{+0.82}_{-0.69}$ & $13.14^{+1.15}_{-1.03}$ & $13.03^{+0.67}_{-0.60}$ & $13.23^{+0.98}_{-1.01}$ \\\hline
$2~M_\odot$ + NICER + $\tilde{\Lambda}$ & $12.59^{+0.61}_{-0.49}$ & $12.73^{+0.84}_{-0.79}$ & $12.63^{+0.59}_{-0.45}$ & $12.75^{+0.82}_{-0.76}$ & $12.76^{+0.52}_{-0.43}$ & $12.85^{+0.74}_{-0.79}$ \\\hline
$2~M_\odot$ + NICER + $\tilde{\Lambda}$ + $M_U$ & $12.47^{+0.59}_{-0.45}$ & $12.38^{+0.81}_{-0.67}$ & $12.51^{+0.57}_{-0.42}$ & $12.42^{+0.79}_{-0.64}$ & $12.61^{+0.53}_{-0.37}$ & $12.44^{+0.74}_{-0.58}$ \\\hline
$2~M_\odot$ + NICER + $\tilde{\Lambda}_\mathrm{BH}$ + $M_\mathrm{gap}$ & $12.70^{+0.57}_{-0.43}$ & $12.95^{+0.59}_{-0.51}$ & $12.78^{+0.54}_{-0.47}$ & $12.98^{+0.58}_{-0.50}$ & $12.86^{+0.48}_{-0.38}$ & $13.10^{+0.52}_{-0.43}$ \\\hline
$2~M_\odot$ + NICER + $\tilde{\Lambda}_\mathrm{BH}$ + HESS & $12.38^{+0.56}_{-0.36}$ & $12.40^{+0.67}_{-0.64}$ & $12.44^{+0.52}_{-0.37}$ & $12.44^{+0.68}_{-0.62}$ & $12.57^{+0.52}_{-0.33}$ & $12.51^{+0.66}_{-0.62}$ \\\hline
\end{NiceTabular}
\caption{\label{tab:RSFHo}Median and $90\%$ bounds for the radii of $1.4~M_\odot$ and $2~M_\odot$ NSs using the SFHo hadronic EoS and various values for the sigma meson masses. Results are shown for different combinations of astrophysical measurements. Results for $m_\sigma=600$~MeV and $m_\sigma=700$~MeV were omitted due to the low statistics of stable and causal EoSs with these parameters.}
\end{table*}

\begin{table*}
\begin{NiceTabular}{|c|c|c|c|c|c|c|}[hvlines,corners=NW]
 & \multicolumn{2}{c|}{$m_\sigma=290$ MeV} & \multicolumn{2}{c|}{$m_\sigma=400$ MeV} & \multicolumn{2}{c|}{$m_\sigma=500$ MeV}\\\hline
Measurement & $R_{1.4}$ [km] & $R_{2.0}$ [km] & $R_{1.4}$ [km] & $R_{2.0}$ [km] & $R_{1.4}$ [km] & $R_{2.0}$ [km] \\\hline\hline
Prior (+pQCD) & $13.14^{+0.89}_{-1.05}$ & $13.70^{+1.05}_{-1.17}$ & $13.16^{+0.82}_{-0.61}$ & $13.67^{+0.96}_{-1.14}$ & $13.23^{+0.69}_{-0.45}$ & $13.58^{+0.93}_{-1.11}$\\\hline
$2~M_\odot$ & $13.48^{+0.69}_{-0.39}$ & $13.72^{+1.04}_{-1.10}$ & $13.47^{+0.62}_{-0.38}$ & $13.69^{+0.95}_{-1.08}$ & $13.43^{+0.57}_{-0.37}$ & $13.60^{+0.92}_{-1.05}$ \\\hline
$2~M_\odot$ + NICER & $13.44^{+0.65}_{-0.33}$ & $13.67^{+0.96}_{-0.93}$ & $13.44^{+0.59}_{-0.33}$ & $13.65^{+0.90}_{-0.92}$ & $13.43^{+0.53}_{-0.33}$ & $13.60^{+0.84}_{-0.92}$ \\\hline
$2~M_\odot$ + NICER + $\tilde{\Lambda}$ & $13.26^{+0.29}_{-0.22}$ & $13.30^{+0.56}_{-0.80}$ & $13.27^{+0.29}_{-0.22}$ & $13.29^{+0.55}_{-0.78}$ & $13.26^{+0.29}_{-0.22}$ & $13.23^{+0.58}_{-0.74}$ \\\hline
$2~M_\odot$ + NICER + $\tilde{\Lambda}$ + $M_U$ & $13.17^{+0.27}_{-0.20}$ & $12.95^{+0.58}_{-0.71}$ & $13.17^{+0.27}_{-0.19}$ & $12.98^{+0.55}_{-0.69}$ & $13.17^{+0.26}_{-0.17}$ & $12.91^{+0.50}_{-0.59}$ \\\hline
\end{NiceTabular}
\caption{\label{tab:RDD2}Median and $90\%$ bounds for the radii of $1.4~M_\odot$ and $2~M_\odot$ NSs using the DD2 hadronic EoS and various values for the sigma meson masses. Results are shown for different combinations of astrophysical measurements. Results for $m_\sigma=600$~MeV and $m_\sigma=700$~MeV and for some combinations of astrophysical measurements were omitted due to the low statistics of stable and causal EoSs with these parameters.}
\end{table*}

\begin{table*}
\begin{NiceTabular}{|c|c|c|c|}[hvlines,corners=NW]
& \multicolumn{3}{c|}{$\vartheta_\mathrm{max}$: ($g_V$, $\bar{n}$ [$n_0$], $\Gamma$ [$n_0$])} \\\hline
Measurement & $m_\sigma=290$ MeV & $m_\sigma=400$ MeV & $m_\sigma=500$ MeV \\\hline\hline
$2~M_\odot$ + NICER &  6.9, 4, 2.5 & 6.3, 5, 3.75 & 6, 5, 3.75 \\\hline
$2~M_\odot$ + NICER + $\tilde{\Lambda}$ & 6.5, 4.5, 2.75 & 5.4, 5, 3.5 & 5, 5, 3.5 \\\hline
$2~M_\odot$ + NICER + $\tilde{\Lambda}$ + $M_U$ & 3.1, 3.5, 2 & 2.6, 3.5, 2 & 2.6, 4.5, 3.25 \\\hline
$2~M_\odot$ + NICER + $\tilde{\Lambda}_\mathrm{BH}$ + $M_\mathrm{gap}$ & 4.9, 4.25, 2.75 & 4.8, 4.5, 3 & 4.5, 4.5, 3.25 \\\hline
$2~M_\odot$ + NICER + $\tilde{\Lambda}_\mathrm{BH}$ + HESS & 4.7, 4.25, 2.5 & 4.8, 4.75, 3 & 4.3, 5, 3.5 \\\hline
\end{NiceTabular}
\caption{\label{tab:parSFHo}Parameters of EoSs with the maximum posterior probability for different combinations of astrophysical measurements, using the SFHo hadronic EoS. Results for $m_\sigma=600$~MeV and $m_\sigma=700$~MeV were omitted due to the low statistics of stable and causal EoSs with these parameters.}
\end{table*}

\begin{table*}
\begin{NiceTabular}{|c|c|c|c|}[hvlines,corners=NW]
& \multicolumn{3}{c|}{$\vartheta_\mathrm{max}$: ($g_V$, $\bar{n}$ [$n_0$], $\Gamma$ [$n_0$])} \\\hline
Measurement & $m_\sigma=290$ MeV & $m_\sigma=400$ MeV & $m_\sigma=500$ MeV \\\hline\hline
$2~M_\odot$ + NICER & 6.7, 4.75, 2.75 & 6.1, 4.75, 2.75 & 5.5, 5, 3 \\\hline
$2~M_\odot$ + NICER + $\tilde{\Lambda}$ & 4.4, 5, 3.75 & 4.1, 5, 3.75 & 5.5, 5, 3 \\\hline
$2~M_\odot$ + NICER + $\tilde{\Lambda}$ + $M_U$ & 3.8, 5, 4 & 3.5, 5, 4 & 3.1, 5, 4 \\\hline
\end{NiceTabular}
\caption{\label{tab:parDD2}Parameters of EoSs with the maximum posterior probability for different combinations of astrophysical measurements, using the DD2 hadronic EoS. Results for $m_\sigma=600$~MeV and $m_\sigma=700$~MeV and for some combinations of astrophysical measurements were omitted due to the low statistics of stable and causal EoSs with these parameters.}
\end{table*}

\begin{figure*}[htbp]
  \centering
  \includegraphics[width=0.45\textwidth]{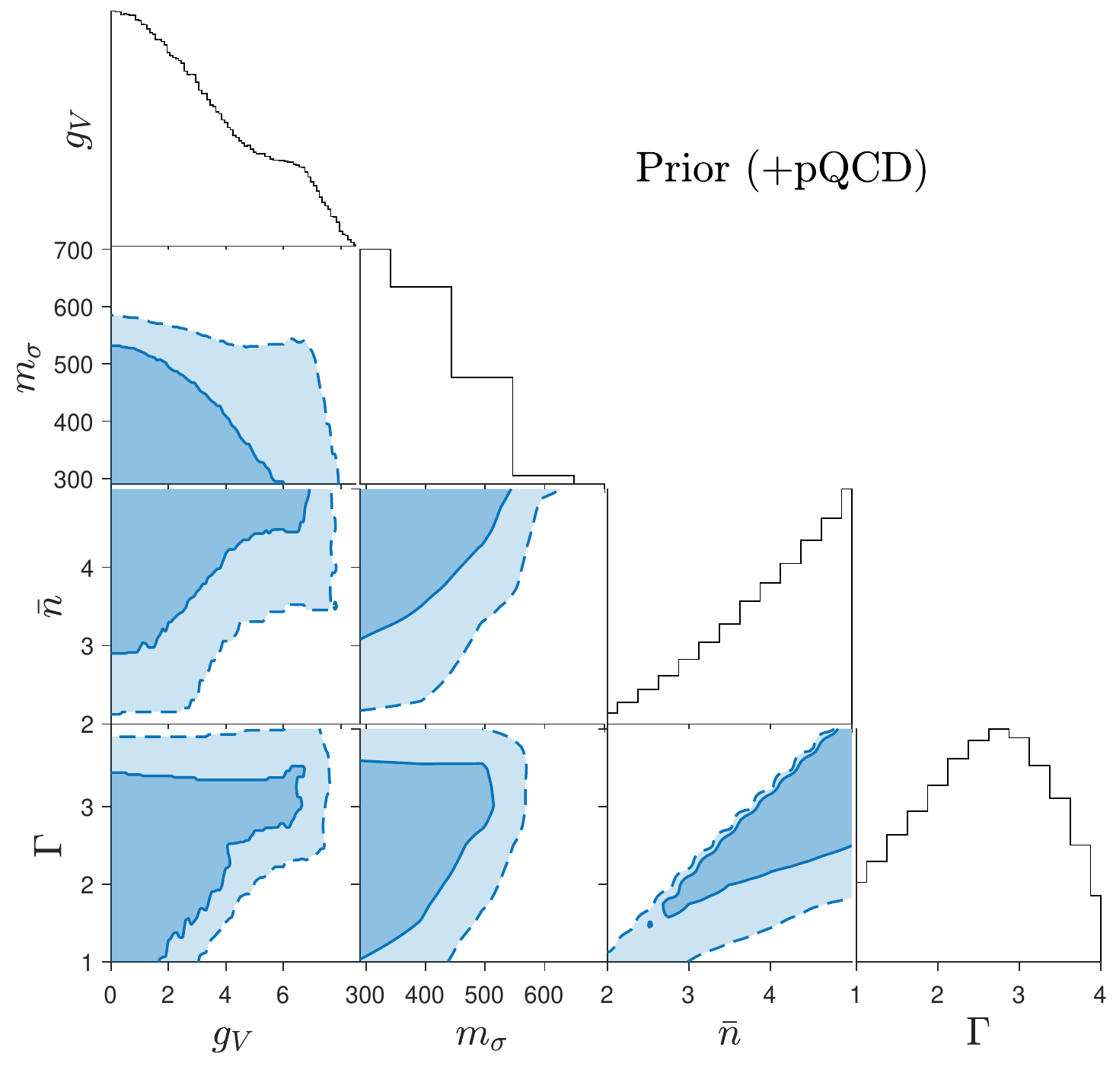}
  \includegraphics[width=0.45\textwidth]{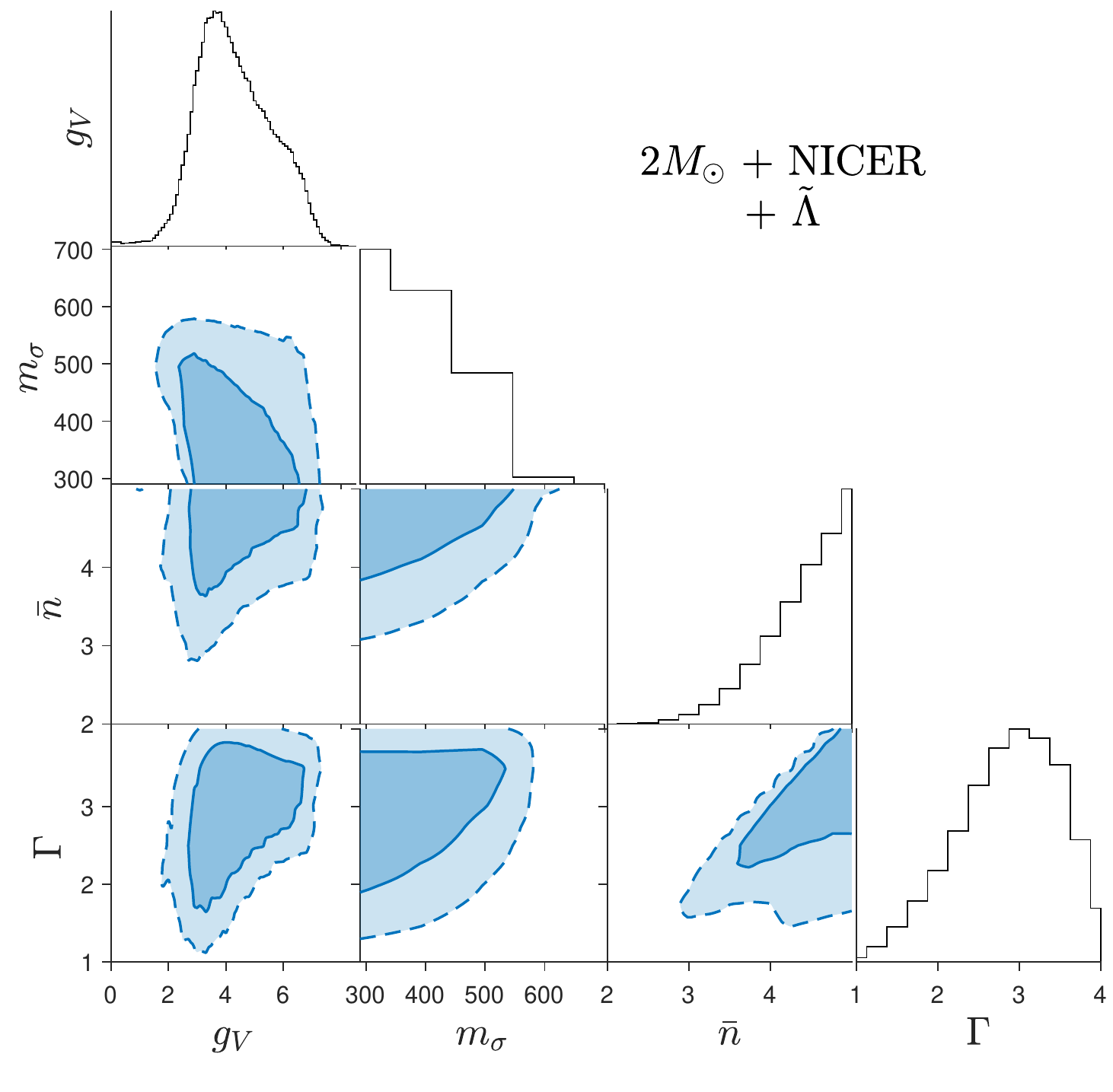}
  \includegraphics[width=0.45\textwidth]{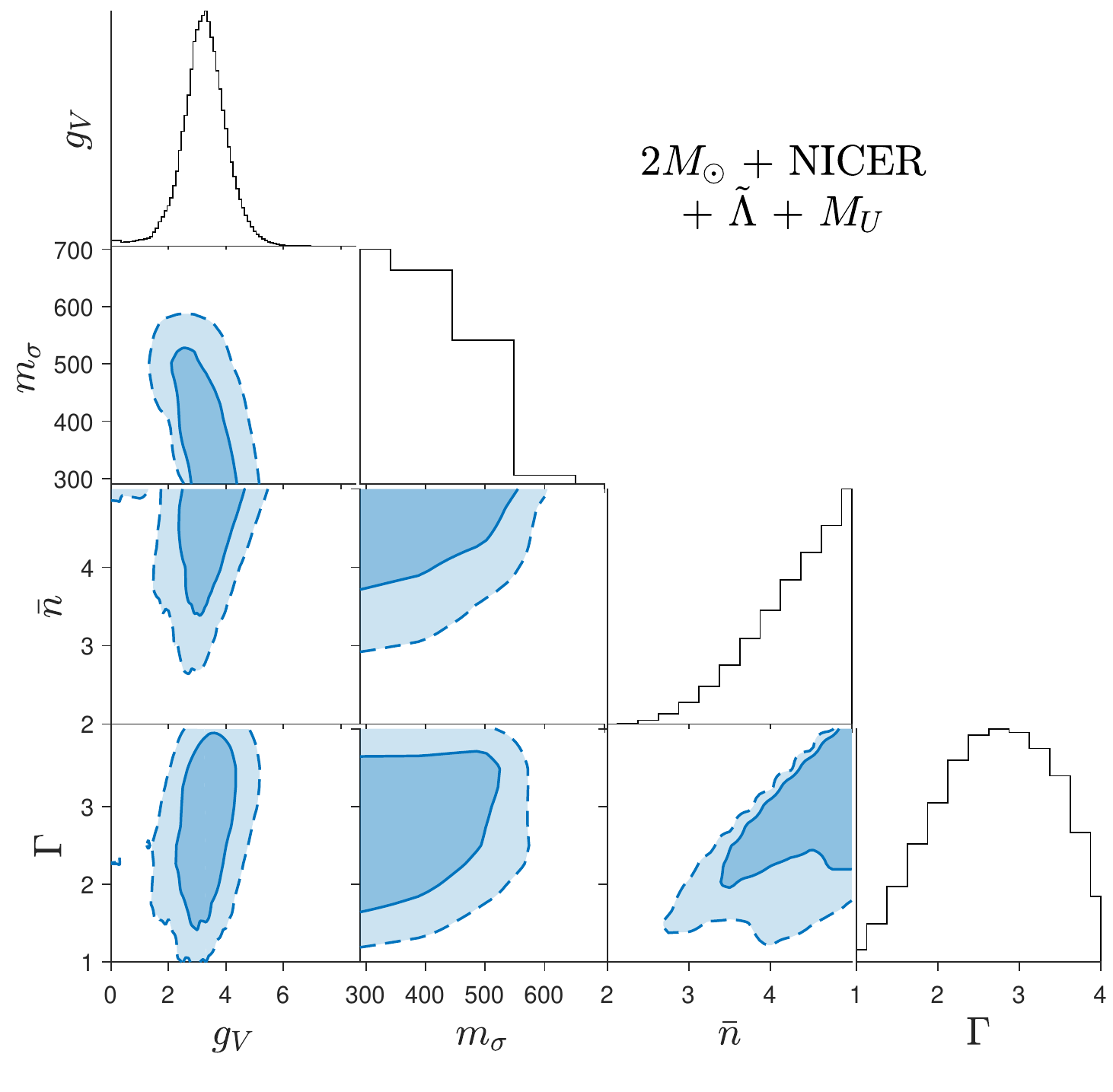}
  \includegraphics[width=0.45\textwidth]{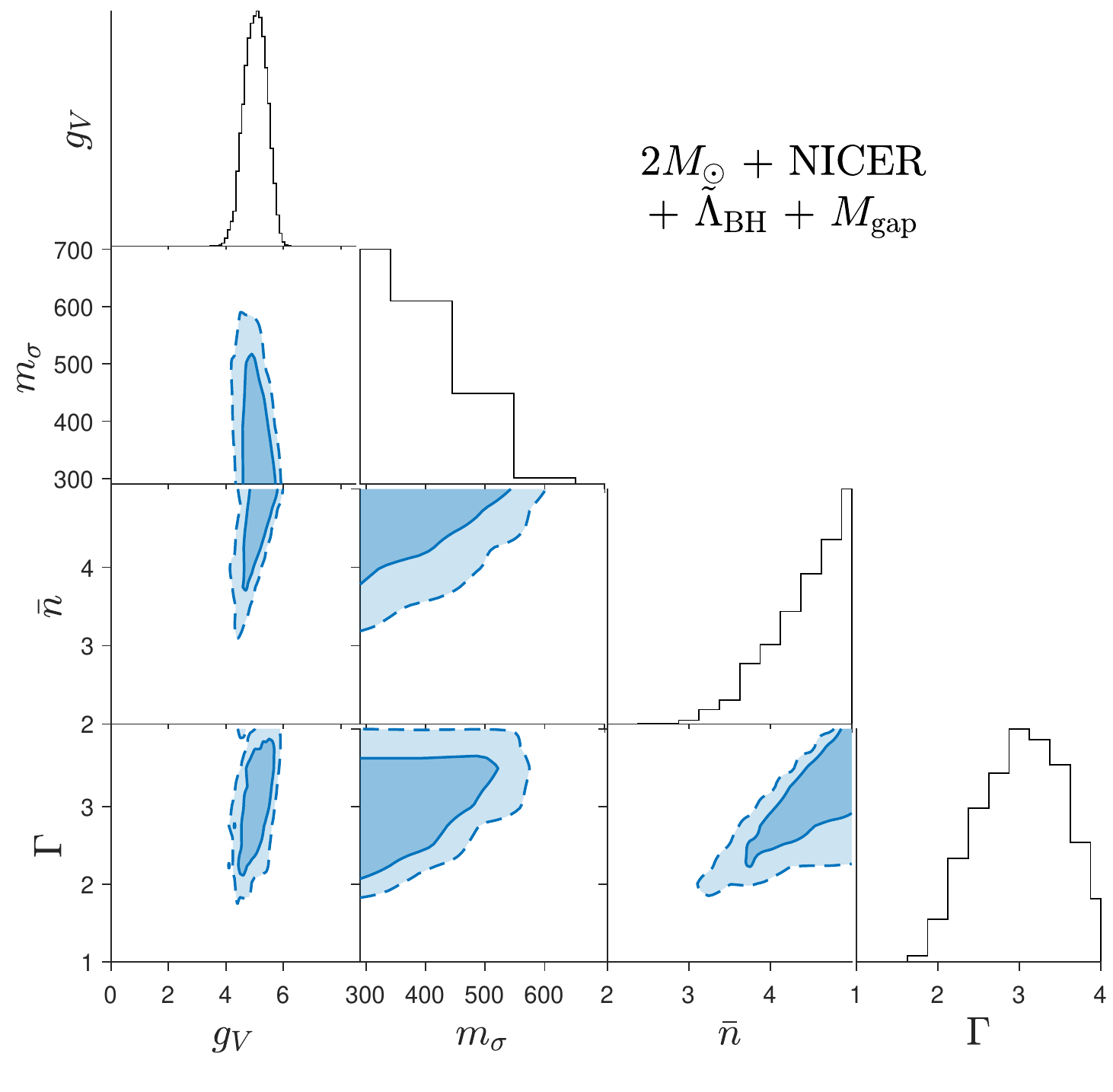}
  \caption{\label{fig:corner_all}Marginalized one- and two-dimensional priors and posteriors for selected combinations of astrophysical measurements. The PDFs were made using our complete EoS ensemble.}
\end{figure*}

\end{document}